\documentclass[11pt]{article}
\pdfoutput=1
\usepackage{graphicx}
\usepackage{amsfonts}
\usepackage{hyperref}
\usepackage{epsfig}
\usepackage{amsmath}
\usepackage{amsthm}
\usepackage{amssymb}
\usepackage{multirow}
\usepackage{color}
\usepackage[nosort]{cite}
\usepackage{hyperref}
\usepackage[fontsize=11.35pt]{scrextend}
\usepackage{braket,mathrsfs,slashed}
\usepackage{float}
\usepackage{setspace}
\usepackage[caption=false]{subfig}
\newlength{\abstractwidth}
\newcommand{\be}{\begin{equation}}
\newcommand{\ee}{\end{equation}}

\renewcommand{\title}[1]{\vbox{\center\bf{\Large{#1}}}\vspace{5mm}}
\renewcommand{\author}[1]{\vbox{\center#1}\vspace{5mm}}
\newcommand{\address}[1]{\vbox{\center\em#1}}
\newcommand{\email}[1]{\vbox{\center\tt#1}\vspace{5mm}}

\usepackage{dsfont}

\usepackage[utf8]{inputenc}
\usepackage{rotating}

\usepackage{dsfont}
\usepackage[paper=a4paper,margin=1.5cm]{geometry}
\usepackage[utf8]{inputenc}
\usepackage{rotating}
\usepackage{soul}

\usepackage{mathrsfs} %beautiful calligraphy
\usepackage{bbm}
\usepackage{etoolbox} % for \appto
\usepackage{multirow}
\renewcommand\[{\begin{equation}}
\renewcommand\]{\end{equation}}

\newcommand{\ba}{\begin{eqnarray}}
\newcommand{\ea}{\end{eqnarray}}

\hypersetup{pdfstartview=FitV,colorlinks=true,linkcolor=midblue,citecolor=midblue,filecolor=midblue,urlcolor=midblue}

\definecolor{midblue}{rgb}{0,0,0.6}

\begin{document}
	
\newgeometry{top=3.1cm,bottom=3.1cm,right=2.4cm,left=2.4cm}

	\begin{titlepage}
	\begin{center}
		\hfill \\
		%\hfill \\
		\vskip 0.5cm

		\title{Ghosts versus Unstable Particles\\[3.5mm] in Quantum Field Theory}

		\author{\large Luca Buoninfante}
		
		\address{Departamento de Física de Partículas,\\ Instituto Galego de Física de Altas Enerxías (IGFAE),\\  Universidade de Santiago de Compostela, Spain}
		\email{\rm \href{mailto:luca.buoninfante@usc.es}{luca.buoninfante@usc.es}}

	\end{center}

\begin{abstract}
We elucidate the physical nature of ghosts above the multi-particle threshold by contrasting them with unstable particles in quantum field theory. We first consider the asymptotic formulation, where ordinary positive-norm one-particle states can be unstable and decay, whereas ghosts survive asymptotically without decaying, yet admit no particle interpretation due to interference with the multi-particle component which masks the negative-norm one-particle state. This distinction originates from two different analytic structures of the dressed propagator, whose complex conjugate poles lie on the first or second Riemann sheet in the ghost or ordinary case, respectively. Ghost resonances are, in principle, phenomenologically distinguishable from ordinary ones, being narrower and exhibiting weaker interference between positive- and negative-energy peaks. We then formulate the quantum field theory in a finite interval of time and, working within a suitable approximation for the dressed propagator, find that finite-time effects amplify differences in the resonant behavior and give rise to new features, such as higher peaks in ghost resonances. Distinct temporal regimes are also identified: for times shorter than the inverse width, an approximate free-particle description is valid; at later times interactions and interference dominate, leading to decay or multi-particle masking. This study supports the absence of freely propagating ghost particles in the asymptotic limit.
\end{abstract}

\end{titlepage}

{
	\hypersetup{linkcolor=black}
	\tableofcontents
}

\baselineskip=17.63pt

%%%%%%%%%%%%%%%%%%%%%%%%%%%%%%%%%%%%%%%%%%%%%%%%%%%%%%%%%%%%%%%%%%%%%%%%

%\doublespacing

\newpage

\section{Introduction}

Ordinary one-particle states have positive norm and can be either stable or unstable, depending on the type of interactions and kinematics. Stable states are Hamiltonian eigenstates, belong to the asymptotic spectrum, and correspond to freely propagating particles asymptotically. On the other hand, unstable states are not Hamiltonian eigenstates and eventually decay, disappearing from the asymptotic spectrum. In the framework of relativistic local Quantum Field Theory~(QFT), these features can be described by resumming radiative corrections and analyzing the dressed propagator. When the particle mass lies above the multi-particle threshold, the real pole moves to the second Riemann sheet of the complex $p^2$ plane and splits into a complex-conjugate pair. Since no poles remain on the first sheet, no asymptotic one-particle state exists~\cite{Brown:1992db,Veltman:1963th}.

If instead the one-particle state has negative norm and is associated with a ghost coupled to ordinary fields, the resummation of radiative corrections leads to a different analytic structure of the dressed propagator above the multi-particle threshold~\cite{Lee:1969fy,Lee:1970iw,Lee:chicago,Cutkosky:1969fq,Coleman:1969xz,Kubo:2024ysu,Buoninfante:2025klm}. In this case, the pair of complex conjugate poles lies on the first Riemann sheet, which, together with unitarity, implies that the one-particle state cannot decay and remains part of the asymptotic spectrum. This behavior differs not only from that of an unstable particle but also from that of an ordinary stable particle, since the interactions and kinematics are such that interference effects between the ghost and the multi-particle sector persists asymptotically, preventing the existence of a free asymptotic one-particle ghost state~\cite{Buoninfante:2026mve}. To emphasize this distinction, we refer to \textit{anti-instability}~\cite{Kubo:2024ysu} rather than stability or instability, and to \textit{multi-particle masking}~\cite{Buoninfante:2026mve} rather than decay.

Ghosts are typically responsible for improving the ultraviolet behavior of a theory thanks to the characteristic ``minus'' sign in their propagator, which induces additional cancellations in the loop expansion; relevant examples are four-derivative theories~\cite{Pais:1950za,Bender:2007wu,Salvio:2015gsi,Holdom:2023usn,Holdom:2024cfq}, in particular Lee-Wick models~\cite{Lee:1969fy,Lee:1970iw,Lee:chicago,Cutkosky:1969fq} and quadratic gravity~\cite{Stelle:1976gc,Tomboulis:1980bs,Avramidi:1985ki,Salvio:2018crh,Anselmi:2018tmf,Donoghue:2021cza,Holdom:2021hlo,Buoninfante:2023ryt,Buoninfante:2025dgy,Kuntz:2024rzu,Oda:2025buc,Kumar:2026qnw}. At the same time, ghost states have raised concerns about a possible breakdown of the probabilistic interpretation of quantum mechanics due to their negative norm, which could lead to observable negative probabilities~\cite{Strumia:2017dvt,Kubo:2023lpz,Holdom:2024onr}. However, as shown in~\cite{Buoninfante:2026mve}, multi-particle masking prevents the existence of freely propagating ghost particles asymptotically, implying that an asymptotic detector cannot observe or localize a ghost excitation.

The aim of this work is to elucidate the physical properties of ghosts above the multi-particle threshold as a new type of quantum objects in QFT. To this end, we perform a detailed comparison with unstable particles, highlighting the main differences and the novel features characterizing ``anti-unstable'' ghosts. We~first work in an infinite interval of time, where initial and final states, or equivalently the boundary conditions, are specified at $t_{\rm i}=-\infty$ and $t_{\rm f}=\infty$. Subsequently, we formulate the QFT in a finite interval of time $\tau\equiv t_{\rm f}-t_{\rm i}<\infty$, identify distinctive finite-time effects, and analyze how ghosts behave across different temporal regimes. In particular, we confirm that a freely propagating ghost is confined to time intervals much shorter than its~inverse~width.

The paper is organized as follows.
\begin{description}
	
	\item[Sec.~\ref{sec:infinite-time}:] We contrast the asymptotic dynamics of unstable particles and ghosts, explaining why the latter cannot decay and how the novel phenomenon of multi-particle masking prevents the existence of a free one-particle ghost state asymptotically. We then compare ordinary and ghost resonances in an infinite interval of time, showing that the latter are narrower and exhibit weaker interference between positive- and negative-energy~peaks.
	
	\item[Sec.~\ref{sec:finite-time}:] Working in a suitable large-$\tau$ approximation, we analyze the dressed propagator in a finite interval of time and find features that are not captured by the infinite-time formulation. For example, ghost resonances exhibit higher peaks. We identify distinct temporal regimes: at times much shorter than the inverse width, both unstable particles and ghosts admit an approximate free-particle interpretation; at later times, interaction and interference effects dominate, leading to decay or multi-particle masking. Our large-$\tau$ approximation is also useful for tracking the emergence of complex conjugate poles asymptotically. Indeed, we show that apparent complex poles emerge at finite but large times and become complex-conjugate pairs only in the asymptotic limit $\tau\rightarrow\infty$.
	
	\item[Sec.~\ref{sec:on-shell-causality}:] We demonstrate that, when the absorptive contributions to the dressed propagator are correctly identified, ghost propagation is consistent with real positive (negative) energies propagating forward (backward) in time, as in the causal Feynman prescription. The QFT formulation in a finite interval of time renders this aspect clear and unambiguous, confirming that ghosts propagate causally on the mass shell. This does not yet guarantee that (micro)causality is preserved off the mass shell, but it at least implies the existence of a single arrow of causality even when ghosts are coupled to ordinary fields.
	
	\item[Sec.~\ref{sec:conclus}:] We summarize the main results, highlight open questions, and conclude with remarks on potential applications and future work, with particular emphasis on four-derivative theories, especially quadratic gravity.
	
	\item[App.~\ref{sec:app-finite-time}:] We review the main ingredients of the QFT formulation in a finite interval of time. In particular, we provide detailed derivations of the free and dressed propagators, obtaining suitable expressions under the assumption that the time interval is finite but sufficiently large, allowing for a useful approximation on which our finite-time approach is based.
	
\end{description}

\paragraph{Conventions and notations.} We work in $D=3+1$ spacetime dimensions, adopt Natural units ($\hbar=1=c$), and choose the mostly-plus metric signature $(-+++)$, so that $p^2=-p_0^2+\vec{p}^{2}$ is positive (negative) for a spacelike (timelike) momentum. Since we are interested in the energy dependence of propagators and cross sections, and because the QFT formulation in a finite interval of time does not allow manifestly Lorentz-covariant expressions in terms of $p^2$, we will work with propagators as functions of $e \equiv p^0$. In particular, whenever the notations $f(-p^2)$, $f(e^2)$, and $f(e)$ are all meaningful, they will denote the same function expressed in terms of different variables.

\section{Infinite-time formulation}\label{sec:infinite-time}

Let us consider the Lagrangian density
\begin{equation}
	\mathcal{L}= -\frac{1}{2}\left(\partial_\mu\chi\partial^\mu\chi +\mu^2\chi^2\right) - a\frac{1}{2} \left(\partial_\mu\phi\partial^\mu \phi+m^2\phi^2\right)+\mathcal{L}_{\rm int}(\chi,\phi)\,,
	\label{lagrangian}
\end{equation}
where $\chi(x)$ is an ordinary scalar field of mass $\mu$, while $\phi(x)$ is an ordinary scalar field for $a=1$ or a ghost field for $a=-1$, both of mass $m$; the masses are assumed to be already renormalized. The term $\mathcal{L}_{\rm int}(\chi,\phi)$ accounts for local interactions between the two fields. We only require the existence of a ``decay channel'' such that the self-energy of $\phi$ develops an imaginary part above the multi-particle threshold; otherwise, the interaction term can be quite general without affecting the qualitative results. For simplicity, we consider
\begin{equation}
\mathcal{L}_{\rm int}(\chi,\phi)=\frac{g}{2}\phi\chi^2\,,
\label{interaction-term}
\end{equation}
and comment on the generality of the results in the conclusions. 

The free propagators in an infinite interval of time associated with $\chi(x)$ and $\phi(x)$ are
\begin{eqnarray}
	D(-p^2) \!\!\!&=&\!\!\! \frac{-i}{p^2+\mu^2-i\epsilon}\,\, \equiv \,\,  \vcenter{\hbox{\includegraphics[scale = 0.35]{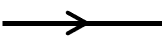}}}\,, 	\label{propagator-phi}\\[1.5mm]
	G(-p^2)\!\!\!&=&\!\!\! \frac{-ia}{p^2+m^2-i\epsilon}\,\, \equiv \,\,  \vcenter{\hbox{\includegraphics[scale = 0.35]{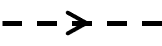}}}\,,
	\label{propagator-chi}
\end{eqnarray}
respectively. The choice of the Feynman prescription with $\epsilon \to 0^+$ for the $\phi$ propagator implies that, for $a=1$, the field theory is perturbatively quantized on a standard Hilbert space with positive-norm states, whereas for $a=-1$ unitarity requires an indefinite-norm vector space. In particular, $n$-particle ghost states have negative (positive) norm for odd (even) $n$~\cite{Buoninfante:2025klm}.

\subsection{Dressed propagators}\label{sec:dressed-propag}

The interaction term~\eqref{interaction-term} induces radiative corrections to the $\phi$ propagator that can be resummed to all orders in perturbation theory. In what follows, we briefly recall the main properties of the dressed propagator and its physical content; for further details we refer to~\cite{Buoninfante:2025klm,Buoninfante:2026mve}.

\medskip

\textbf{Self-energy.} For our purposes, it is sufficient to focus on the one-loop self-energy $\Sigma(-p^2)$, obtained by evaluating the bubble diagram with $\chi$ propagators on the two internal lines. As a function of the complex momentum squared, $-p^2 \to z \in \mathbb{C}$, it reads
\begin{equation}
	\begin{aligned}
		i\Sigma(z)\equiv  \,\vcenter{\hbox{\includegraphics[scale = 0.23]{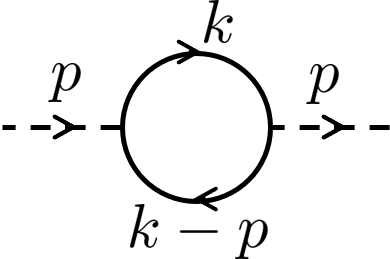}}} =&\,  \frac{(-ig)^2}{2}\int \frac{{\rm d}^4k}{(2\pi)^4} D\left(-(k-p)^2\right)D\left(-k^2\right)\\[1mm]
		=&\,i\frac{g^2}{32\pi^2}\left[2-\log\left(\frac{\mu^2}{\Lambda^2}\right)-2\sqrt{\frac{4\mu^2-z}{z}}\arctan\sqrt{\frac{z}{4\mu^2-z}}\right]\,,
	\end{aligned}
	\label{self-energy}
\end{equation}
where $\Lambda$ is the renormalization scale. For physical momentum squared, $z\to -p^2>0$, we have
\begin{equation}
	{\rm Re}\left[\Sigma(-p^2)\right]=\frac{g^2}{32\pi^2}\left[ 2-\log\left(\frac{\mu^2}{\Lambda^2}\right)+\sqrt{1+\frac{4\mu^2}{p^2}}\log\left(\frac{1-\sqrt{1+4\mu^2/p^2}}{1+\sqrt{1+4\mu^2/p^2}}\right) \right]
	\label{real-self-energy}
\end{equation}
and
\begin{equation}
	{\rm Im}\left[\Sigma(-p^2\pm i\epsilon)\right]=-{\rm Im}\left[\Sigma^{II}(-p^2\pm i\epsilon)\right]=\pm\frac{g^2}{32\pi}\sqrt{1+\frac{4\mu^2}{p^2}}\,\theta\left(-p^2-4\mu^2\right)\,,
	\label{imag-self-en}
\end{equation}
where $\Sigma^{II}(p^2 \pm i\epsilon)$ denotes the self-energy evaluated on the second Riemann sheet of the complex $z$ plane, whose continuation to the real axis has the opposite sign to that on the first sheet~\cite{Buoninfante:2025klm}.

\medskip

\textbf{Resummation.} The dressed propagator $\bar{G}(z)$ of the $\phi$ field, as a function of $-p^2 \to z \in \mathbb{C}$, is given by the following geometric series:
\begin{eqnarray}
	\bar{G}(z)\!\!\!&\equiv&\!\!\!\vcenter{\hbox{\includegraphics[scale = 0.23]{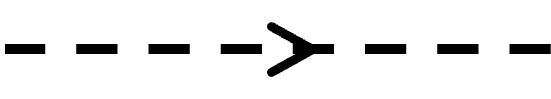}}}\,+\, \vcenter{\hbox{\includegraphics[scale = 0.23]{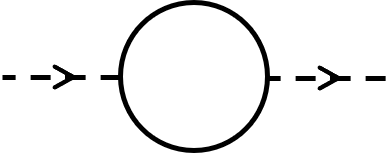}}}\,+\,\vcenter{\hbox{\includegraphics[scale = 0.23]{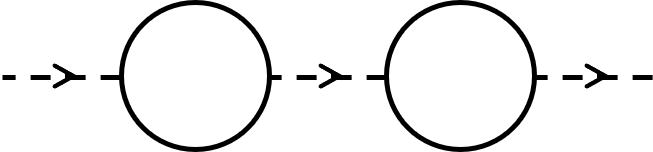}}}\,+\cdots\nonumber\\[1mm]
	&=&\!\!\!  G(z)\sum_{n=0}^\infty \left[i\Sigma(z)G
	(z) \right]^n=\frac{G(z)}{1-i\Sigma(z)G(z)}=\frac{ia}{z-m^2+a\Sigma(z)}\,.
	\label{dressed-propagator}
\end{eqnarray}
Above the multi-particle threshold, $m>2\mu$, the convergence condition $|\Sigma(z)G(z)|<1$ is violated near the real pole $z\simeq m^2$, where the nonvanishing imaginary part of the self-energy contributes with the divergent term ${\rm Im}[\Sigma(m^2)]/(z-m^2)$. Nevertheless, the geometric series can still be resummed by exploiting analyticity: one first performs the resummation in regions where the convergence condition holds and then analytically continues the result to the real axis. This procedure is consistent because the Feynman prescription is analytic and the denominator of the dressed propagator, $z - m^2 + a\Sigma(z)$, has no real zeros above the multi-particle threshold~\cite{Anselmi:2020lfx,Buoninfante:2025klm}.

\medskip

\textbf{Modified perturbation theory.} The divergence near $z\simeq m^2$ also signals a breakdown of the perturbation theory built in terms of bare $\phi$ propagators, as higher-order terms become larger than the lower-order ones. This issue can be resolved by reorganizing the diagrammatic expansion so that all bare propagators are replaced by the dressed ones, with no additional self-energy insertion~\cite{Veltman:1963th}. In this way, the series of diagrams contributing to a given physical process can be consistently resummed without hitting spurious poles on the real axis. Since the self-energy, and hence the dressed propagator, can only be computed up to finite order in the interaction coupling, the resulting modified perturbation theory can be systematically improved recursively.

\medskip

\textbf{Complex conjugate poles.} The dressed propagator~\eqref{dressed-propagator} develops complex poles for $m>2\mu$. In the ordinary case, $a=1$, radiative corrections move the real pole through the branch cut into the second Riemann sheet, where it splits into a pair of complex conjugate poles. In the ghost case, $a=-1$, the same poles instead appear on the first sheet. Their complex-conjugate nature follows from the reflection relations $\Sigma^*(z)=\Sigma(z^*)$ and $G^*(z)=-G(z^*).$ Denoting the poles by $M^2=m^2+im\Gamma$ and $M^{*2}=m^2-im\Gamma$, their real and imaginary parts are determined by the renormalization condition ${\rm Re}[\Sigma(m^2\pm im\Gamma)]=0$ and the~pole~equations~\cite{Buoninfante:2025klm}
\begin{eqnarray}
\text{First sheet:}&&  \pm im\Gamma =-ai {\rm Im}\big[\Sigma(m^2\pm im\Gamma)\big]\,,\label{1st-sheet-eq}\\
\text{Second sheet:}&&  \pm i m\Gamma =-ai {\rm Im}\big[\Sigma^{II}(m^2\pm im\Gamma)\big]\,.\label{2nd-sheet-eq}
\end{eqnarray}
In the narrow-width approximation, $\Gamma/m\ll1$, we can write 
\begin{eqnarray}
{\rm Im}\big[\Sigma(m^2\pm im\Gamma)\big]=-{\rm Im}\big[\Sigma^{II}(m^2\pm im\Gamma)\big]\simeq	\pm\frac{g^2}{32\pi}\sqrt{1-\frac{4\mu^2}{m^2}}\,,
\label{self-energy-relations}
\end{eqnarray}
and analytically solve the pole equations up to order $\mathcal{O}(g^2)$. The relations in~\eqref{self-energy-relations} imply that~\eqref{1st-sheet-eq} admits solutions (first-sheet poles) only for $a=-1$, whereas~\eqref{2nd-sheet-eq} admits solutions (second-sheet poles) only for $a=1$. The width at order~$\mathcal{O}(g^2)$~reads
\begin{eqnarray}
\Gamma\simeq \frac{g^2}{32\pi m}\sqrt{1-\frac{4\mu^2}{m^2}}\,.
\label{width-expression}
\end{eqnarray}

\medskip

\textbf{Two absorptive contributions.} The absorptive part of the dressed propagator for physical momentum squared, i.e. $z= -p^2+i\epsilon$ with $\epsilon\rightarrow 0^+,$ can be written as
\begin{eqnarray}
{\rm Re}\left[\bar{G}(-p^2)\right]={\rm Im}\left[i\bar{G}(-p^2)\right]\!\!\!&=&\!\!\!\frac{a\epsilon+{\rm Im}[\Sigma(-p^2+i\epsilon)]}{\big(p^2+m^2-a{\rm Re}[\Sigma(-p^2)]\big)^2+\big(\epsilon+a{\rm Im}[\Sigma(-p^2+i\epsilon)]\big)^2}  \nonumber\\ [1mm]
\!\!\!&=&\!\!\!A(-p^2)+B(-p^2)\,,
\label{absorpt-propag}
\end{eqnarray}
where we have identified two different terms:
\begin{eqnarray}
A(-p^2)\!\!\!&\equiv&\!\!\! \bar{G}(-p^2)\,a\epsilon\, \bar{G}^*(-p^2)\,, \label{A-abs}\\[1mm]
B(-p^2)\!\!\!&\equiv&\!\!\! \bar{G}(-p^2)\,{\rm Im}[\Sigma(-p^2+i\epsilon)]\, \bar{G}^*(-p^2)\,.\label{B-abs} 
\end{eqnarray}
The function $A(-p^2)$ describes processes in which the $\phi$ particle propagates with negligible interference with the multi-particle sector. In contrast, $B(-p^2)$ captures processes where interference between one- and multi-particle states is significant.
Diagrammatically,~\eqref{A-abs} and~\eqref{B-abs} correspond to different cuts~\cite{tHooft:1973wag}: the former arises from cutting the $\phi$~lines, while the latter corresponds to cutting the self-energy bubble. Indeed,~defining~the~cuts
\begin{eqnarray}
	\vcenter{\hbox{\includegraphics[scale = 0.23]{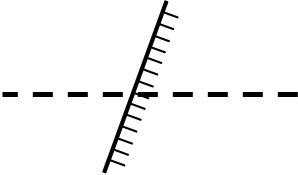}}}\!\!\!&\equiv&\!\!\! G(-p^2)+G^*(-p^2)\,,\\[1mm]
	\vcenter{\hbox{\includegraphics[scale = 0.23]{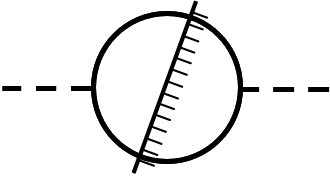}}}\!\!\!&\equiv&\!\!\! G(-p^2) 2{\rm Im}[\Sigma(-p^2+i\epsilon)] G^*(-p^2)\,,
\end{eqnarray}
it can be shown that~\cite{Anselmi:2020lfx,Buoninfante:2025klm}
\begin{eqnarray}
2A(-p^2)\!\!\!&=&\!\!\!\vcenter{\hbox{\includegraphics[scale = 0.23]{cut-line-right}}}\,+\,\vcenter{\hbox{\includegraphics[scale = 0.23]{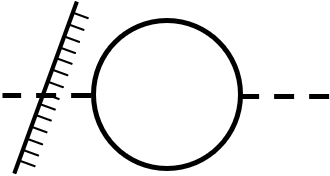}}}\,+\,\vcenter{\hbox{\includegraphics[scale = 0.23]{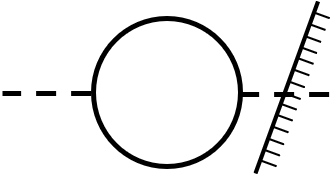}}}\,+\,\vcenter{\hbox{\includegraphics[scale = 0.23]{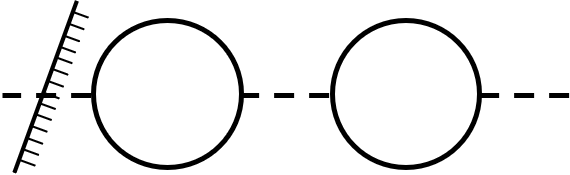}}}\,+\,\cdots\,,
\label{A-function-diagram} \\[1mm]
2B(-p^2)\!\!\!&=&\!\!\!\vcenter{\hbox{\includegraphics[scale = 0.23]{cut-line-bubble}}}\,+\,\vcenter{\hbox{\includegraphics[scale = 0.23]{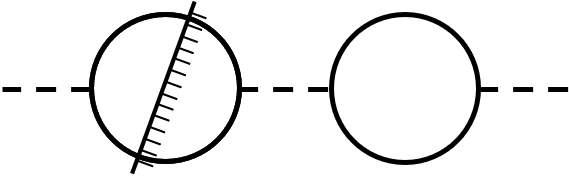}}}\,+\,\vcenter{\hbox{\includegraphics[scale = 0.23]{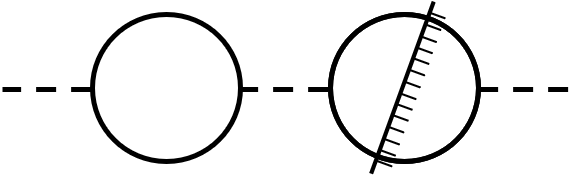}}}\,+\,\vcenter{\hbox{\includegraphics[scale = 0.23]{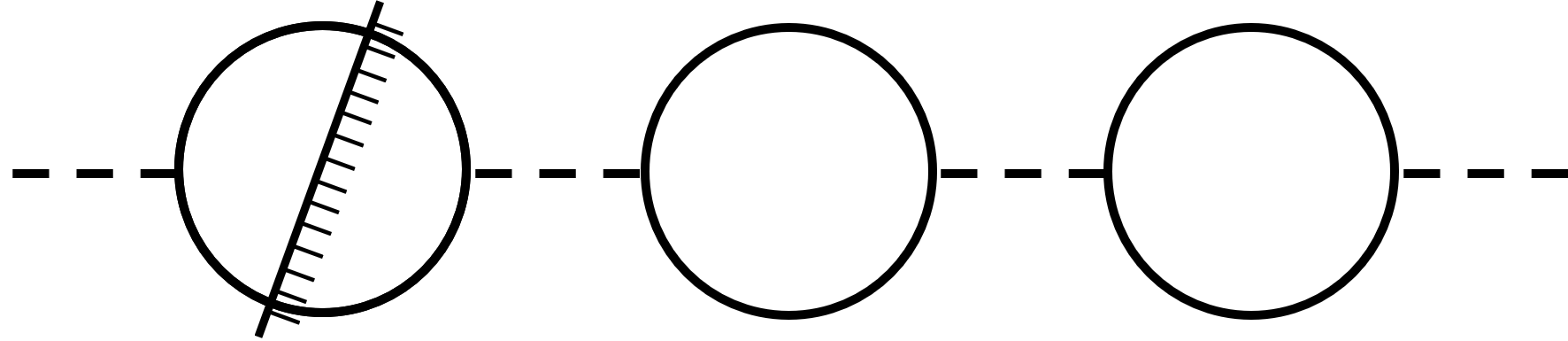}}}\,+\,\cdots\,.
\label{B-function-diagram}
\end{eqnarray}
Strictly speaking, since $\epsilon$ is a mathematical artefact, in a QFT formulated in an infinite interval of time the term $A(-p^2)$ vanishes above the multi-particle threshold, i.e. when ${\rm Im}[\Sigma(-p^2)]\neq 0$, as it goes to zero in the limit $\epsilon\to 0^+$. Physically, this reflects the fact that the infinite-time formulation cannot capture regimes in which an unstable particle is still alive or a ghost is not yet masked by the multi-particle states. In Secs.~\ref{sec:finite-time} and~\ref{sec:on-shell-causality}, we will show that a QFT formulation in a finite interval of time can instead consistently describe regimes where $A(-p^2)$ is non-zero~and~dominant.

\medskip

\textbf{Spectral representation.} From the analytic structure of the dressed propagator~\eqref{dressed-propagator} above the multi-particle threshold, i.e. a pair of complex conjugate poles on either the first or second sheet and a branch cut from $4\mu^2$ to $\infty$, one derives the first-sheet spectral representation~\cite{Kubo:2024ysu,Buoninfante:2025klm,Buoninfante:2026mve}
\begin{equation}
	\bar{G}(-p^2)= \frac{1-a}{2}\left[\frac{iZ}{p^2+M^2}+\frac{iZ^*}{p^2+M^{*2}}\right]+\int_{4\mu^2}^\infty{\rm d}\sigma \rho(\sigma)\frac{-i}{p^2+\sigma-i\epsilon}\,,
	\label{spectral-represent}
\end{equation}
where $\rho(-p^2) = {\rm Im}\big[i\bar{G}(-p^2+i\epsilon)\big]/\pi > 0$ is the spectral density; an analogous expression on the second sheet can also be derived, see~\cite{Buoninfante:2025klm}. In position space we have
\begin{equation}
\left\langle\bar{0}|{\rm T}[\phi(x)\phi(y)]|\bar{0} \right\rangle =\frac{1-a}{2}\Big[\!-Z\Delta_{\rm F}(x-y;M^2)-Z^*\Delta_{\rm F}(x-y;M^{*2})\Big]+\int_{4\mu^2}^\infty{\rm d}\sigma \rho(\sigma)\Delta_{\rm F}(x-y;\sigma)\,,
\label{spectral-represent-position}
\end{equation}
where $|\bar{0}\rangle$ denotes the interacting vacuum, the complex spectral weight $Z \in \mathbb{C}$ is defined by
\begin{eqnarray}
	Z^{-1}=1- \left.\text{Re}\left[\frac{{\rm d}\Sigma(z)}{{\rm d}z}\right]\right|_{z=M^2} -  i \left.\text{Im}\left[ \frac{{\rm d}\Sigma(z)}{{\rm d}z}\right]\right|_{z=M^2}\,,
\end{eqnarray}
and we have introduced the Feynman-like propagators
\begin{equation}
	\Delta_{\rm F}(x-y;s)=\int_{\mathcal{C}} \frac{{\rm d}p_0}{2\pi}\int\frac{{\rm d}^3p}{(2\pi)^3} {\rm e}^{ip\cdot (x-y)} \frac{-i}{p^2+s-i\epsilon}\,,\qquad s=M^2, M^{*2}, \sigma\,.
	\label{definition-propags}
\end{equation}
The integration contour $\mathcal{C}$ is the so-called Lee-Wick contour~\cite{Lee:1969fy,Lee:1970iw,Lee:chicago}, which is obtained within QFT by tracking how the causal Feynman contour must be deformed so as to avoid being crossed when radiative corrections split the real pole into the complex conjugate pair. In particular, it coincides with the standard Feynman contour for $s=\sigma, M^{*2}$; see~\cite{Buoninfante:2026mve} for further details.

\medskip

\textbf{Sum rule.} Knowing the spectral representation of the propagator, we can also readily derive that of the commutator, which reads
\begin{equation}
	\left\langle\bar{0}|[\phi(x),\phi(y)]|\bar{0} \right\rangle =\frac{1-a}{2}\Big[-Z\Delta(x-y;M^2)-Z^*\Delta(x-y;M^{*2})\Big]+\int_{4\mu^2}^\infty{\rm d}\sigma \rho(\sigma)\Delta(x-y;\sigma)\,,
	\label{spectral-represent-commutator}
\end{equation}
where 
\begin{equation}
	\Delta(x-y;s)= i\int\frac{{\rm d}^3p}{(2\pi)^3} \frac{\sin\left(\vec{p}\cdot (\vec{x}-\vec{y})-\sqrt{\vec{p}^2+s}\, (x^0-y^0)\right)}{\sqrt{\vec{p}^2+s}}\,,\qquad s=M^2, M^{*2}, \sigma\,.
	\label{definition-Pauli-Jordan}
\end{equation}
The expression in~\eqref{spectral-represent-commutator} is useful for deriving an important sum rule. Indeed, using the relations $\partial_{x^0}\langle\bar{0}|T[\phi(x)\phi(y)]|\bar{0}\rangle\big|_{x^0=y^0} = -a i\delta^{(3)}(\vec{x}-\vec{y})$ and $\partial_{x^0}\Delta(x-y;s)\big|_{x^0=y^0} = -i\delta^{(3)}(\vec{x}-\vec{y})$, we~get~\cite{Kubo:2024ysu,Buoninfante:2025klm}
\begin{eqnarray}
	a = -\frac{1-a}{2}(Z+Z^*) + C\,, \qquad C\equiv \int_{4\mu^2}^\infty {\rm d}\sigma\rho(\sigma)>0\,,
	\label{sum-rule}
\end{eqnarray}
whose physical meaning will be discussed in Secs.~\ref{sec:unstable} and~\ref{sec:anti-unstable}.

\medskip

\textbf{Multi-particle component.} Another important object is the composite field $\chi^2(x)$, associated with the positive-norm multi-particle states, for which one can also obtain a spectral representation of the propagator. Using the field equation for the field $\phi(x)$, namely
\begin{equation}
(\Box-m^2)\phi(x)=a\frac{g}{2}\chi^2(x)\,, 
\end{equation}
and the relations
\begin{eqnarray}
	\!\!\!\!\!\!\!\!\!\!\!\!\!\!\!\!\!\!\!\!\!\!&&(\Box_x-m^2)(\Box_y-m^2)\big\langle\bar{0}\big|{\rm T}\big[\phi(x)\phi(y)\big]\big|\bar{0} \big\rangle = \frac{g^2}{4}\big\langle\bar{0}\big|{\rm T}\big[\chi^2(x)\chi^2(y)\big]\big|\bar{0} \big\rangle-(\Box_x-m^2)i\delta^{(4)}(x-y)\,,	\nonumber\\[1.5mm]
	\!\!\!\!\!\!\!\!\!\!\!\!\!\!\!\!\!\!\!\!\!\!&&(\Box_x-m^2)(\Box_y-m^2)\Delta_{\rm F}(x-y;s)= (s-m^2)^2\Delta_{\rm F}(x-y;s)+(\Box_x+s-2m^2)i\delta^{(4)}(x-y)\,,
\end{eqnarray}
we can derive the following spectral representation by acting with $(\Box_x - m^2)(\Box_y - m^2)$ on~\eqref{spectral-represent-position}:
\begin{eqnarray}
\frac{g^2}{4m^2\Gamma^2}\big\langle\bar{0}\big|{\rm T}\big[\chi^2(x)\chi^2(y)\big]\big|\bar{0} \big\rangle \!\!\!&=&\!\!\!\frac{1-a}{2}\Big[Z\Delta_{\rm F}(x-y;M^2)+Z^*\Delta_{\rm F}(x-y;M^{*2})\Big]\nonumber \\[1mm]
&&\!\!\!+\int_{4\mu^2}^\infty{\rm d}\sigma \rho(\sigma)\frac{(\sigma-m^2)^2}{m^2\Gamma^2}\Delta_{\rm F}(x-y;\sigma) + g(x,y)\,,
\label{spectral-represent-composite-field}
\end{eqnarray}
where the last piece accounts for the contact terms,
\begin{eqnarray}
g(x,y)\equiv\frac{1}{m^2\Gamma^2}\left[(1+a)(\Box_x-m^2)+2{\rm Im}[Z]m\Gamma +\int_{4\mu^2}^\infty{\rm d}\sigma \rho(\sigma)(\sigma-m^2)\right]i\delta^{(4)}(x-y)\,,
\label{contact-terms}
\end{eqnarray}
which is irrelevant for the purposes of this work. The presence or absence of first-sheet complex conjugate poles in the $\chi^2$ propagator~\eqref{spectral-represent-composite-field} is in one-to-one correspondence with the presence or absence of the same poles in the $\phi$ propagator~\eqref{spectral-represent-position}; this fact has important physical implications, as shown in the next subsections.

\subsection{Unstable particles and decay}\label{sec:unstable}

In the ordinary case, $a=1$, the spectral representations of the $\phi$ and $\chi^2$ propagators~\eqref{spectral-represent-position} and~\eqref{spectral-represent-composite-field} above the multi-particle threshold and in the first Riemann sheet read
\begin{eqnarray}
\big\langle\bar{0}\big|{\rm T}\big[\phi(x)\phi(y)\big]\big|\bar{0} \big\rangle \!\!\!&=&\!\!\!\int_{4\mu^2}^\infty{\rm d}\sigma \rho(\sigma)\Delta_{\rm F}(x-y;\sigma)\,,\label{spectral-represent-position-ordinary}\\[1mm]
\frac{g^2}{4m^2\Gamma^2}\big\langle\bar{0}\big|{\rm T}\big[\chi^2(x)\chi^2(y)\big]\big|\bar{0} \big\rangle \!\!\!&=&\!\!\!\int_{4\mu^2}^\infty{\rm d}\sigma \rho(\sigma)\frac{(\sigma-m^2)^2}{m^2\Gamma^2}\Delta_{\rm F}(x-y;\sigma)+g(x,y)\,,
\label{spectral-represent-composite-field-odinary}
\end{eqnarray}
where
\begin{eqnarray}
	g(x,y)=\frac{1}{m^2\Gamma^2}\left[2(\Box_x-m^2)+2{\rm Im}[Z]m\Gamma +\int_{4\mu^2}^\infty{\rm d}\sigma \rho(\sigma)(\sigma-m^2)\right]i\delta^{(4)}(x-y)\,.
	\label{contact-terms-ordinary}
\end{eqnarray}
Both propagators have no poles on the first sheet, while a pair of complex conjugate poles at $m^2 \pm i m\Gamma$ appears in the second sheet~\cite{Brown:1992db,Buoninfante:2025klm}. Since poles of the propagator on the first sheet are associated with asymptotic one-particle states, their absence implies that there is no asymptotic field associated with the Heisenberg field $\phi(x)$ in the case $a=1$.

\medskip

\textbf{Instability.} The analytic structure of the two propagators in the first sheet admits a clear physical interpretation: above the multi-particle threshold (i.e. for $m>2\mu$), the one-particle state associated with the ordinary field $\phi(x)$ decays and disappears from the set of asymptotic states. Unitarity then requires these unstable states to be projected out of the physical Hilbert space through the so-called Veltman projection~\cite{Veltman:1963th}, leaving only stable one-particle states associated with the elementary field $\chi(x)$ in the asymptotic spectrum. This picture is further supported by the sum rule~\eqref{sum-rule} for $a=1$, namely
\begin{equation}
1=C\,,
\label{sum-rule-ordinary}
\end{equation}
which indicates that the particle has decayed and what remains is the continuum component. In other words, the initial probability of having a one-particle $\phi$ state (equal to one) is entirely converted into the probability of having multi-particle states at the end of the decay process.

\medskip

\textbf{Asymptotic field and propagator.}  In the case $a=1$, the only asymptotic field is the one associated with the elementary Heisenberg field $\chi(x)$, i.e.
\begin{equation}
\chi(x)\xrightarrow[x^0\rightarrow \pm \infty]{} Z_\chi^{1/2} \chi^{\rm as}(x)\,,
\label{chi-asymp}
\end{equation}
where the limit is understood in a weak sense, and $Z_\chi \in \mathbb{R}$ is the corresponding wave-function renormalization constant. The asymptotic Lagrangian reads
\begin{equation}
\mathcal{L}^{\rm as}=\mathcal{L}^{\rm as}_\chi=-\frac{1}{2}\left(\partial_\mu \chi^{\rm as}\partial^\mu \chi^{\rm as}-\mu^2 \chi^{{\rm as}\,2}\right)\,,
\label{asymp-lagr-ordin}
\end{equation}
and the associated field equation is
\begin{equation}
(\Box-\mu^2)\chi^{\rm as}(x)=0\,.
\label{asymp-EOM-ordin}
\end{equation}
The propagator has a real pole at $p^2=-\mu^2$ corresponding to a stable one-particle state: 
\begin{equation}
	Z_\chi\big\langle0\big|{\rm T}\big[\chi^{\rm as}(x)\chi^{\rm as}(y)\big]\big|0 \big\rangle= \int \frac{{\rm d}^4p}{(2\pi)^4}{\rm e}^{ip \cdot (x-y)}\frac{-iZ_\chi}{p^2+\mu^2-i\epsilon}\,,
	\label{chi-asymp-propag}
\end{equation}
where $\mu$ is the renormalized mass, and $|0\rangle$ is the asymptotic vacuum, i.e. the state annihilated by the Hamiltonian associated with the asymptotic Lagrangian~\eqref{asymp-lagr-ordin}.

\medskip

\textbf{Asymptotic states.} The asymptotic one-particle state associated with $\chi(x)$ admits a free-particle interpretation and can be attached to external legs of Feynman diagrams to compute scattering amplitudes between the asymptotic times $t_{\rm i}=-\infty$ and $t_{\rm f}=\infty.$ In particular, the standard LSZ construction~\cite{Lehmann:1954rq} works and can be  applied to the $\chi$~states. 

\medskip

\textbf{Resonance and complex mass.} In the narrow-width approximation, $\Gamma/m\ll 1$, the $\phi$~propagator~\eqref{spectral-represent-position-ordinary}, in momentum space, can be approximated around $-p^2\simeq m^2$ as
\begin{equation}
	 \bar{G}(-p^2)\simeq \frac{-iZ_{\rm R}}{p^2+m^2-im\Gamma}\simeq  \frac{-iZ_{\rm R}}{p^2+(m-i\Gamma/2)^2} \,,
	 \label{propag-ordinary-narrow}
\end{equation}
where $Z_{\rm R}\equiv {\rm Re}[Z]$. Despite its pole-like form,~\eqref{propag-ordinary-narrow} does not signal an actual pole on the first sheet, but rather a bump in the spectral density around $-p^2 \simeq m^2$~\cite{Buoninfante:2025klm}. The complex mass $m-i\Gamma/2$ is interpreted as follows. The real part $m$ corresponds to the physical mass of the quasi-free unstable particle before decay and, in resonance language, sets the position of the Breit-Wigner peak. The imaginary part $-\Gamma/2$ represents (minus) the inverse lifetime of the unstable particle in its rest frame and $\Gamma$ also determines the width of the Breit-Wigner function at half maximum.

\subsection{Anti-unstable ghosts and multi-particle masking}\label{sec:anti-unstable}

In the ghost case, $a=-1$, the spectral representations of the $\phi$ and $\chi^2$ propagators~\eqref{spectral-represent-position} and~\eqref{spectral-represent-composite-field} above the multi-particle threshold and in the first Riemann sheet are given by
\begin{eqnarray}
\!\!\!\!\!\!\!\!\!\big\langle\bar{0}\big|{\rm T}\big[\phi(x)\phi(y)\big]\big|\bar{0} \big\rangle \!\!\!&=&\!\!\!-Z\Delta_{\rm F}(x-y;M^2)-Z^*\Delta_{\rm F}(x-y;M^{*2})+\int_{4\mu^2}^\infty{\rm d}\sigma \rho(\sigma)\Delta_{\rm F}(x-y;\sigma)\,,\label{spectral-represent-position-ghost}\\[1mm]
\!\!\!\!\!\!\!\!\!\big\langle\bar{0}\big|{\rm T}\big[\tilde{\phi}(x)\tilde{\phi}(y)\big]\big|\bar{0} \big\rangle \!\!\!&=&\!\!\! Z\Delta_{\rm F}(x-y;M^2)+Z^*\Delta_{\rm F}(x-y;M^{*2})\nonumber\\[1mm]
\!\!\!\!\!\!\!\!\!\!\!\!&&\!\!\!+\int_{4\mu^2}^\infty{\rm d}\sigma \rho(\sigma)\frac{(\sigma-m^2)^2}{m^2\Gamma^2}\Delta_{\rm F}(x-y;\sigma)+g(x,y)\,,
\label{spectral-represent-composite-field-ghost}
\end{eqnarray}
where the contact term now reads
\begin{eqnarray}
g(x,y)=\frac{1}{m^2\Gamma^2}\left[2{\rm Im}[Z]m\Gamma +\int_{4\mu^2}^\infty{\rm d}\sigma \rho(\sigma)(\sigma-m^2)\right]i\delta^{(4)}(x-y)\,,
\label{contact-terms-ghost}
\end{eqnarray}
and for convenience we have redefined the composite field as
\begin{eqnarray}
	\tilde{\phi}(x)\equiv \frac{g}{2m\Gamma}\chi^2(x)\,.
	\label{redef-composite-field}
\end{eqnarray}
Both propagators~\eqref{spectral-represent-position-ghost} and~\eqref{spectral-represent-composite-field-ghost} exhibit complex conjugate poles at $m^2 \pm i m\Gamma$ in the first Riemann sheet, while no poles appear in the second sheet~\cite{Kubo:2024ysu,Buoninfante:2025klm,Buoninfante:2026mve}. This analytic structure differs significantly from that of an ordinary unstable particle and leads to distinct physical implications. We anticipate that the presence of first-sheet poles in the propagator of the composite field does not signal a bound state, as the poles are complex and appear only above the multi-particle threshold.

\medskip

\textbf{Anti-instability.} Since the norm is conserved under unitary time evolution, a negative-norm one-particle state associated with $\phi(x)$ cannot decay and disappear into a purely positive-norm two-particle state associated with $\chi(x)$. This feature is further supported by the sum rule~\eqref{sum-rule}~for~$a=-1$,
\begin{equation}
-1=-(Z+Z^*)+C\,,
\label{sum-rule-ghost}
\end{equation}
where $-(Z+Z^*)$ can be interpreted as the ``probability'' (more precisely, the norm contribution) of having a one-particle ghost state in the asymptotic spectrum. Thus,~\eqref{sum-rule-ghost} can be read as stating that the initial probability of having a one-particle $\phi$ state (equal to minus one) is only partially converted into the probability of having positive-norm multi-particle states, while the remaining part still carries a one-particle ghost contribution. In other words, rewriting~\eqref{sum-rule-ghost} as $Z+Z^*=1+C$, the larger the probability of finding the multi-particle component is, the larger (in modulus) the probability of finding the ghost will be. This property implies that a ghost cannot decay and was termed \textit{anti-instability} in~\cite{Kubo:2024ysu}; accordingly, we can use the term \textit{anti-unstable ghost} to distinguish it from both stable and unstable particles.

\medskip

\textbf{Asymptotic fields.} Unlike the ordinary case of an unstable particle, in addition to the asymptotic field $\chi^{\rm as}(x)$ introduced in~\eqref{chi-asymp}, we now also have the asymptotic fields $\phi^{\rm as}(x)$ and $\tilde{\phi}^{\rm as}(x)$ associated with the elementary field $\phi(x)$ and the composite field $\tilde{\phi}(x)$:
\begin{equation}
	\phi(x)\xrightarrow[x^0\rightarrow \pm \infty]{} Z_\phi^{1/2} \phi^{\rm as}(x)\,,\qquad 	\tilde{\phi}(x)\xrightarrow[x^0\rightarrow \pm \infty]{} Z_{\tilde{\phi}}^{1/2} \tilde{\phi}^{\rm as}(x)\,,
	\label{phi-phitilde-asymp}
\end{equation}
where the limits are understood in a weak sense, and $Z_\phi,\, Z_{\tilde{\phi}}\in \mathbb{R}$ are the respective wave-function renormalization constants. The propagators of these two asymptotic fields, i.e. the expectation values of the time-ordered products in the asymptotic vacuum, must reproduce the correct pole parts in~\eqref{spectral-represent-position-ghost} and~\eqref{spectral-represent-composite-field-ghost}. 

\medskip

\textbf{Asymptotic dynamics.} It is easy to see that no local and Hermitian decoupled free Lagrangians can be constructed for the individual fields $\phi^{\rm as}(x)$ and $\tilde{\phi}^{\rm as}(x)$, whose propagators reproduce the pole terms in~\eqref{spectral-represent-position-ghost} and~\eqref{spectral-represent-composite-field-ghost}. Instead, quadratic couplings between the two fields are required for a consistent asymptotic description. One can show that the asymptotic~Lagrangian~is
\begin{equation}
	\mathcal{L}^{\rm as}=\mathcal{L}^{\rm as}_\chi+\mathcal{L}^{\rm as}_{\phi\tilde{\phi}}\,,
	\label{asymp-lagr-ghost}
\end{equation}
where $\mathcal{L}^{\rm as}_\chi$ is given in~\eqref{asymp-lagr-ordin}, while the second and highly nontrivial term reads~\cite{Buoninfante:2026mve}
\begin{eqnarray}
	\mathcal{L}^{\rm as}_{\phi\tilde{\phi}}\!\!\!&=&\!\!\!\frac{1}{2}\Big(\partial_\mu\phi^{\rm as}\partial^\mu\phi^{\rm as}+m^2\phi^{{\rm as}\,2}\Big)-\frac{1}{2}\Big(\partial_\mu\tilde{\phi}^{\rm as}\partial^\mu\tilde{\phi}^{\rm as}+m^2\tilde{\phi}^{{\rm as}\,2}\Big)-m\Gamma\, \phi^{\rm as} \,\tilde{\phi}^{\rm as}\nonumber \\[1mm]
	&&\!\!\!+ \tan \theta_{\rm Z}\left[\partial_\mu \phi^{\rm as}\partial^\mu \tilde{\phi}^{\rm as}+m^2\,\phi^{\rm as} \,\tilde{\phi}^{\rm as}+\frac{m\Gamma}{2}\left(\phi^{{\rm as}\,2}-\tilde{\phi}^{{\rm as}\,2}\right) \right]\,.
	\label{asymp-phiphitilde-lagran}
\end{eqnarray}
We have introduced the polar angle $\theta_{\rm Z} \equiv \arctan({\rm Im}[Z]/{\rm Re}[Z])$, and used the relations
\begin{equation}
Z_\phi=Z_{\tilde{\phi}}=\frac{2|Z|}{\cos \theta_{\rm Z}}\,
\label{Zphiphitilde-rel}
\end{equation}
to ensure canonical normalization of the kinetic term. Note that, while $\phi^{\rm as}(x)$ is a ghost, $\tilde{\phi}^{\rm as}(x)$ is not, owing to the standard sign of its kinetic term in the first line of~\eqref{asymp-phiphitilde-lagran}. The corresponding field equations can be shown to be equivalent to the coupled system
\begin{eqnarray}
	(\Box-m^2)\phi^{\rm as}(x)=-m\Gamma \tilde{\phi}^{\rm as}(x)\,, \qquad 
	(\Box-m^2)\tilde{\phi}^{\rm as}(x) = m\Gamma\phi^{\rm as}(x)\,.
	\label{EOM-phiphitilde}
\end{eqnarray}
From~\eqref{asymp-phiphitilde-lagran} and~\eqref{EOM-phiphitilde}, it follows that, unlike $\chi^{\rm as}(x)$, the asymptotic fields $\phi^{\rm as}(x)$ and $\tilde{\phi}^{\rm as}(x)$ remain quadratically coupled at asymptotic times rather than becoming truly free.

\medskip

\textbf{Asymptotic propagators.} Interference effects between the two fields are also reflected in the non-diagonal form of the propagator matrix, whose momentum-space expression is~\cite{Buoninfante:2026mve}
\begin{eqnarray}
	\hat{G}_{\left(\phi^{\rm as},\tilde{\phi}^{\rm as}\right)}(-p^2)= 
	\left(\begin{array}{cc}
		G_{\phi^{\rm as}}(-p^2) & G_{\tilde{\phi}^{\rm as}\phi^{\rm as}}(-p^2)\\[2.5mm]
		G_{\phi^{\rm as}\tilde{\phi}^{\rm as}}(-p^2)	& G_{\tilde{\phi}^{\rm as}}(-p^2)
	\end{array}
	\right)\,,
	\label{propag-matrix}\,
\end{eqnarray}
where 
\begin{eqnarray}
	G_{\phi^{\rm as}}(-p^2)=-G_{\tilde{\phi}^{\rm as}}(-p^2)\!\!\!&=&\!\!\!Z_\phi^{-1}\left[\frac{iZ}{p^2+M^2}+\frac{iZ^*}{p^2+M^{*2}}\right]\,,\label{propag-phiphitilde}\\[1mm]
	G_{\phi^{\rm as} \tilde{\phi}^{\rm as}}(-p^2)=G_{\tilde{\phi}^{\rm as}\phi^{\rm as}}(-p^2)\!\!\!&=&\!\!\!Z_\phi^{-1}\left[\frac{Z}{p^2+M^2}-\frac{Z^*}{p^2+M^{*2}}\right]\,.
	\label{propagators-phiphitilde-mixed}
\end{eqnarray}
The relative minus sign between the $\phi^{\rm as}$ and $\tilde{\phi}^{\rm as}$ propagators consistently matches the corresponding sign difference in the pole parts of the spectral representations~\eqref{spectral-represent-position-ghost} and~\eqref{spectral-represent-composite-field-ghost}. In fact, carrying over the quantization procedure as done in~\cite{Buoninfante:2026mve}, one can show that the correct pole structures are reproduced asymptotically:
\begin{eqnarray}
Z_\phi\big\langle 0\big|{\rm T}\big[\phi^{\rm as}(x)\phi^{\rm as}(y)\big]\big|0\big\rangle \!\!\!&=&\!\!\!-Z\Delta_{\rm F}(x-y;M^2)-Z^*\Delta_{\rm F}(x-y;M^{*2})\,,\label{asympt-phi-propag-ghost}\\[1mm]
Z_\phi \big\langle 0\big|{\rm T}\big[\tilde{\phi}^{\rm as}(x)\tilde{\phi}^{\rm as}(y)\big]\big|0 \big\rangle \!\!\!&=&\!\!\! Z\Delta_{\rm F}(x-y;M^2)+Z^*\Delta_{\rm F}(x-y;M^{*2})\,,
\label{asympt-composit-propag-ghost}
\end{eqnarray}
where the asymptotic vacuum $|0\rangle$ is the state annihilated by the Hamiltonian associated with the asymptotic Lagrangian~\eqref{asymp-lagr-ghost}.

\medskip

\textbf{Asymptotic states.} The nontrivial quadratic dynamics of the fields $\phi^{\rm as}(x)$ and $\tilde{\phi}^{\rm as}(x)$ also induce nontrivial interference effects between their respective ``one-particle'' states.\footnote{It is worth clarifying that a two-particle state associated with the elementary field $\chi(x)$ is effectively seen as a one-particle state from the point of view of the composite field $\tilde{\phi}(x)=\frac{g}{2m\Gamma}\chi^2(x)$.} Denoting these states by $|\phi^{\rm as}(\vec{p};x^0)\rangle$ and $|\tilde{\phi}^{\rm as}(\vec{p};x^0)\rangle$, and assuming they are already normalized to $-1$ and $+1$, respectively, we can show that, in the narrow-width approximation and in the rest frame ($\vec{p}=0$), their unequal-time inner product squared reads~\cite{Buoninfante:2026mve}
\begin{equation}
\big|\big\langle \phi^{\rm as}(0;0)\big|\tilde{\phi}^{\rm as}(0;t)\big\rangle\big|^2\simeq \sinh^2\left(\frac{\Gamma t}{2}\right)\,.
\label{phi-mixed-innerproduct-squared-approx}
\end{equation}
Physically, this means that after a time $t\sim 2/\Gamma$, the negative-norm one-particle ghost state becomes non-orthogonal to a superposition of positive-norm multi-particle states; in a boosted frame ($\vec{p}\neq 0$) this timescale is given by $(\omega_{\vec{p}}/m)(2/\Gamma),$ with $\omega_{\vec{p}}=\sqrt{\vec{p}^2+m^2}.$ A key implication is that no free asymptotic one-particle ghost state exists, and thus no particle interpretation can be given to ghosts at asymptotic times. In fact, one can show that the true free asymptotic states are linear combinations of the negative- and positive-norm states above, with zero norm and hence vanishing localization probability. The physical picture that emerges is the following: the nontrivial asymptotic dynamics induced by first-sheet complex conjugate poles increases the number of asymptotic fields by one, giving rise to the additional asymptotic multi-particle state $|\tilde{\phi}^{\rm as}(\vec{p};x^0)\rangle$, which interferes with and \textit{masks} the one-particle ghost state $|\phi^{\rm as}(\vec{p};x^0)\rangle$, rendering any potential negative probability associated with the ghost unobservable~\cite{Buoninfante:2026mve}. The rest-frame timescale for this \textit{multi-particle masking} mechanism to become efficient is $2/\Gamma$. Consequently, despite the enlarged sets of asymptotic fields and states compared to the ordinary case of an unstable particle, the only well-defined LSZ states that can be attached to external legs of Feynman diagrams between $t_{\rm i}=-\infty$ and $t_{\rm f}=\infty$ remain those associated with the asymptotic field $\chi^{\rm as}(x)$.  

\medskip

\textbf{Resonance and complex mass.} In the narrow-width approximation, $\Gamma/m\ll 1$, the $\phi$ and $\chi^2$ propagators~\eqref{spectral-represent-position-ghost} and~\eqref{spectral-represent-composite-field-ghost}, in momentum space, can be approximated around $-p^2\simeq m^2$ as~\cite{Buoninfante:2025klm}
\begin{eqnarray}
\bar{G}(-p^2)\!\!\!&\simeq&\!\!\! \frac{+iZ_{\rm R}}{p^2+m^2+im\Gamma}\simeq  \frac{+iZ_{\rm R}}{p^2+(m+i\Gamma/2)^2} \,, \label{propag-ghost-narrow}\\[1mm]
\tilde{\bar{G}}(-p^2)\!\!\!&\simeq&\!\!\! \frac{-iZ_{\rm R}}{p^2+m^2+im\Gamma}\simeq  \frac{-iZ_{\rm R}}{p^2+(m+i\Gamma/2)^2} \,,
\label{propag-composite-ghost-narrow}
\end{eqnarray}
where we denoted by $\tilde{\bar{G}}(-p^2)$ the $\chi^2$ propagator in momentum space. These pole-like expressions indicate actual poles in the first Riemann sheet. The complex mass $m + i\Gamma/2$ is interpreted as follows~\cite{Buoninfante:2026mve}. The real part $m$ corresponds to the physical mass of the quasi-free ghost particle before it becomes masked by the multi-particle sector and, in resonance language, sets the position of the Breit-Wigner peak. The imaginary part $\Gamma/2$ is interpreted as the inverse of the rest-frame timescale (the \textit{masking time}) after which interference effects between the one-particle ghost state and the multi-particle component become significant, so that masking becomes order one; in resonance language, $\Gamma$ corresponds to the width of the Breit-Wigner function~at~half~maximum.

\subsection{Comparison of resonant behaviors}

We now discuss key differences between the ordinary ($a=1$) and ghost ($a=-1$) cases, which could in principle lead to distinct phenomenological signatures in scattering experiments. In particular, we consider the $\chi\chi \to \chi\chi$ process in the $s$-channel and analyze the resonant behavior associated with an intermediate $\phi$ state above the multi-particle threshold, $m>2\mu$. 

Resumming the self-energies, the corresponding amplitude reads
\begin{eqnarray}
\mathcal{M}_{\chi\chi\rightarrow \chi\chi}(-p^2)\!\!\!&\equiv&\!\!\!(-i)\,\vcenter{\hbox{\includegraphics[scale = 0.23]{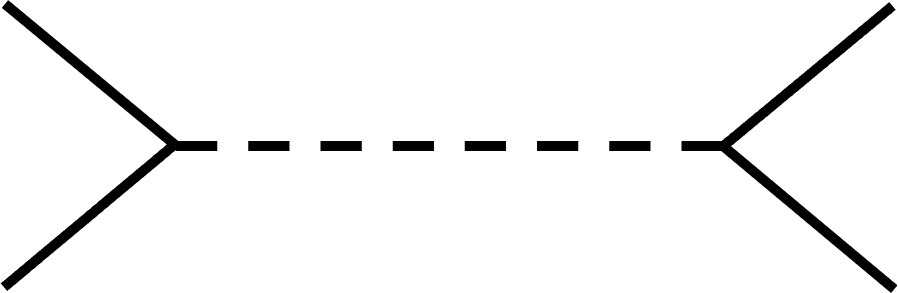}}}\,+\, (-i)\,\vcenter{\hbox{\includegraphics[scale = 0.23]{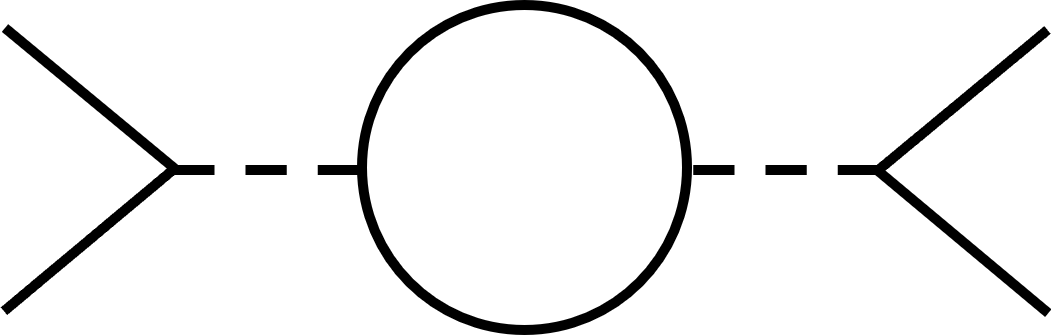}}}\,+\,(-i)\,\vcenter{\hbox{\includegraphics[scale = 0.23]{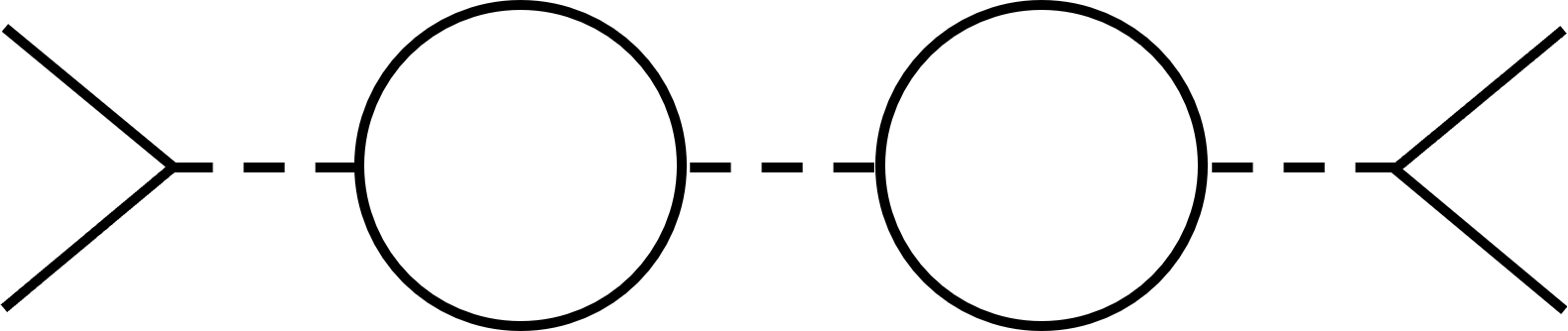}}}\,+\cdots\nonumber\\[1mm]
&=&\!\!\! (-i)(ig)^2 G(-p^2)\sum_{n=0}^\infty \left[i\Sigma(-p^2+i\epsilon)G
(-p^2) \right]^n\nonumber\\[1mm]
&=&\!\!\! g^2\frac{a}{p^2+m^2-i\epsilon-a\Sigma(-p^2+i\epsilon)}\,,
\label{chichi-scattering}
\end{eqnarray}
where $p^2$ is the total ingoing/outgoing momentum squared. The relevant physical quantity is the cross section, which is proportional to $|\mathcal{M}_{\chi\chi\rightarrow \chi\chi}|^2$ up to kinematic factors, and hence to $|\bar{G}(-p^2)|^2$. It is therefore sufficient to focus on the modulus squared of the dressed $\phi$ propagator to capture the main differences in the resonant behavior between the cases $a=1$ and $a=-1$.  

In what follows, we consider functions of the energy $e \equiv p^0$, rather than of $p^2 = -e^2 + \vec{p}^{2}$; namely, we analyze $|\bar{G}(e)|^2.$ The poles in the complex $e$ plane are given by
\begin{equation}
 \pm \sqrt{\vec{p}^2+M^2}\,,\qquad \pm \sqrt{\vec{p}^2+M^{*2}}\,.
\label{complex-energy-poles}
\end{equation}

One typically studies cross sections, and hence dressed propagators, near the resonance peak at $-p^2 \simeq m^2 \Leftrightarrow e \simeq \pm \omega_{\vec{p}}=\pm \sqrt{\vec{p}^2+m^2}$, in the narrow-width approximation $\Gamma/m \ll 1$. Taking the modulus squared of~\eqref{propag-ordinary-narrow} and~\eqref{propag-ghost-narrow}, we obtain the Breit-Wigner form
\begin{equation}
\left|\bar{G}(e\simeq \pm \omega_{\vec{p}})\right|^2\simeq \frac{a^2}{(p^2+m^2)^2+a^2m^2\Gamma^2}=\frac{1}{(e^2-\omega_{\vec{p}}^2)^2+m^2\Gamma^2}\equiv \left|\bar{G}_{\rm BW}(e)\right|^2\,,
\label{BW-function-infinite}
\end{equation}
where we have neglected the wave-function renormalization constant for simplicity, without loss of generality. The previous expression depends on $a$ only through $a^2 = (\pm 1)^2 = 1$; therefore, the approximate form~\eqref{BW-function-infinite} cannot distinguish between an ordinary and a ghost resonance and, in particular, fails to capture interference effects between positive- and negative-energy peaks.

%%%%%%%%%%%%%%%%%%%%%%%%%%%%%%%%%%%%%%%%

\begin{figure}[t!]
	\centering
	\subfloat[Subfigure 1 list of figures text][]{
		\includegraphics[scale=0.345]{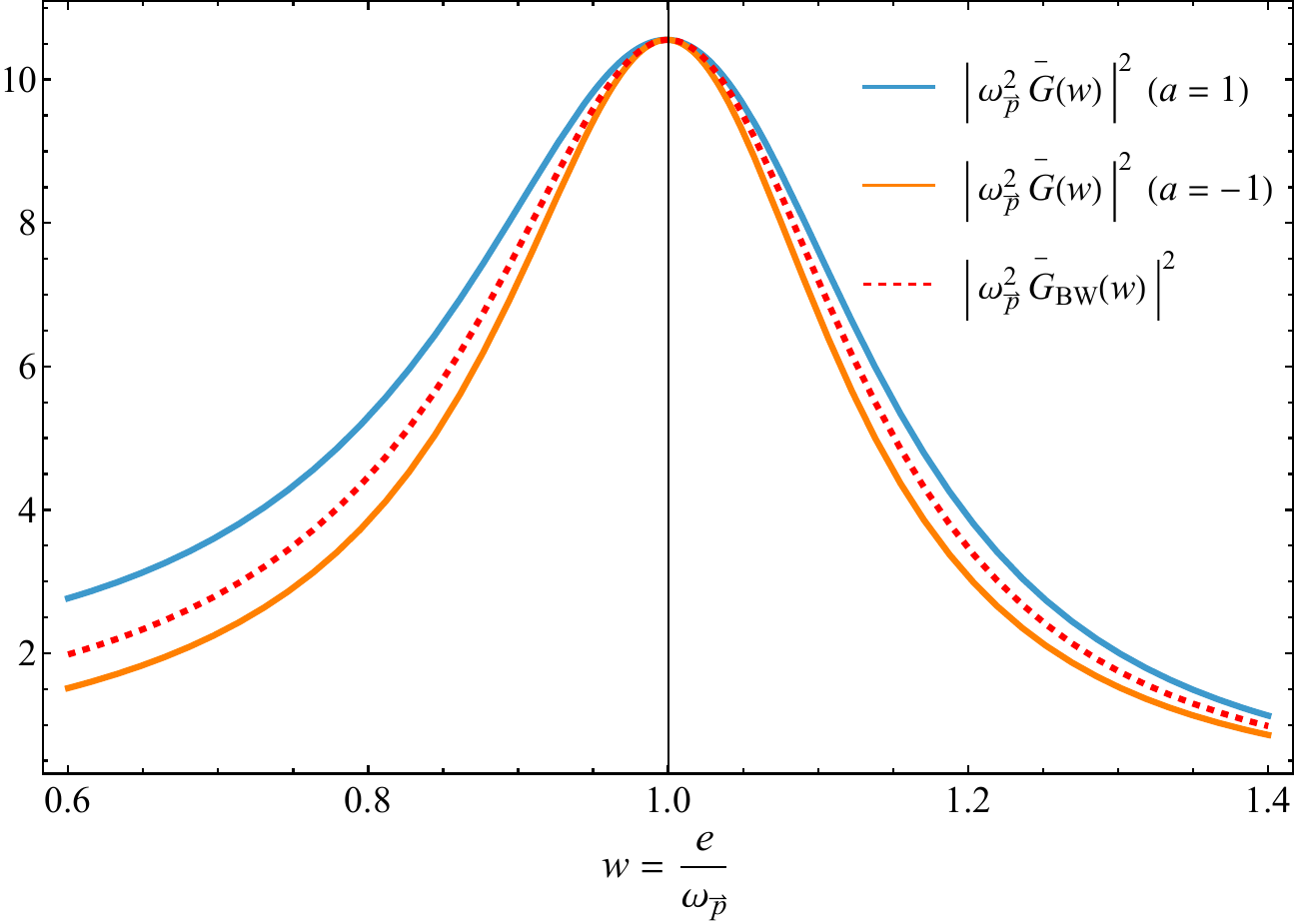}\label{fig1.a}}\quad\,
	\subfloat[Subfigure 2 list of figures text][]{
		\includegraphics[scale=0.3448]{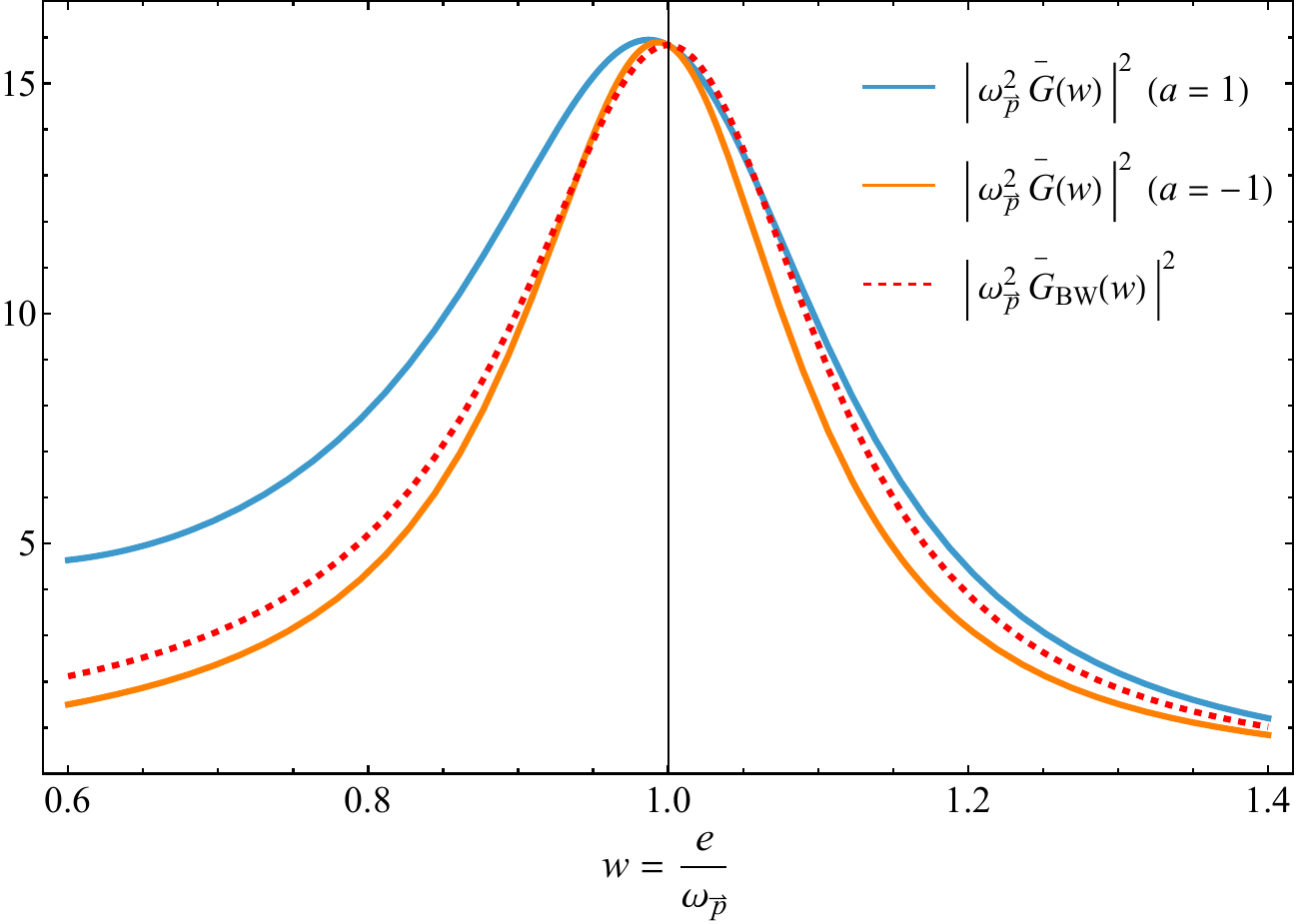}\label{fig1.b}}
	\protect\caption{The behavior of $|\omega_{\vec{p}}^2 \bar{G}(w)|^2$ as a function of $w=e/\omega_{\vec{p}}$ is shown for both the ordinary ($a=1$, blue line) and ghost ($a=-1$, orange  line) cases, and compared with the Breit-Wigner function $|\omega_{\vec{p}}^2 \bar{G}_{\rm BW}(w)|^2$ (red dashed line). For illustrative purposes, (a)~in the left panel we set $g^2/(32\pi^2\omega_{\vec{p}}^2)=0.1$, $m/\omega_{\vec{p}}=1$, and $\mu/\omega_{\vec{p}}=0.1$, which give $\Gamma/\omega_{\vec{p}}\simeq 0.31$; (b)~in the right panel the same values of parameters are chosen except for $\mu/\omega_{\vec{p}}=0.3$, giving $\Gamma/\omega_{\vec{p}}\simeq 0.25$.}
	\label{fig1}
\end{figure}

%%%%%%%%%%%%%%%%%%%%%%%%%%%%%%%%%%%%%%%%

In Fig.~\ref{fig1}, we show the modulus squared of the dressed $\phi$ propagator for both cases $a=1$ and $a=-1$, together with the approximate Breit-Wigner form~\eqref{BW-function-infinite}. We plot the dimensionless quantities $|\omega_{\vec{p}}^2\bar{G}(w)|^2$ and $|\omega_{\vec{p}}^2\bar{G}_{\rm BW}(w)|^2$ as functions of $w\equiv e/\omega_{\vec{p}}$ and, for simplicity, focus only on the positive-energy peak; an analogous behavior holds for the negative-energy peak. Note that the renormalization condition ${\rm Re}[\Sigma(m^2 \pm i m \Gamma)] \simeq {\rm Re}[\Sigma(m^2)] = 0$ is employed to eliminate the renormalization scale $\Lambda$ from the self-energy expression up to order $\mathcal{O}(g^2)$.

We find that the Breit-Wigner approximation underestimates the width at half maximum in the ordinary case, while it overestimates it in the ghost case. This implies that ghost resonances are generically narrower. In Fig.~\ref{fig1.b}, the parameters are chosen to enhance the effects of the superposition of the two peaks: the positive- (negative-) energy peak is slightly shifted to the left (right). In particular, this effect is stronger in the ordinary case, leading to a larger separation between the positive- and negative-energy peaks than in the ghost case.

\section{Finite-time formulation}\label{sec:finite-time}

All the properties discussed so far have been derived within QFT formulated in an infinite interval of time, where the initial and final states, or more generally the boundary conditions, are defined at asymptotic times $t_{\rm i} = -\infty$ and $t_{\rm f} = \infty$. Strictly speaking, this approach cannot describe the physics of an unstable particle while it is still alive, nor the propagation of a ghost before it is masked by the multi-particle sector. We now formulate the QFT in a finite interval of time, $0 < \tau \equiv t_{\rm f} - t_{\rm i} < \infty,$ to clarify several statements made in the previous section and uncover features that are not accessible in the asymptotic formulation.

When $\tau$ is finite, time-translation invariance is broken and the propagator in Fourier space depends on two independent energy variables, $e$ and $e'$, as shown in App.~\ref{sec:app-finite-time}. Nevertheless, for sufficiently large $\tau$, one can isolate the relevant leading finite-time effects and derive a suitable approximate expression for the dressed propagator as a function of a single Fourier energy $e$, despite the absence of full time-translation invariance. This large-$\tau$ approximation greatly simplifies the analysis and provides a qualitatively correct understanding of the physical implications of finite-time effects. This approach was first studied in~\cite{Anselmi:2023phm,Anselmi:2023wjx}; App.~\ref{sec:app-finite-time} reviews its main ingredients, while here we discuss additional aspects and extend the analysis to the ghost case.

In what follows, we show that finite-time effects amplify the differences between ordinary and ghost resonances discussed in the previous section and reveal additional distinctive features. We identify distinct temporal regimes and trace the onset of decay and multi-particle masking. For simplicity, here we work with a symmetric time interval, i.e. we choose $t_{\rm f}=\tau/2$ and $t_{\rm i}=-\tau/2.$

\subsection{Dressed propagators}

We start directly from the approximate expression of the Fourier transform of the dressed propagator in a finite interval of time~\eqref{dressed-propag-suitable}, whose derivation is detailed in App.~\ref{sec:app-finite-time}:
\begin{eqnarray}
	\bar{G}_\tau (e)= G_\tau (e) \sum\limits_{n=0}^\infty \left[i\Sigma(e^2)G_\tau(e)\right]^n= \frac{G_\tau(e)}{1-iG_\tau(e)\Sigma(e^2)}= \frac{1}{G^{-1}_\tau(e)-i\Sigma(e^2)}\,,
	\label{dressed-propag-suitable-main}
\end{eqnarray}
where $\Sigma(e^2)$ is the one-loop self-energy at $\tau=\infty$ discussed in the previous section, now expressed as a function of $e^2$ rather than $-p^2=e^2-\vec{p}^2$, while $G_\tau(e)$ is the Fourier transform of the free propagator in a finite time interval, given in~\eqref{free-propag-suitable}. For clarity, we report its explicit form here,
\begin{equation}
	G_\tau(e)= \frac{a\tau}{2\omega_{\vec{p}}}\left[\frac{{\rm e}^{i\kappa_+}-i\kappa_+-1}{(i\kappa_+)^2}+\frac{{\rm e}^{i\kappa_-}-i\kappa_--1}{(i\kappa_-)^2}\right]\,, \qquad \kappa_\pm\equiv (\pm e-\omega_{\vec{p}})\tau\,,
	\label{free-propag-suitable-main}
\end{equation}
and refer to App.~\ref{sec:suitable-free} for further details on its properties and derivation. Let us just recall that the free propagator~\eqref{free-propag-suitable-main} is an entire function, i.e. it has no poles for $\tau<\infty$, and exhibits a Breit-Wigner-like shape with effective width $\Gamma_{\rm eff}=6/\tau$, as shown in~\eqref{BW-form-finite-free}. As a consistency check, we can verify that the limit $\tau \to \infty$ of~\eqref{dressed-propag-suitable-main} recovers the standard dressed propagator in~\eqref{dressed-propagator}. This follows from the limits in~\eqref{limits-tau-inf} and~\eqref{recovery-feynm-propag}, which imply $G_\tau(e) \to G_\infty(e) = G(e)$.
Let us make some remarks on~\eqref{dressed-propag-suitable-main}. 

\medskip

\textbf{Series convergence.} As discussed in Sec.~\ref{sec:dressed-propag}, when $\tau=\infty$ the geometric series does not converge near the peaks if the self-energy has a non-zero imaginary part (i.e. above the multi-particle threshold), and analyticity is required to resum it everywhere except at the poles of the dressed propagator, which for $a=1$ ($a=-1$) lie on the second (first) Riemann sheet~\cite{Buoninfante:2025klm}. However, inspecting~\eqref{dressed-propag-suitable-main} we notice that there exist values of $\tau$ for which the series converges for any $e$ and $\omega_{\vec{p}}$. Using the leading-order expression of the self-energy near the peaks in the first sheet, i.e. $\Sigma(e^2 \simeq \omega_{\vec{p}}^2 + i\epsilon)=\Sigma(-p^2\simeq m^2+i\epsilon) \simeq i m \Gamma$, and the bounds~\eqref{real-imag-bounds}, the condition $|\Sigma(e^2)G_\tau(e)| < 1$ is satisfied if
\begin{equation}
\tau< \frac{2\pi}{\sqrt{4+\pi^2}}\frac{\omega_{\vec{p}}}{m}\frac{1}{\Gamma}\simeq 1.7 \frac{\omega_{\vec{p}}}{m}\frac{1}{\Gamma} \quad \Leftrightarrow \quad \Gamma_{\rm eff}>6\frac{\sqrt{4+\pi^2}}{2\pi}\frac{m}{\omega_{\vec{p}}}\Gamma\simeq 3.6\frac{m}{\omega_{\vec{p}}}\Gamma\,.
\label{convergence-condition}
\end{equation}
In the rest frame ($\vec{p} = 0$), the convergence condition reads $\tau < 1.7/\Gamma \Leftrightarrow \Gamma_{\rm eff} > 3.6 \,\Gamma$. The inequality~\eqref{convergence-condition} is compatible with the large-$\tau$ approximation ($\tau \gg 1/m > 1/\omega_{\vec{p}}$) underlying our approach and discussed in~\ref{sec:suitable-free} and~\ref{sec:suitable-dressed}, provided the resonance is narrow ($\Gamma \ll m < \omega_{\vec{p}}$):
\begin{equation}
1.7 \frac{\omega_{\vec{p}}}{m}\frac{1}{\Gamma}>1.7\frac{1}{\Gamma}>\tau \gg \frac{1}{m} >\frac{1}{\omega_{\vec{p}}}\,.
\label{convergence-condition-compatibility}
\end{equation}
This set of inequalities defines a particularly relevant temporal regime, allowing us to study an ordinary unstable particle before decay ($a=1$) and a ghost before it is masked by the multi-particle sector ($a=-1$), as discussed in Sec.~\ref{sec:two-diff-time-regime}.

\medskip

\textbf{Total width.} Radiative corrections introduce the width $\Gamma$ in addition to $\Gamma_{\rm eff} = 6/\tau$. An estimation of the total width $\Gamma_{\rm tot}$ of the dressed propagator can be obtained by analyzing the behavior of~\eqref{dressed-propag-suitable-main} near the peaks. Expanding around $e \simeq \pm \omega_{\vec{p}}$, we~get\footnote{\label{foot-limit}Note that naively taking the limit $\tau \to \infty$ in~\eqref{BW-form-finite-dressed} does not reproduce $\frac{\pm ia}{2\omega_{\vec{p}}}\frac{1}{\pm e-\omega_{\vec{p}}+ia\Gamma/2}$. This is because the infinite-$\tau$ result requires $(e\pm \omega_{\vec{p}})\tau \to \infty$, which corresponds to a different asymptotic region of the multi-variable function $\bar{G}_\tau (e)\equiv f(e,\tau)$. Here, instead, we are interested in the region $|\pm e-\omega_{\vec{p}}|\tau \ll 1$, with $|e|\tau,\,\omega_{\vec{p}}\tau \gg 1$. Within the large-$\tau$ approximation under consideration, according to which the free propagator is given by the oscillatory approximate expression in~\eqref{free-propag-suitable-main}, the $3/2$ mismatch in the coefficient is therefore expected. Nevertheless, the Breit-Wigner-like behavior and the total width $|\Gamma_{\rm tot}|$ derived in~\eqref{BW-form-finite-dressed} provide good qualitative approximations to the modulus squared of the dressed propagator~\eqref{dressed-propag-suitable-main} near the peaks and to its width at half maximum.}
\begin{eqnarray}
\bar{G}_\tau (e\simeq \pm\omega_{\vec{p}})\simeq \frac{3}{4\omega_{\vec{p}}}\frac{\pm i a}{e\mp \omega_{\vec{p}}\pm i\Gamma_{\rm tot}/2}\equiv \bar{G}^\pm_{\tau{\rm BW}}(e)\,,\qquad \Gamma_{\rm tot}\equiv \frac{6}{\tau}+a\frac{3}{2}\frac{m}{\omega_{\vec{p}}}\Gamma\,,
\label{BW-form-finite-dressed}
\end{eqnarray}
where, for simplicity and without loss of generality, we set the wave-function renormalization constant to one. The Breit-Wigner-like functions $\bar{G}^\pm_{\tau{\rm BW}}(e)$ are defined only near their respective peaks; in the rest frame, the total width reads $6/\tau + 3a\Gamma/2$. For $\Gamma_{\rm eff} > 3m\Gamma/(2\omega_{\vec{p}})$ the total width is always positive, whereas for sufficiently large $\tau$ one has $\Gamma_{\rm eff} < 3m\Gamma/(2\omega_{\vec{p}})$, in which case the total width is positive (negative) for an ordinary (a ghost) resonance. Note that the actual width at half maximum is given by the modulus $|\Gamma_{\rm tot}|$.

\medskip

\textbf{Height of the peaks.} The total width is directly related to the maximum of the modulus squared of the dressed propagator, which is given by
\begin{equation}
\max \big|\bar{G}_\tau(e)\big|^2\simeq \max \big|\bar{G}^{\pm}_{\tau{\rm BW}}(e)\big|^2=  \frac{9}{4\Gamma_{\rm tot}^2}\,.
\label{BW-maximum}
\end{equation}
This implies that the resonance peaks around $e \simeq \pm \omega_{\vec{p}}$ are higher in the ghost case ($a=-1$) than in the ordinary case ($a=1$). This distinctive feature is not visible when $\tau=\infty.$

%%%%%%%%%%%%%%%%%%%%%%%%%%%%%%%%%%%%%%%%

\begin{figure}[t!]
	\centering
	\subfloat[Subfigure 1 list of figures text][]{
		\includegraphics[scale=0.34]{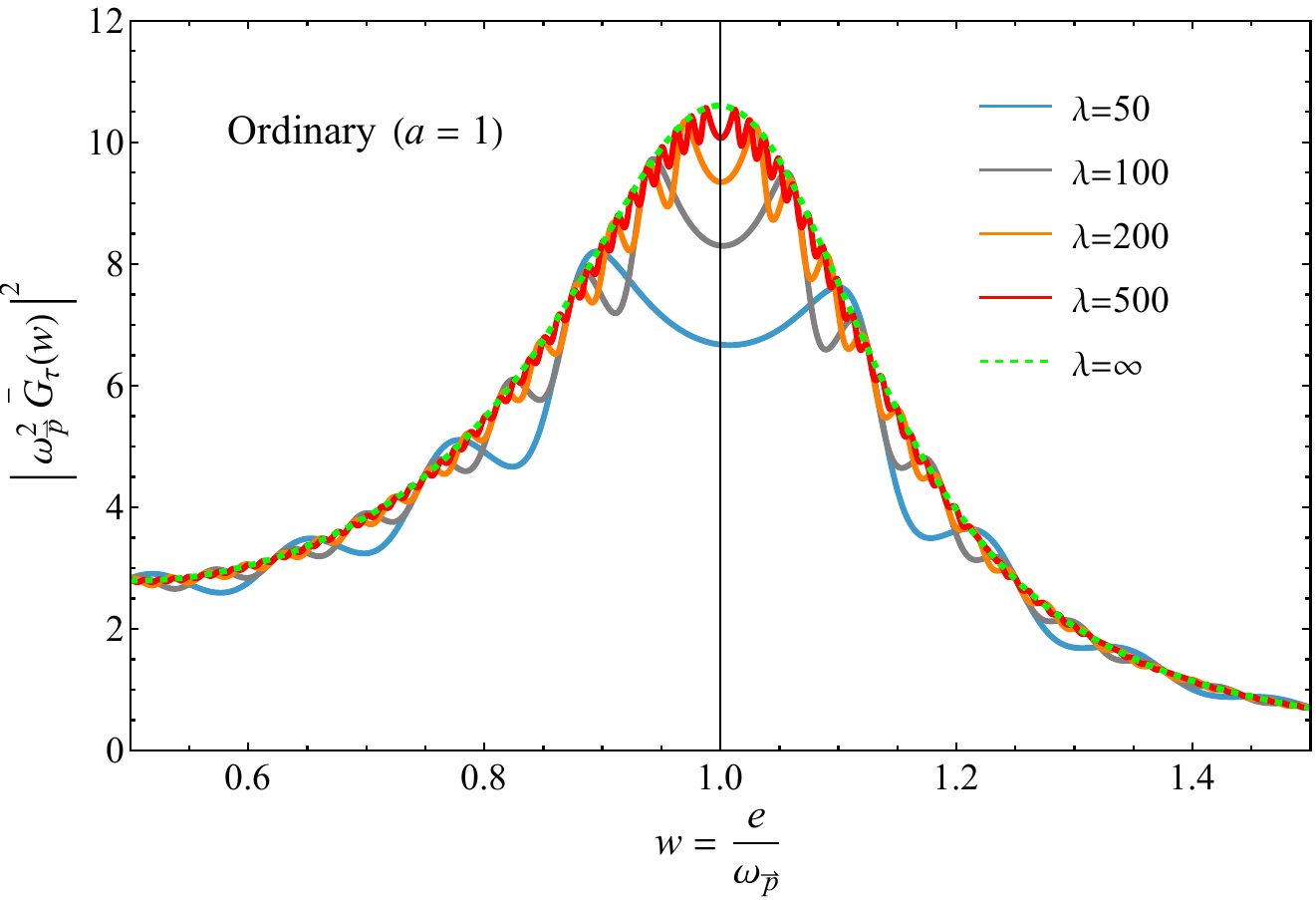}\label{fig2.a}}\quad\,
	\subfloat[Subfigure 2 list of figures text][]{
		\includegraphics[scale=0.34]{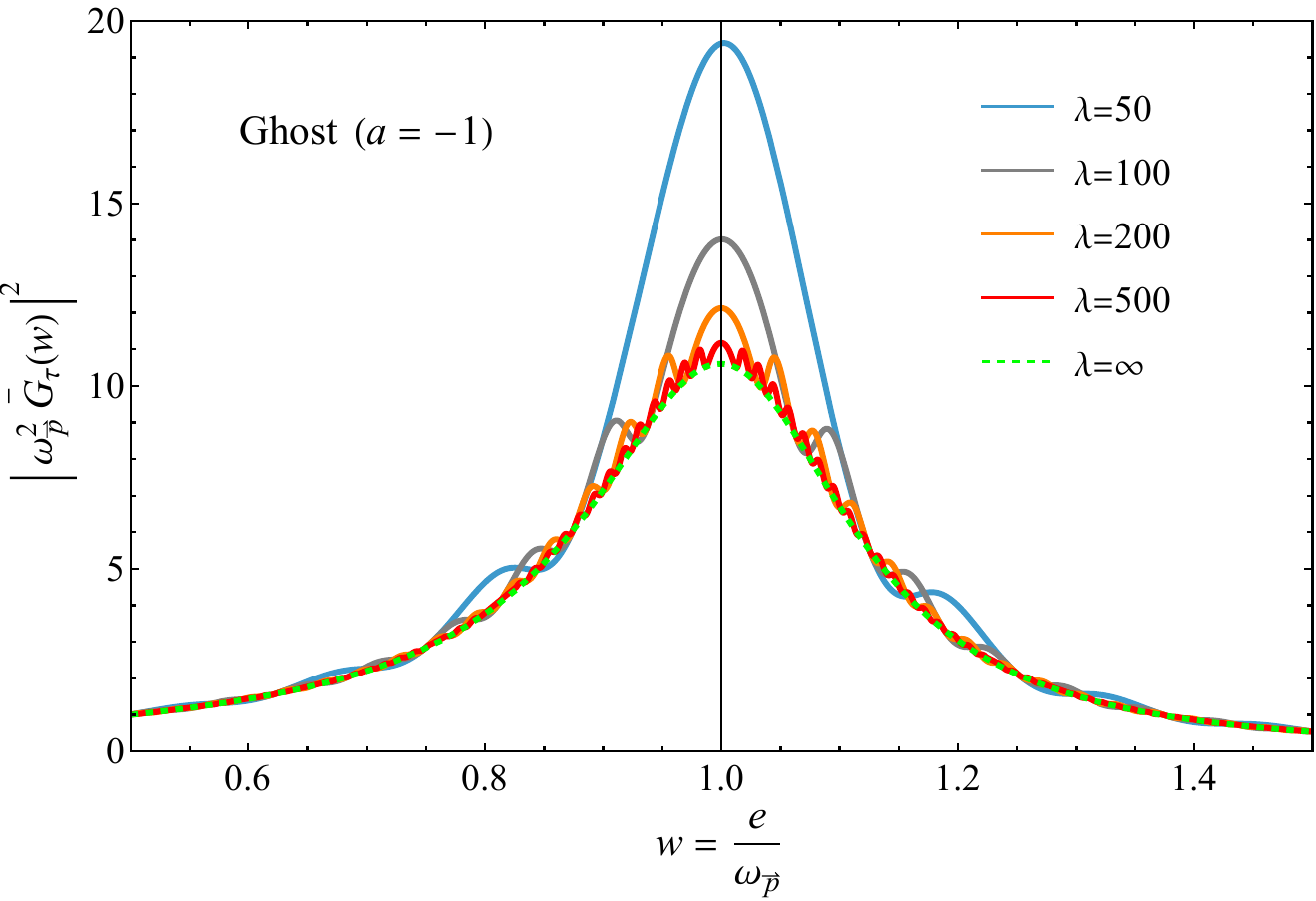}\label{fig2.b}}\,
	\subfloat[Subfigure 1 list of figures text][]{
		\includegraphics[scale=0.34]{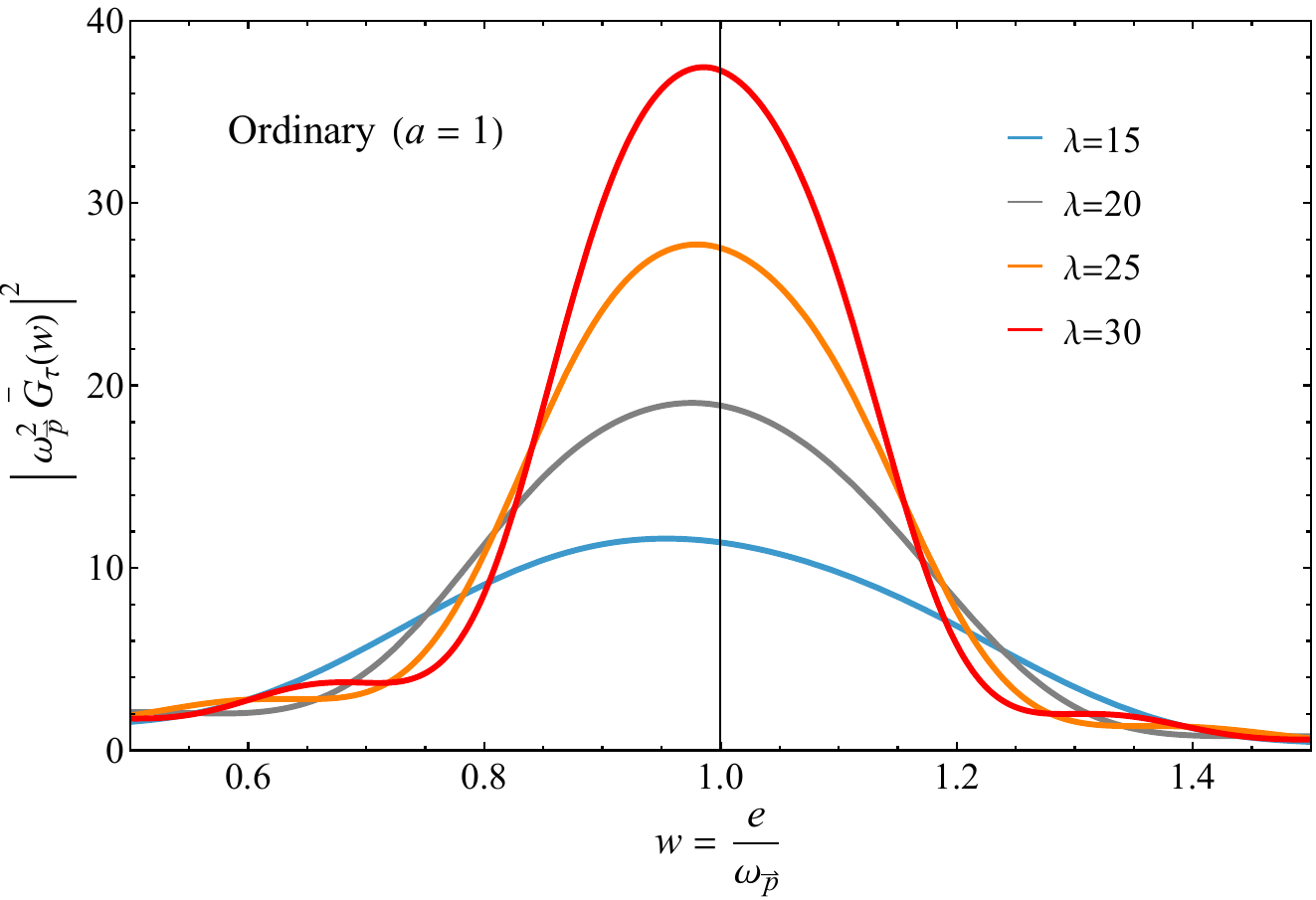}\label{fig2.c}}\quad\,
	\subfloat[Subfigure 2 list of figures text][]{
		\includegraphics[scale=0.34]{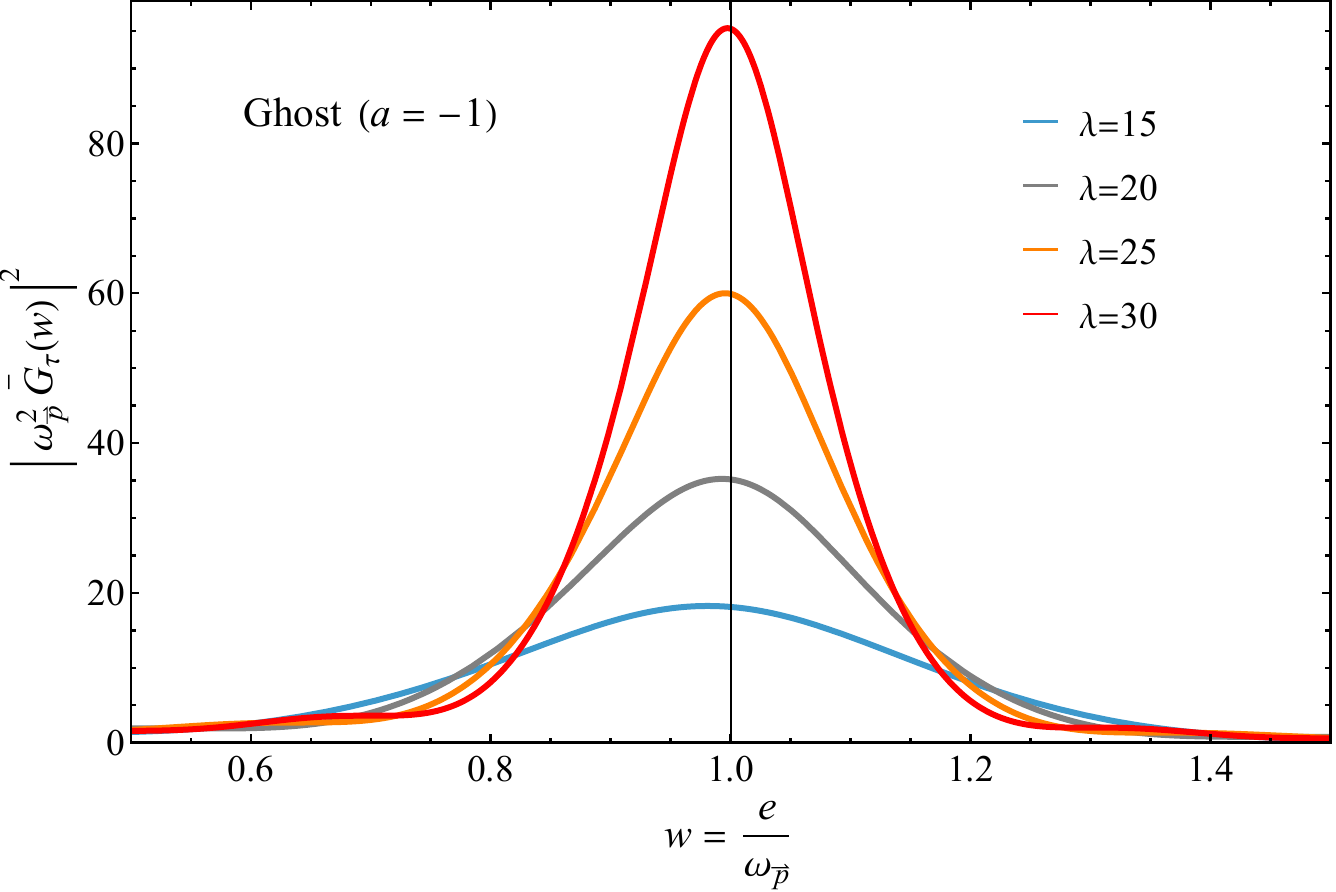}\label{fig2.d}}
	\protect\caption{The behavior of $|\omega_{\vec{p}}^2 \bar{G}_\tau(w)|^2$ as a function of $w = e/\omega_{\vec{p}}$ is shown for different values of $\lambda \equiv \omega_{\vec{p}}\tau$. Panels (a) and (b) contain the plots with $\lambda = 50$ (blue line), $\lambda = 100$ (gray line), $\lambda = 200$ (orange line), $\lambda = 500$ (red line), and $\lambda = \infty$ (green dashed line), for $g^2/(32\pi^2 \omega_{\vec{p}}^2) = 0.1$. Panels (c) and (d) contain the plots with $\lambda = 15$ (blue line), $\lambda = 20$ (gray line), $\lambda = 25$ (orange line), and $\lambda = 30$ (red line), for $g^2/(32\pi^2 \omega_{\vec{p}}^2) = 0.01$. The left panels correspond to the ordinary case, while the right panels to the ghost case. In all plots we set $m/\omega_{\vec{p}} = 0.95$ and $\mu/\omega_{\vec{p}} = 0.1$.}
	\label{fig2}
\end{figure}

%%%%%%%%%%%%%%%%%%%%%%%%%%%%%%%%%%%%%%%%

\medskip

\textbf{Oscillatory behavior.} In Fig.~\ref{fig2}, we plot $|\bar{G}_\tau(w)|^2$ as a function of $w=e/\omega_{\vec{p}}$ for different values of $\lambda\equiv\omega_{\vec{p}}\tau$, separately for $a=1$ (Figs.~\ref{fig2.a},~\ref{fig2.c}) and $a=-1$ (Figs.~\ref{fig2.b},~\ref{fig2.d}), focusing on the positive-energy peak. The oscillations visible in Figs.~\ref{fig2.a} and~\ref{fig2.b} originate from the oscillatory terms in~\eqref{free-propag-suitable-main} and arise for sufficiently large $\tau$ when the convergence condition~\eqref{convergence-condition} is violated, namely when $\Gamma_{\rm eff}<3.6\,\Gamma$. By contrast, no such oscillations appear in Figs.~\ref{fig2.c} and~\ref{fig2.d}, where the convergence condition is satisfied, i.e. $\Gamma_{\rm eff}>3.6\,\Gamma$, since the corresponding values of $\tau$ lie within a single period of the sine and cosine functions. As a consistency check of the finite-time approach, Figs.~\ref{fig2.a} and~\ref{fig2.b} show that the dressed propagator approaches the $\tau=\infty$ expression in the limit $\tau\to\infty$, despite the presence of oscillations.

\medskip

\begin{figure}[t!]
	\centering
	\subfloat[Subfigure 1 list of figures text][]{
		\,\,\,\includegraphics[scale=0.323]{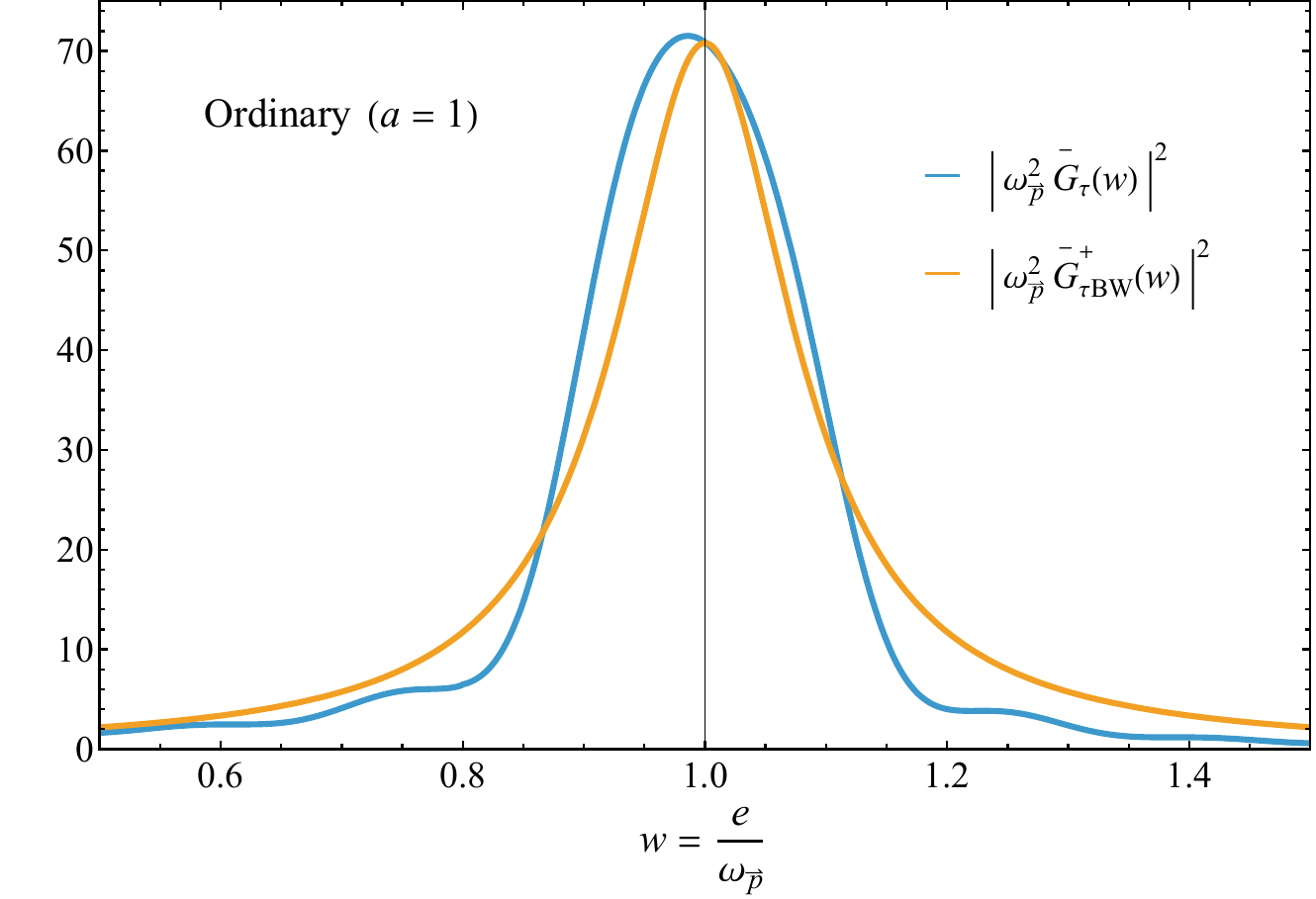}\label{fig3.a}}\quad\,\,\,\,\,\,
	\subfloat[Subfigure 2 list of figures text][]{
		\includegraphics[scale=0.33]{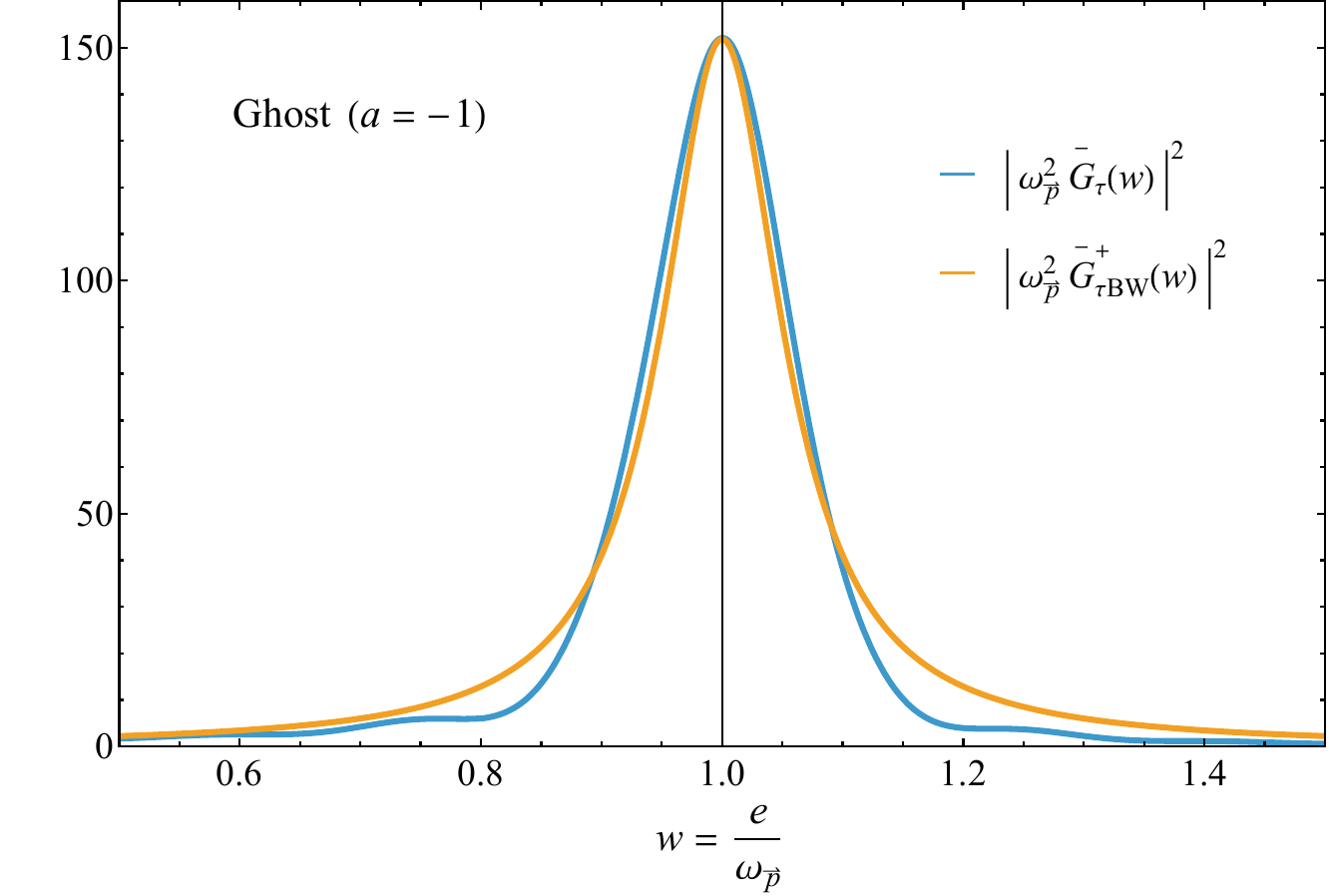}\label{fig3.b}}\,
	\!\!\!	\subfloat[Subfigure 1 list of figures text][]{
		\includegraphics[scale=0.3357]{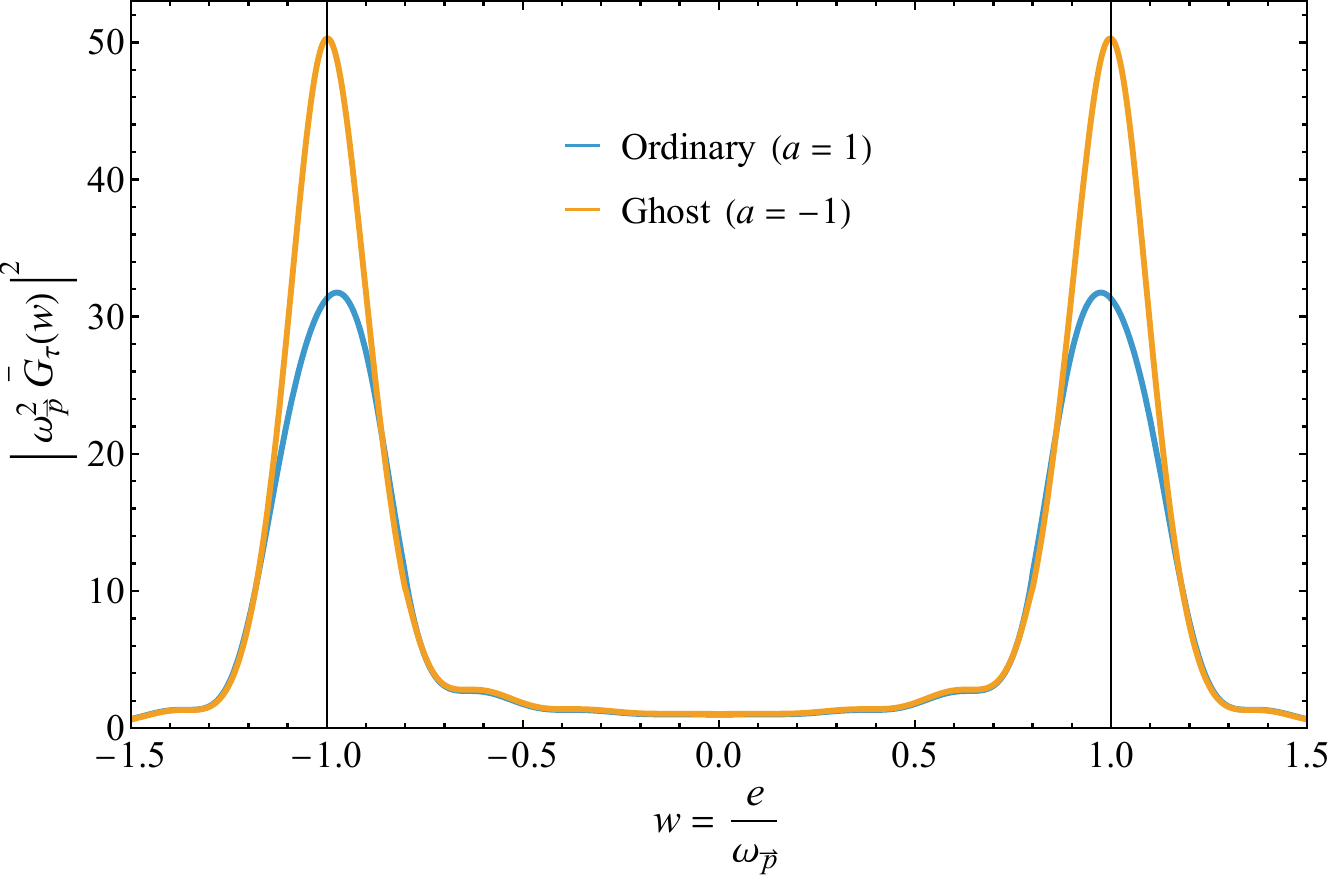}\label{fig3.c}}\quad
	\subfloat[Subfigure 2 list of figures text][]{
		\includegraphics[scale=0.345]{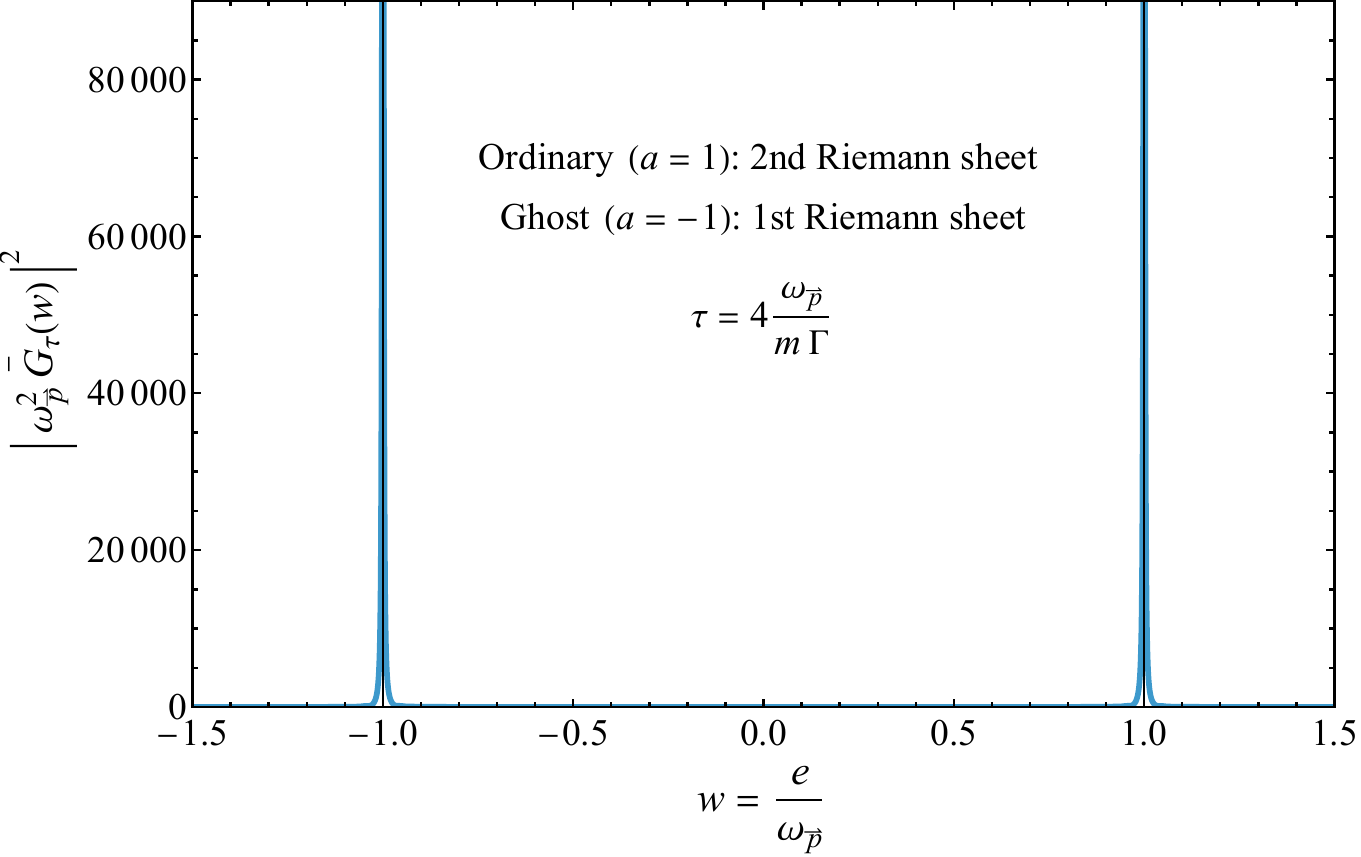}\label{fig3.d}}
	\protect\caption{In the upper panels, the behaviors of $|\omega_{\vec{p}}^2 \bar{G}_\tau(w)|^2$ (blue line) and $|\omega_{\vec{p}}^2 \bar{G}^+_{\tau {\rm BW}}(w)|^2$ (orange line) around the positive-energy peak are compared for $\lambda\equiv \omega_{\vec{p}}\tau=40;$ (a) corresponds to the ordinary case, while  (b) to the ghost case.
	Panel (c)~shows $|\omega_{\vec{p}}^2 \bar{G}_\tau(w)|^2$  as a function of $w = e/\omega_{\vec{p}}$ in the ordinary (blue line) and ghost (orange line) cases, for $\lambda = \omega_{\vec{p}}\tau = 25$. Panel (d)~shows $|\omega_{\vec{p}}^2 \bar{G}_\tau(w)|^2$ in the ordinary (ghost) case in the second (first) Riemann sheet, for $\tau = 4\omega_{\vec{p}}/(m\Gamma)\Leftrightarrow\Gamma_{\rm eff}= 3m\Gamma/(2\omega_{\vec{p}})$. In all plots we set $m/\omega_{\vec{p}} = 1$, $\mu/\omega_{\vec{p}} = 0.4$, and $g^2/(32\pi^2 \omega_{\vec{p}}^2) = 0.01$.}
	\label{fig3}
\end{figure}

\textbf{Peaks separation.} From Figs.~\ref{fig2.c} and~\ref{fig2.d}, we also learn that in the ordinary case the two peaks around $e\simeq \pm \omega_{\vec{p}}$ are closer, with the positive- (negative-) energy peak slightly shifted to the left (right). In the ghost case, the effect of this peak superposition is weaker due to the negative radiative contribution to the total width. In Figs.~\ref{fig3.a} and~\ref{fig3.b}, we compare $|\bar{G}_\tau(w)|^2$ with the corresponding Breit-Wigner-like function $|\bar{G}^+_{\tau{\rm BW}}(w)|^2$ introduced in~\eqref{BW-form-finite-dressed}. Since the latter is defined only around a single peak, it does not capture the interference between the positive- and negative-energy peaks. Consequently, the Breit-Wigner approximation is more accurate in the ghost case, where superposition effects are weaker. In Fig.~\ref{fig3.c}, we show the ordinary and ghost cases in a single plot, highlighting differences in the height, width, and separation of the peaks.

%%%%%%%%%%%%%%%%%%%%%%%%%%%%%%%%%%%%%%%%

%%%%%%%%%%%%%%%%%%%%%%%%%%%%%%%%%%%%%%%%

\subsection{Transition point and poles emergence}\label{sec:transition-point}

Although we have not been able to derive the dressed propagator in a finite time interval without relying on approximations, its exact time-dependent expression is expected to have compact temporal support. The Paley-Wiener theorem then implies that its Fourier transform is an entire function of the energy variables. Thus, real or complex energy poles can arise only asymptotically, i.e. as $\tau\rightarrow\infty$. However, our large-$\tau$ approximation introduces apparent complex energy poles even at finite times. Despite being artefacts of the approximation, these poles, if properly interpreted, provide a useful way to track the emergence of the complex conjugate poles of the asymptotic formulation, as they become complex-conjugate pairs only in the limit $\tau\rightarrow\infty$.

\subsubsection*{Ghost case} 

For values of $\tau$ satisfying~\eqref{convergence-condition}, the approximate form of the dressed propagator~\eqref{dressed-propag-suitable-main} has no poles, since $i\Sigma(e^2)G_\tau(e)\neq 1$. On the other hand, for $a=-1$, there exists a value of $\tau$ beyond the convergence condition~\eqref{convergence-condition} at which the total width in~\eqref{BW-form-finite-dressed} vanishes, $\Gamma_{\rm tot}=0$, namely
\begin{equation}
\tau=4\frac{\omega_{\vec{p}}}{m\Gamma}\quad \Leftrightarrow \quad \Gamma_{\rm eff}=\frac{3}{2}\frac{m\Gamma}{\omega_{\vec{p}}} \,.
\label{zero-total-width}
\end{equation}
At this point, the dressed propagator develops a pair of real energy poles, as shown in Fig.~\ref{fig3.d}. This feature can be understood analytically by expanding~\eqref{dressed-propag-suitable-main} around the peaks and analyzing the zeros of the denominator. Focusing, for instance, on the positive-energy peak, we have
\begin{equation}
	\bar{G}^{-1}_\tau (e\simeq \omega_{\vec{p}})\simeq  -\frac{4\omega_{\vec{p}}}{3i}\left[e-\omega_{\vec{p}}+\frac{i}{2}\Gamma_{\rm eff}-\frac{3}{4\omega_{\vec{p}}}\Sigma(e^2\simeq \omega_{\vec{p}}^2+i\epsilon) \right]\,,
	\label{pole-ghost}
\end{equation}
where the usual $\epsilon \to 0^+$ indicates that the self-energy is continued onto the real axis from above. Using the renormalization condition ${\rm Re}[\Sigma(-p^2\simeq m^2)]={\rm Re}[\Sigma(e^2\simeq \omega_{\vec{p}}^2)]=0$ and the relation
\begin{equation}
	{\rm Im}\left[\Sigma(e^2\simeq \omega_{\vec{p}}^2 + i\epsilon)\right]={\rm Im}\left[\Sigma(-p^2\simeq m^2+ i\epsilon)\right]\simeq m\Gamma\,,
	\label{imag-self-en-energy-first}
\end{equation}
valid in the first Riemann sheet and in the narrow-width approximation, we obtain
\begin{equation}
	\bar{G}^{-1}_\tau (e\simeq \omega_{\vec{p}})\simeq  -\frac{4\omega_{\vec{p}}}{3i}\left[e-\omega_{\vec{p}}+\frac{i}{2}\left(\Gamma_{\rm eff}-\frac{3}{2}\frac{m}{\omega_{\vec{p}}}\Gamma\right) \right]\,,
	\label{pole-ghost-2}
\end{equation}
which has a zero at $e=\omega_{\vec{p}}$ when~\eqref{zero-total-width} holds; analogous arguments apply to $e=-\omega_{\vec{p}}$. 

These real poles lie on the first sheet of the complex energy plane and act as precursors of the complex conjugate poles appearing in the infinite-$\tau$ formulation. Indeed, for time intervals $\tau > 4\omega_{\vec{p}}/(m\Gamma)$, the denominator~\eqref{pole-ghost-2} develops the following complex zeros:
\begin{eqnarray}
\nu_{1}^\pm \!\!\!&\equiv&\!\!\!\pm \left(\omega_{\vec{p}} +\frac{i}{2}\gamma_1 \right)\,,\qquad \gamma_1\equiv  -\Gamma_{\rm eff}+\frac{3}{2}\frac{m\Gamma}{\omega_{\vec{p}}}>0\,,\label{complex-poles-positive}\\[1mm]
\nu_{2}^\pm\!\!\!&\equiv&\!\!\! \pm \left(\omega_{\vec{p}} -\frac{i}{2}\gamma_2 \right)\,,\qquad \gamma_2\equiv  \Gamma_{\rm eff}+\frac{3}{2}\frac{m\Gamma}{\omega_{\vec{p}}}>0\,,
\label{complex-poles-negative}
\end{eqnarray}
which can be found in the narrow-width approximation $\gamma_1/\omega_{\vec{p}}\ll 1$ and $\gamma_2/\omega_{\vec{p}}\ll 1$ by using ${\rm Im}[\Sigma(\omega_{\vec{p}}^2+i\omega_{\vec{p}}\gamma_1)]\simeq m\Gamma$ and ${\rm Im}[\Sigma(\omega_{\vec{p}}^2-i\omega_{\vec{p}}\gamma_2)]\simeq -m\Gamma$.
Note that for finite values of $\tau$ these poles are not related by complex conjugation, i.e. $\gamma_1 \neq \gamma_2$ and $\nu_1^\pm \neq \nu_2^{\pm *}$; this feature is expected because the free propagator in a finite interval of time does not satisfy the reflection property, namely $G_\tau^*(e) \neq -G_\tau(e^*)$ for $\tau < \infty$. Only in the limit $\tau\to\infty \Leftrightarrow \Gamma_{\rm eff}\rightarrow 0$, we recover $G_{\tau=\infty}^*(e) = -G_{\tau=\infty}(e^*)$ and $\gamma_1=\gamma_2$, so that~\eqref{complex-poles-positive} and~\eqref{complex-poles-negative} become complex-conjugate pairs.\footnote{\label{footnote-complex}Within the large-$\tau$ approximation under consideration, the exact expressions for the complex conjugate poles~\eqref{complex-energy-poles} of the infinite-$\tau$ formulation are recovered only if the limit $\tau \to \infty$ is taken before expanding around the peaks; see also the remark in Footnote~\ref{foot-limit}. An improved large-$\tau$ approximation will be studied in future work.}

\subsubsection*{Ordinary case} 

An analogous pole structure arises for $a=1$ on the second Riemann~sheet, for example the real poles are shown in Fig.~\ref{fig3.d}. In this case,~\eqref{pole-ghost} and~\eqref{imag-self-en-energy-first} are replaced by
\begin{equation}
	\bar{G}^{II-1}_\tau (e\simeq \omega_{\vec{p}})\simeq  \frac{4\omega_{\vec{p}}}{3i}\left[e-\omega_{\vec{p}}+\frac{i}{2}\Gamma_{\rm eff}+\frac{3}{4\omega_{\vec{p}}}\Sigma^{II}(e^2\simeq \omega_{\vec{p}}^2+i\epsilon) \right]\,,
	\label{pole-ordinary}
\end{equation}
and
\begin{equation}
	{\rm Im}\left[\Sigma^{II}(e^2\simeq \omega_{\vec{p}}^2 + i\epsilon)\right]={\rm Im}\left[\Sigma^{II}(-p^2\simeq m^2+ i\epsilon)\right]\simeq -m\Gamma\,,
	\label{imag-self-en-energy-second}
\end{equation}
respectively. Consequently, we obtain
\begin{equation}
	\bar{G}^{II-1}_\tau (e\simeq \omega_{\vec{p}})\simeq  \frac{4\omega_{\vec{p}}}{3i}\left[e-\omega_{\vec{p}}+\frac{i}{2}\left(\Gamma_{\rm eff}-\frac{3}{2}\frac{m}{\omega_{\vec{p}}}\Gamma\right) \right]\,,
	\label{pole-ordinary-2}
\end{equation}
which has a zero at $e=\omega_{\vec{p}}$ when~\eqref{zero-total-width} holds; analogous arguments apply to $e=-\omega_{\vec{p}}$. 

The discussion surrounding~\eqref{complex-poles-positive} and~\eqref{complex-poles-negative} carries over straightforwardly to the ordinary case, with the only difference being that the complex poles now lie on the second Riemann sheet. We therefore do not repeat it here.

\subsubsection*{Transition point} 

The emergence of real poles on the second or first Riemann sheet when~\eqref{zero-total-width} is satisfied marks a transition between an early-time regime, $\tau \ll 4\omega_{\vec{p}}/(m\Gamma)$, in which an unstable particle or a ghost propagates quasi-freely, and a late-time resonant regime, $\tau \gg 4\omega_{\vec{p}}/(m\Gamma)$, in which the unstable particle has decayed or the ghost is effectively masked by the multi-particle sector. In what follows, we analyze these two regimes and show that they correspond to two distinct absorptive contributions of the dressed~propagator.

\subsection{Distinct temporal regimes}\label{sec:two-diff-time-regime}

The absorptive part of the dressed propagator~\eqref{dressed-propag-suitable-main} has two different contributions:
\begin{equation}
{\rm Re}\left[\bar{G}_\tau(e)\right]= {\rm Im}\left[i\bar{G}_\tau(e)\right] = A_\tau(e)+B_\tau(e)\,,
\label{dressed-absorptive}
\end{equation}
where 
\begin{eqnarray}
A_\tau(e)\!\!\!&\equiv&\!\!\! \frac{{\rm Re}[G_\tau(e)]}{|1-iG_\tau(e)\Sigma(e^2+i\epsilon)|^2}=\bar{G}_\tau(e)\,\frac{{\rm Re}[G_\tau(e)]}{|G_\tau(e)|^2}\,\bar{G}^*_\tau(e) \,,\label{dress-abs-A}\\[1.5mm]
B_\tau(e)\!\!\!&\equiv&\!\!\! \frac{|G_\tau(e)|^2\, {\rm Im}[\Sigma(e^2+i\epsilon)]}{|1-iG_\tau(e)\Sigma(e^2+i\epsilon)|^2}=\bar{G}_\tau(e)\,{\rm Im}[\Sigma(e^2+i\epsilon)]\,\bar{G}^*_\tau(e) \label{dress-abs-B}
\end{eqnarray}
are the finite-$\tau$ analogues of the expressions~\eqref{A-abs} and~\eqref{B-abs} as functions of the energy $e$, namely $A_{\tau=\infty}(e)=A(e)$ and $B_{\tau=\infty}(e)=B(e),$ respectively. However, an important difference is that $A(e)$ in~\eqref{A-abs} vanishes, since $\epsilon$ is merely a mathematical artefact, whereas $A_\tau(e)$ in~\eqref{dress-abs-A} is non-zero and contributes non-trivially to the absorptive part of the dressed propagator~when~$\tau <\infty.$

Working in the peak region and in the narrow-width approximation, we can identify two distinct early- and late-time regimes corresponding to $|A_\tau(e)| \gg |B_\tau(e)|$ and $|A_\tau(e)| \ll |B_\tau(e)|$, respectively. We now analyze them in detail.

\subsubsection*{Early-time regime}

Let us first consider the early-time regime in which the time interval remains shorter than the lifetime of an ordinary unstable particle or the masking time of a ghost ($\tau < 1/\Gamma < \omega_{\vec{p}}/(m\Gamma)$), while the large-$\tau$ ($\omega_{\vec{p}}\tau \gg 1$) and narrow-width ($\Gamma/\omega_{\vec{p}} < \Gamma/m \ll 1$) approximations are still valid. To simplify the analysis, we restrict to
\begin{equation}
\frac{\omega_{\vec{p}}}{m}\frac{1}{\Gamma}>\frac{1}{\Gamma}\gg \tau \gg \frac{1}{m}>\frac{1}{\omega_{\vec{p}}}\,.
\label{ineq-A-dominant}
\end{equation}

Using the approximations $|{\rm Re}[G_\tau(e)]|\simeq \tau/(4\omega_{\vec{p}})$ and $|G_\tau(e)|^2\simeq \tau^2/(16\omega^2_{\vec{p}})$ valid in the peak region, and the inequalities~\eqref{ineq-A-dominant}, we obtain
\begin{equation}
	\left|\frac{A_\tau(e)}{B_\tau(e)}\right|= \frac{\left|{\rm Re}[G_\tau(e)]\right|}{\left|G_\tau(e)\right|^2\left|{\rm Im}[\Sigma(e^2)]\right|}\simeq \frac{\omega_{\vec{p}}}{m}\frac{4}{\tau\Gamma}\gg 1\,.
	\label{A-dominant}
\end{equation}
This confirms that in the finite-time interval~\eqref{ineq-A-dominant}, the absorptive term $A_\tau(e)$ dominates over $B_\tau(e)$, thus describing processes admitting an approximate free-particle interpretation, both for unstable particles before decay and for ghosts before multi-particle masking. 
In other words, $|G_\tau(e)\Sigma(e^2)| \lesssim (m/\omega_{\vec{p}})\tau\Gamma \ll 1$ implies $|1 - iG_\tau(e)\Sigma(e^2)| \simeq 1$, so that the dressed propagator can be approximated by its free behavior and $A_\tau(e) \simeq \mathrm{Re}[G_\tau(e)]$.

In this regime, the geometric series in~\eqref{dressed-propag-suitable-main} converges and the dressed propagator has no poles. Nevertheless, an on-shell Dirac delta can still be obtained approximately. Given the form of ${\rm Re}[G_\tau(e)]$ in~\eqref{real-part-free-propag-finite-t} and using the following distributional limit~\cite{Anselmi:2023wjx},
\begin{equation}
\lim\limits_{\lambda\rightarrow \infty}\frac{1-\cos (x \lambda)}{x^2 \lambda}=\pi \delta(x)\,,\qquad x\equiv\bigg(\pm \frac{e}{\omega_{\vec{p}}}-1\bigg)\,, \quad \lambda\equiv \omega_{\vec{p}}\tau\,,
\label{distr-limit}
\end{equation}
we obtain the approximate expression
\begin{equation}
A_\tau(e)\simeq {\rm Re}[G_\tau(e)] \simeq   \frac{a\pi}{2\omega_{\vec{p}}}\left[\delta(e-\omega_{\vec{p}})+\delta(e+\omega_{\vec{p}})\right]=a\pi \delta(e^2-\omega^2_{\vec{p}})=a\pi \delta(p^2+m^2)\,.
\label{dirac-delta-limit}
\end{equation}
The previous equation does not mean that we are taking the strict limit $\tau \to \infty$; in fact, the latter would yield $A_{\tau=\infty}(e) = A(e)$, which is proportional to $\epsilon \to 0^+$ and therefore vanishes; see~\eqref{A-abs} and the discussion below it. Instead,~\eqref{dirac-delta-limit} can be justified by taking $\Gamma/m \to 0$ and  $\omega_{\vec{p}}\tau \to \infty$ in an approximate sense, while keeping $\tau < \infty$ and the product $(\omega_{\vec{p}}\tau)(\Gamma/m)$ fixed.
Moreover, $|e|\tau$ must be large and comparable to $\omega_{\vec{p}}\tau$ in order for the Dirac delta to have non-vanishing support.

\subsubsection*{Late-time regime}

Let us now consider the late-time regime described by the set of inequalities
\begin{equation}
\tau \gg \frac{\omega_{\vec{p}}}{m}\frac{1}{\Gamma}>\frac{1}{\Gamma}\gg \frac{1}{m}>\frac{1}{\omega_{\vec{p}}}\,,
\label{ineq-B-dominant}
\end{equation}
which remains consistent with the large-$\tau$ and narrow-width approximations. 

For $\tau = \infty$, it follows from~\eqref{A-abs} and~\eqref{B-abs} that $A_{\tau=\infty}(e)=A(e)$ vanishes, while $B_{\tau=\infty}(e)=B(e)$ is the only non-zero contribution to the absorptive part of the dressed propagator. Moreover, we can also show that $B_\tau(e)$ dominates for $\infty > \tau \gg 1/\Gamma$. Indeed, working in the peak region and using~\eqref{ineq-B-dominant}, we obtain
\begin{equation}
\left|\frac{A_\tau(e)}{B_\tau(e)}\right|= \frac{\left|{\rm Re}[G_\tau(e)]\right|}{\left|G_\tau(e)\right|^2\left|{\rm Im}[\Sigma(e^2)]\right|}\simeq \frac{\omega_{\vec{p}}}{m}\frac{4}{\tau\Gamma}\ll 1\,.
\label{B-dominant}
\end{equation}
This result confirms that when the inequalities~\eqref{ineq-B-dominant} are satisfied, the absorptive term $B_\tau(e)$ dominates over $A_\tau(e),$ describing a resonant scenario in which interaction and interference effects between the one-particle $\phi$ state and the multi-particle sector are relevant, either through instability and decay for $a=1$, or anti-instability and multi-particle masking for $a=-1$.

In such a late-time regime, no particle interpretation is possible, not even approximately, and the approximate form of the dressed $\phi$ propagator~\eqref{dressed-propag-suitable-main} acquires complex poles in the second (first) Riemann sheet for $a=1$ ($a=-1$), which become complex-conjugate pairs only in the strict asymptotic limit, i.e.  $\tau \to \infty$.

\subsubsection*{Intermediate regime}

A third relevant regime is an intermediate one between the two discussed above, namely 
\begin{equation}
\tau \approx \frac{\omega_{\vec{p}}}{m}\frac{1}{\Gamma}>\frac{1}{\Gamma}\gg \frac{1}{m}>\frac{1}{\omega_{\vec{p}}}\,.
\label{ineq-intermediate}
\end{equation}
This describes a transient during which $|A_\tau(e)| \approx |B_\tau(e)|$, just before the onset of decay or multi-particle masking, and coincides with the timescale at which the transition point discussed in the previous subsection occurs. In fact, working in the peak region, we have
\begin{equation}
	\left|\frac{A_\tau(e)}{B_\tau(e)}\right|\simeq \frac{\omega_{\vec{p}}}{m}\frac{4}{\tau\Gamma}=1\quad \Leftrightarrow \quad \tau=4\frac{\omega_{\vec{p}}}{m\Gamma}\,,
	\label{B-A-equal}
\end{equation}
which exactly matches the value of $\tau$ in~\eqref{zero-total-width}. At this point, the dressed $\phi$ propagator develops real poles at $e \simeq \pm \omega_{\vec{p}}$ in the second (first) Riemann sheet  for $a=1$ ($a=-1$).

Within the validity of our large-$\tau$ approximation, the appearance of these real poles could be physically interpreted as signaling some kind of discontinuity in the number of degrees of freedom. Indeed, due to the late-time emergence of first- or second-sheet complex poles that become complex-conjugate pairs asymptotically, the number of asymptotic fields and one-particle states decreases by one for an unstable particle, while it increases by one for a ghost.

\section{On-shell propagation and causality}\label{sec:on-shell-causality}

We have shown that for sufficiently short time intervals satisfying~\eqref{ineq-A-dominant}, the absorptive part of the propagator is dominated by $A_\tau(e)$, which can be approximated by a Dirac delta. Such an approximate on-shell contribution is typically extracted in the infinite-$\tau$ formulation by assuming a narrow resonance and taking suitable limits. However, this procedure lacks a clear physical justification~\cite{Anselmi:2020lfx,Buoninfante:2025klm}, since at $\tau=\infty$ an unstable particle has already decayed and a ghost has already been masked by the multi-particle sector. Let us elaborate more~on~this~point.

In the narrow-width approximation, we can write the absorptive terms~\eqref{A-abs} and~\eqref{B-abs} of the dressed propagator $\bar{G}_{\tau=\infty}(e)=\bar{G}(e)$ as
\begin{eqnarray}
	A(e)\simeq  \frac{a\epsilon}{(e^2-\omega_{\vec{p}}^2)^2+(\epsilon +a m \Gamma)^2}\,,\qquad
	B(e)\simeq  \frac{m\Gamma}{(e^2-\omega_{\vec{p}}^2)^2+(\epsilon+a m \Gamma)^2}\,.
	\label{A-B-func-narrow-appr}
\end{eqnarray}
We can then consider two distinct double limits~\cite{Buoninfante:2025klm}:
\begin{eqnarray}
\lim\limits_{\Gamma\rightarrow 0^+}\lim\limits_{\epsilon\rightarrow 0^+} {\rm Re}\left[\bar{G}(e)\right]&=&\!\!\!\lim\limits_{\Gamma\rightarrow 0^+} B(e) = \pi \delta(e^2-\omega_{\vec{p}}^2)\,,\label{B-limit}\\[1mm]
\lim\limits_{\epsilon\rightarrow 0^+}\lim\limits_{\Gamma\rightarrow 0^+} {\rm Re}\left[\bar{G}(e)\right]\!\!\!&=&\!\!\!\lim\limits_{\epsilon\rightarrow 0^+} A(e) = a\pi \delta(e^2-\omega_{\vec{p}}^2)\,.
\label{A-limit}
\end{eqnarray}
The double limit~\eqref{B-limit} is usually taken in the case of unstable resonances; however, it is physically incorrect, since the on-shell one-particle contribution must arise from $A(e)$ rather than $B(e)$. This fact is even more evident in the ghost case, where the limits $\epsilon \to 0^+$ and $\Gamma \to 0^+$ do not commute for $a=-1$, and the correct negative on-shell Dirac delta arises only from the double limit~\eqref{A-limit}. Nevertheless, strictly speaking, even this limit cannot describe an unstable particle before decay or a ghost before being masked, because at $\tau=\infty$ multi-particle effects necessarily dominate. This motivates the need for a QFT formulation in a finite interval of time~\cite{Anselmi:2020lfx,Anselmi:2023wjx,Buoninfante:2025klm}.

Despite this caveat, in the ordinary case, $a=1,$ the double limit~\eqref{B-limit} happens to provide a successful physical description of narrow unstable resonances. In contrast, for $a=-1$ the same limit leads to incorrect conclusions about the on-shell propagation of a ghost. In particular, it yields a positive absorptive one-particle contribution to the ghost propagator, implying that the latter behaves as if it were defined by the acausal anti-Feynman prescription, or equivalently by an anti-time-ordered product, due to the $+im\Gamma$ shift with $\Gamma \to 0^+$ in the denominator of~\eqref{propag-ghost-narrow}.

This line of reasoning has been used to argue that a ghost behaves acausally on the mass shell, with a reversed arrow of causality, i.e. with real positive (negative) frequencies propagating backward (forward) in time~\cite{Grinstein:2008bg,Salvio:2018kwh,Donoghue:2019ecz,Aoki:2025uff}. However, as explained above, once the correct double limit and the appropriate diagrammatic cuts are taken into account or, more precisely, when a finite-time approach is used, one finds that the approximate on-shell Dirac delta arises from the correct absorptive term $A(e)$, which is negative for a ghost. This implies that the propagator is still given by the causal time-ordered product, that the ghost propagates causally whenever the on-shell approximation is valid, and that only a single arrow of causality is present.

At the same time, it remains unclear whether complex conjugate poles on the first Riemann sheet may, in principle, induce acausal effects off the mass shell. For example, such an acausal behavior could manifest through loop effects in transition amplitudes. Even if this occurs, it would then be interesting to understand whether potential (micro)causality violations are masked by interference with the multi-particle sector. These issues will be investigated in future work.

\paragraph{Remark.} It is worth mentioning that, by introducing modified diagrammatic rules, one can render the ghost purely virtual so that it never appears on the mass shell; this is realized in the \textit{fakeon} quantization, where analyticity and causality are violated already at tree level~\cite{Anselmi:2018bra,Anselmi:2018kgz,Anselmi:2021hab}. In this approach, ${\rm Re}[G_\tau(e)]=0$ identically, implying $A_\tau(e)=0$, so that the only non-vanishing absorptive contribution is $B_\tau(e)$. Moreover, due to the non-analyticity of the fakeon prescription, the geometric series~\eqref{dressed-propag-suitable-main} was argued to be resummable only within its convergence radius, i.e. for $\tau\Gamma \ll 1$~\cite{Anselmi:2020lfx,Anselmi:2023wjx}. In this regime, however, $B_\tau(e)$ is small, so the dominant contribution to $|\bar{G}_\tau(e)|^2$ comes from the dispersive part, i.e. ${\rm Im}[\bar{G}_\tau(e)]$. A distinctive feature of the dressed fakeon propagator is that each peak at $e \simeq \pm \omega_{\vec{p}}$ is replaced by two lower twin peaks~\cite{Anselmi:2020lfx,Anselmi:2023wjx}. Although compatible with unitarity, the fakeon quantization lies outside the standard framework of local QFT and is therefore not considered~here.

\section{Conclusions}\label{sec:conclus}

Working within the framework of relativistic local QFT, we have carried out a thorough comparison between ordinary unstable particles and ghosts above the multi-particle threshold, clarifying various mathematical and physical properties and uncovering new features of ghosts. Let us summarize the main results.
\begin{itemize}

\item We contrasted the phenomenon of decay with that of multi-particle masking, showing that these correspond to distinct asymptotic dynamics. While an unstable particle disappears from the asymptotic spectrum after decaying, a ghost survives asymptotically and continues to interact with the composite field associated with the multi-particle states. Consequently, in the former case the number of asymptotic fields decreases by one, whereas in the latter it effectively increases by one. This crucial difference is encoded in the presence of first- or second-sheet complex conjugate poles in the $\phi$ and $\chi^2$ propagators in~\eqref{spectral-represent} and~\eqref{spectral-represent-composite-field}. 

\item We found that ordinary and ghost resonances can, in principle, be distinguished phenomenologically: the latter are narrower and exhibit weaker interference between positive- and negative-energy peaks. Moreover, working in a finite interval of time, we showed that finite-time effects amplify these differences and introduce new ones. In particular, ghost resonances have peaks higher than those of ordinary resonances.

\item The QFT formulation in a finite interval of time allowed us to analyze the dressed $\phi$ propagator as a function of the time interval $\tau=t_{\rm f}-t_{\rm i}$ and identify distinct temporal regimes, by working within a suitable large-$\tau$ approximation. At early times, $\tau\ll (\omega_{\vec{p}}/m)(1/\Gamma)$, the approximate form of the dressed propagator~\eqref{dressed-propag-suitable-main} has no poles, and both unstable particles and ghosts admit an approximate free-particle interpretation, since their interaction with the multi-particle sector is negligible. At late times, $\tau\gg (\omega_{\vec{p}}/m)(1/\Gamma)$, the dressed propagator develops (apparent) complex poles in the second or first Riemann sheet and interactions dominate, leading to either decay or multi-particle masking. The complex poles become complex-conjugate pairs only asymptotically, i.e. in the limit $\tau\rightarrow \infty.$

\item We also identified an intermediate temporal regime, $\tau\approx (\omega_{\vec{p}}/m)(1/\Gamma)$, in which the dressed propagator exhibits real energy poles at $e=\pm\omega_{\vec{p}}$, located on the second Riemann sheet for ordinary unstable particles and on the first sheet for ghosts. These poles act as transition points between the absence and emergence of complex poles and physically mark the onset of decay or multi-particle masking.

\end{itemize}

These findings not only provide a clear way to distinguish ordinary and ghost resonances, but also shed new light on the true nature of ghosts as a new type of quantum objects. In particular, the QFT formulation in a finite interval of time confirms our previous conclusion that no free asymptotic one-particle ghost state exists or, in other words, that a detector cannot isolate and observe a ghost particle asymptotically~\cite{Buoninfante:2026mve}. Indeed, a ghost is effectively confined to times much shorter than the inverse width, where an approximate free-particle description is valid, as reflected by the fact that an on-shell Dirac delta approximately emerges only in the early-time regime.

The results of this work support the physical consistency of QFTs containing ghosts quantized compatibly with unitarity in an indefinite-norm vector space. In fact, a negative-norm one-particle state does not give rise to observable negative probabilities asymptotically, since a ghost particle is always masked by its interference with the multi-particle sector. This implies that the standard LSZ construction~\cite{Lehmann:1954rq} cannot be applied to ghost states, which therefore cannot be used as isolated states attached to external legs of Feynman diagrams. At the same time, several questions remain open; let us briefly mention some of them before concluding.

First, the finite-time approach considered here is based on a large-$\tau$ approximation, which allowed us to derive suitable expressions for the free and dressed propagator while capturing the leading finite-time effects. Although this provides a qualitatively correct description of the underlying physics, a more quantitative analysis requires a refined approximation scheme incorporating higher-order finite-time corrections. In particular, such an improved approximation is expected to avoid the appearance of poles at finite but large times in the dressed propagator, suppress the oscillations originating from the exponential factor in the free propagator~\eqref{free-propag-suitable-main}, and resolve the issues discussed in Footnotes~\ref{foot-limit} and~\ref{footnote-complex}.

Second, while any experimental apparatus operating over a time interval $\tau \gg 1/\Gamma$ is unable to isolate and observe a ghost, it remains unclear whether a ghost particle and its associated negative probability could, in principle, be probed in sufficiently fast experiments over time scales shorter than $1/\Gamma$. In this regime, finite-time measurements are characterized by an energy uncertainty $\Delta E \gtrsim \Gamma$, so observability may depend on the value of $\Gamma$ in the specific model. The finite-time approach to scattering proposed in~\cite{Collins:2019ozc} may be useful in addressing this question. Furthermore, in some theories with ghosts, such as four-derivative models, at high energies the relevant cross sections are expected to remain positive due to intermediate cancellations between positive and negative contributions~\cite{Holdom:2021hlo,Holdom:2023usn,Holdom:2024cfq}. This fact may suggest an alternative mechanism for masking negative probabilities in the short-time regime.

Third, the study carried out here is based on the model described by the Lagrangian~\eqref{lagrangian}. However, as already mentioned at the beginning of Sec.~\ref{sec:finite-time}, the conclusions are quite general. The only requirement is an interaction term that generates a width for the field $\phi$ above the multi-particle threshold. Nevertheless, physically relevant applications must be studied explicitly, for instance in four-derivative theories~\cite{Pais:1950za,Bender:2007wu,Salvio:2015gsi,Holdom:2023usn,Holdom:2024cfq}, such as Lee-Wick models~\cite{Lee:1969fy,Lee:1970iw,Lee:chicago,Cutkosky:1969fq} and quadratic gravity~\cite{Stelle:1976gc,Tomboulis:1980bs,Avramidi:1985ki,Salvio:2018crh,Anselmi:2018tmf,Donoghue:2021cza,Holdom:2021hlo,Buoninfante:2023ryt,Buoninfante:2025dgy,Kuntz:2024rzu,Oda:2025buc}, which will be the subject of future work. It is worth noting that, in quadratic gravity, the width of the spin-two massive ghost is expected to satisfy $\Gamma \gtrsim \mathcal{O}(10^3)\,$GeV~\cite{Buoninfante:2025dgy}, implying that the early-time regime where an approximate free-particle description is valid corresponds to time scales much shorter than $1/\Gamma \lesssim \mathcal{O}(10^{-28})\,$s.

These open questions and potential applications, together with others discussed in~\cite{Buoninfante:2026mve}, will be addressed in separate works. It is important to emphasize that a deeper understanding of ghosts in QFT and their quantum dynamics across different time scales is not merely an academic exercise, but can shed new light on quantum gravity in the sub-Planckian regime~\cite{Buoninfante:2024yth,Basile:2024oms}, given that quadratic gravity provides a unique strictly renormalizable, local, metric, four-dimensional QFT of gravity~\cite{Buoninfante:2025dgy}. These issues are therefore pressing and will be investigated in the~near~future.

%%%%%%%%%%%%%%%%%%%%%%%%%%%%%%%%%%%%%%%%%%%%%%%%%%%%%%%%%%%%%%%%%%

\subsection*{Acknowledgements}
I thank Luca Smaldone for useful discussions.  I acknowledge financial support from the
Xunta de Galicia (CIGUS Network), the EU through the Galicia Feder 2021-2027 Program, and
the Grant CEX2023-001318-M funded by MICIU/AEI/10.13039/501100011033.

%%%%%%%%%%%%%%%%%%%%%%%%%%%%%%%%%%%%%%%%%%%%%%%%%%%%%%%%%%%%%%

\appendix

\section{Propagator in a finite interval of time}\label{sec:app-finite-time}

The aim of this Appendix is to present the details underlying the derivation of the approximate form of the dressed propagator in a finite time interval given in~\eqref{dressed-propag-suitable-main}. Following~\cite{Anselmi:2023phm,Anselmi:2023wjx}, we first review the key elements of the finite-time QFT formulation and then derive the free and dressed propagators within a suitable large-time-interval approximation.

\subsection{Elements of QFT in a finite interval of time}

To illustrate the main features of the QFT formulation in a finite interval of time, it is enough to consider a single scalar field and assume that all its interactions with other fields are effectively encoded in a single interaction term. The dressed propagator can then be derived by incorporating the effects of the interactions through the self-energy insertions.

\subsubsection*{Action}

The action for a scalar field $\phi'$ in a finite interval of time $\infty>\tau=t_{\rm f}-t_{\rm i}>0$ reads
\begin{equation}
S(\phi')=\int_{t_{\rm i}}^{t_{\rm f}} {\rm d}t \int {\rm d}^3x \mathcal{L}(\phi')\,,\quad
\mathcal{L}(\phi')=\mathcal{L}_0(\phi')+\mathcal{L}_{\rm int}(\phi')\,,\quad \mathcal{L}_0(\phi')=\frac{a}{2}\left[-\partial_\mu \phi' \partial^\mu \phi'-m^2 \phi'^2\right]\,,
\label{action-phi-prime-finite-t}
\end{equation}
where $\mathcal{L}_{\rm int}(\phi')$ contains the interaction terms; $a=1$ corresponds to an ordinary field, while $a=-1$ corresponds to a ghost field. 

Let us introduce the fluctuation field $\phi(x)$ around the classical solution of the free field equation, i.e. $\phi'(x)=\phi_0(x)+\phi(x)$ where $\phi_0$ satisfies $(\Box-m^2)\phi_0(x)=0,$ and choose the boundary conditions $\phi_0(t_{\rm i},\vec{x})=\phi'(t_{\rm i},\vec{x})\equiv \phi'_{\rm i}(\vec{x}),$ $\phi_0(t_{\rm f},\vec{x})=\phi'(t_{\rm f},\vec{x})\equiv \phi'_{\rm f}(\vec{x}),$ that imply $\phi(t_{\rm i},\vec{x})=0=\phi(t_{\rm f},\vec{x})$. In terms of $\phi_0(x)$ and $\phi(x)$, we can recast the action~\eqref{action-phi-prime-finite-t} as
\begin{eqnarray}
S(\phi')=S(\phi_0)+S(\phi,\phi_0)\,,\qquad S(\phi,\phi_0)=S_0(\phi)+S_{\rm int}(\phi,\phi_0)\,,
\end{eqnarray}
where   
\begin{eqnarray}
S(\phi_0)\!\!\!&=&\!\!\! \int{\rm d}^3x \frac{a}{2}\left[ \phi'_{\rm f}(\vec{x})\dot{\phi}_0(t_{\rm f},\vec{x})-\phi'_{\rm i}(\vec{x})\dot{\phi}_0(t_{\rm i},\vec{x}) \right]+\int^{t_{\rm f}}_{t_{\rm i}}{\rm d}t \int {\rm d}^3x \mathcal{L}_{\rm int}(\phi_0)\,,\\
S_0(\phi)\!\!\!&=&\!\!\! \int^{t_{\rm f}}_{t_{\rm i}}{\rm d}t \int {\rm d}^3x \frac{a}{2} \left[-\partial_\mu \phi \partial^\mu \phi-m^2 \phi^2\right]\,,\\
S(\phi,\phi_0)\!\!\!&=&\!\!\! \int^{t_{\rm f}}_{t_{\rm i}}{\rm d}t \int {\rm d}^3x \left[\mathcal{L}_{\rm int}(\phi_0+\phi)-\mathcal{L}_{\rm int}(\phi_0)\right]\,. 
\label{action-position}
\end{eqnarray}

\subsubsection*{Generating functional}

The transition amplitude between the initial field configuration/state $\phi'_{\rm i}(\vec{x})$ and the final field configuration/state $\phi'_{\rm f}(\vec{x})$ reads
\begin{eqnarray}
\left\langle \phi'_{\rm f}, t_{\rm f} \big|\phi'_{\rm i}, t_{\rm i}\right\rangle = \int\limits_{\phi'(t_{\rm i},\vec{x})=\phi'_{\rm i}(\vec{x})}^{\phi'(t_{\rm f},\vec{x})=\phi'_{\rm f}(\vec{x})}\!\!\!\!\!\!\!\!\! \mathcal{D}\phi'\, {\rm e}^{iS(\phi')+i\int_{t_{\rm i}}^{t_{\rm f}}{\rm d}t\int {\rm d}^3x J(x)\phi'(x)}=Z[J] {\rm e}^{iS(\phi_0)+i\int_{t_{\rm i}}^{t_{\rm f}}{\rm d}t\int {\rm d}^3x J(x)\phi_0(x)}\,,
\label{amplitude-position}
\end{eqnarray}
where the functional integral is given by
\begin{eqnarray}
Z[J] = \int\limits_{\phi(t_{\rm i},\vec{x})=\phi(t_{\rm f},\vec{x})=0}\!\!\!\!\!\!\!\!\! \mathcal{D}\phi\,  {\rm e}^{iS(\phi,\phi_0)+i\int_{t_{\rm i}}^{t_{\rm f}}{\rm d}t\int {\rm d}^3x J(x)\phi(x)}\,,
\label{functional-integral-position}
\end{eqnarray}
and $J(x)$ is the external source coupled to background and fluctuation field, i.e. $J\phi'=J\phi_0+J\phi$.

The functional $Z[J]$ generates the correlation functions that include all disconnected and reducible diagrams through the formula
\begin{eqnarray}
\left\langle \phi(x_1)\cdots \phi(x_n)\right\rangle = \frac{1}{Z[0]}\left.\frac{\delta^n Z[J]}{i\delta J(x_1)\cdots i\delta J(x_n) }\right|_{J=0} \,.
\label{Z-correlation-function}
\end{eqnarray}
Moreover, the functional $W[J]=-i\log Z[J]$ generates all connected diagrams, while its Legendre transform generates all amputated, one-particle irreducible diagrams. 

As usual, one can show that the functional integral of a functional derivative vanishes~\cite{Anselmi:2023phm}:
\begin{eqnarray}
\int\limits_{\phi(t_{\rm i},\vec{x})=\phi(t_{\rm f},\vec{x})=0} \!\!\!\!\!\!\!\!\!\mathcal{D}\phi \, \frac{\delta}{\delta \phi(x)}\left[ f(\phi) {\rm e}^{iS(\phi,\phi_0)+i\int_{t_{\rm i}}^{t_{\rm f}}{\rm d}t'\int {\rm d}^3x' J(x')\phi(x')}\right]=0\,,
\label{functional-derivative-identity}
\end{eqnarray}
where $t_{\rm f}>x^0>t_{\rm i}$ and $f(\phi)$ is a product of local functionals.

\subsubsection*{Free propagator}

Using formula~\eqref{functional-derivative-identity} with $f(\phi)=\phi(y)$ and $S(\phi,\phi_0)\rightarrow S_0(\phi),$ we obtain the differential equation for the free propagator $G(x,y)=\left\langle \phi(x) \phi(y)\right\rangle_0$ with $t_{\rm f}>x^0,y^0>t_{\rm i}:$
\begin{equation}
(\Box_x-m^2)G(x,y)=i\,a\,\delta^{(4)}(x-y)\,,\quad G(x,y)=G(y,x)\,,\quad \left.G(x,y)\right|_{x^0=t_{\rm i}}=0= \left.G(x,y)\right|_{x^0=t_{\rm f}}\,.
\label{diff-eq-propag-position}
\end{equation}
To find the explicit solution, it is convenient to Fourier transform to spatial momentum:\footnote{To simplify the notation, throughout the entire paper, we make an abuse of notation by denoting both the function and its Fourier transform with the same symbol; the distinction will be clear from the different arguments of the functions.\label{footnote-abuse}}
\begin{equation}
G(x,y)=\int \frac{{\rm d}^3p}{(2\pi)^3}{\rm e}^{i\vec{p}\cdot(\vec{x}-\vec{y})} G(x^0,y^0;\vec{p})\,.
\end{equation}
Calling $x^0=t,$ $y^0=t',$ and $G(x^0,y^0;\vec{p})= G(t,t'),$ for simplicity, we can recast the differential equation~\eqref{diff-eq-propag-position} as
\begin{equation}
	\left(\frac{{\rm d^2}}{{\rm d}t^2}+\omega^2_{\vec{p}}\right)G(t,t')=-i\,a\,\delta(t-t')\,,\quad G(t,t')=G(t',t)\,,\quad G(t_{\rm i},t')=0= G(t_{\rm f},t')\,,
	\label{diff-eq-propag-position-QM}
\end{equation}
where $\omega^2_{\vec{p}}=\vec{p}^2+m^2.$ The full solution for the free propagator will be the sum of the particular and homogeneous solutions, respectively given by
\begin{equation}
G_{{\rm part}}(t,t')=a\frac{{\rm e}^{-i\omega_{\vec{p}}|t-t'|}}{2\omega_{\vec{p}}}\,, \qquad G_{{\rm hom}}(t,t')=C_1(t') {\rm e}^{-i\omega_{\vec{p}} t}+C_2(t') {\rm e}^{i\omega_{\vec{p}}t}\,.
\end{equation}
The former corresponds to the standard Feynman propagator in an infinite interval of time, while the latter is sensitive to finite-time effects through the boundary conditions at $|t_{\rm i}|<\infty$ and $t_{\rm f}<\infty,$ that fix the expressions of the two integration constants (with respect to $t$):
\begin{eqnarray}
C_1(t')\!\!\!&=&\!\!\!\frac{-a i}{4\omega_{\vec{p}}\sin \omega_{\vec{p}}\tau}\left[{\rm e}^{-i\omega_{\vec{p}}(t_{\rm f}-t_{\rm i}-t')}-{\rm e}^{i\omega_{\vec{p}}(t_{\rm f}+t_{\rm i}-t')}\right] \,, \\
C_2(t')\!\!\!&=&\!\!\!\frac{a i}{4\omega_{\vec{p}}\sin \omega_{\vec{p}}\tau}\left[{\rm e}^{-i\omega_{\vec{p}}(t_{\rm f}+t_{\rm i}-t')}-{\rm e}^{i\omega_{\vec{p}}(t_{\rm i}-t_{\rm f}-t')}\right] \,.
\end{eqnarray}
Therefore, the homogeneous solution reads
\begin{equation}
G_{{\rm hom}}(t,t')= \frac{-a i}{2\omega_{\vec{p}}\sin \omega_{\vec{p}}\tau}\left[ {\rm e}^{-i\omega_{\vec{p}}\tau}\cos\big(\omega_{\vec{p}}(t-t')\big)-\cos\big(\omega_{\vec{p}}(t+t'-t_{\rm i}-t_{\rm f})\big)\right]\,,
\label{homog-sol}
\end{equation}
which implies
\begin{eqnarray}
G(t,t')\!\!\!&=&\!\!\! G^+(t,t')+G^-(t,t') \nonumber\\[1mm]
\!\!\!&=&\!\!\!\frac{a i}{\omega_{\vec{p}}\sin \omega_{\vec{p}}\tau}\Big[\theta(t-t')\sin \big(\omega_{\vec{p}}(t_{\rm f}-t)\big)\sin \big(\omega_{\vec{p}}(t'-t_{\rm i})\big)\nonumber\\
&&\!\!\!\qquad\qquad\quad+\theta(t'-t)\sin \big(\omega_{\vec{p}}(t_{\rm f}-t')\big)\sin \big(\omega_{\vec{p}}(t-t_{\rm i})\big)\Big]\,,
\label{space-fourier-propag-position}
\end{eqnarray}
where $t_{\rm f}>t,t'>t_{\rm i}$; $G^+(t,t')$ and $G^-(t,t')$  correspond to the second and third line in~\eqref{space-fourier-propag-position}, namely to the $\theta(t-t')$ and $\theta(t'-t)$ contributions, respectively.

Furthermore, the temporal Fourier transform of~\eqref{space-fourier-propag-position} in a finite interval of time is given by
\begin{eqnarray}
	G_{\tau}(e,e')\!\!\!\!&=&\!\!\!\! \int_{-\infty}^{\infty}{\rm d}t \int_{-\infty}^{\infty}{\rm d}t' {\rm e}^{iet} {\rm e}^{ie't'} G_\tau(t,t') = \int_{t_{\rm i}}^{t_{\rm f}}\!{\rm d}t \int_{t_{\rm i}}^{t_{\rm f}}\!{\rm d}t' {\rm e}^{iet} {\rm e}^{ie't'} G(t,t')\nonumber\\
	\!\!\!\!&=&\!\!\!\! \frac{ia}{\sin \omega_{\vec{p}}\tau}\frac{1}{e^2-\omega^2_{\vec{p}}}\frac{1}{e'^2-\omega^2_{\vec{p}}}\frac{1}{e+e'}\left[ \left({\rm e}^{i(et_{\rm i}+e't_{\rm f})}\!+\!{\rm e}^{i(et_{\rm f}+e't_{\rm i})}\right)\!(e+e')\omega_{\vec{p}}\right. \nonumber\\
	\!\!\!\!&&\!\!\!\!\left. -\left({\rm e}^{i(e+e')t_{\rm f}}\!+\!{\rm e}^{i(e+e')t_{\rm i}}\right)\!(e\!+\!e')\omega_{\vec{p}}\cos \omega_{\vec{p}}\tau +i\left({\rm e}^{i(e+e')t_{\rm f}}\!-\!{\rm e}^{i(e+e')t_{\rm i}}\right)\!(ee'\!+\!\omega^2)\sin \omega_{\vec{p}}\tau\right]\,,\nonumber\\
	\label{time-fourier-propag-position}
\end{eqnarray}
where the subscript $\tau$ indicates that we are taking the temporal Fourier transform of the projected propagator $G_\tau(t,t')\equiv \Pi_\tau(t)G(t,t')\Pi_\tau(t')$, with the projector defined as $\Pi_\tau(t)\equiv \theta(t_{\rm f}-t)\theta(t-t_{\rm i})$. We again note the abuse of notation associated with the Fourier transform (see Footnote~\ref{footnote-abuse}).

Assuming, as usual, that $\omega_{\vec{p}}$ has a small negative imaginary part, namely $\omega_{\vec{p}}\rightarrow \omega_{\vec{p}}-i\epsilon$ with $\epsilon\rightarrow 0^+$, it is straightforward to show that, in the double limit $t_{\rm i}\rightarrow -\infty$ and $t_{\rm f}\rightarrow \infty$, the homogeneous solution vanishes and the Feynman propagator is recovered:
\begin{eqnarray}
	\lim\limits_{t_{\rm i}\rightarrow-\infty}\lim\limits_{t_{\rm f}\rightarrow\infty} G(t,t')\!\!\!&=&\!\!\! \lim\limits_{t_{\rm i}\rightarrow-\infty}\left[\frac{a}{2\omega_{\vec{p}}}\theta(t-t'){\rm e}^{-i\omega_{\vec{p}}(t-t_{\rm i})}\left( {\rm e}^{-i\omega_{\vec{p}}(t'-t_{\rm i})}-{\rm e}^{i\omega_{\vec{p}}(t'-t_{\rm i})}\right)+(t\leftrightarrow t')\right]\nonumber\\
	\!\!\!&=&\!\!\! a\frac{e^{-i\omega_{\vec{p}}|t-t'|}}{2\omega_{\vec{p}}}\,.
\end{eqnarray}
Similarly, for the temporal Fourier transform we have
\begin{eqnarray}
	\lim\limits_{t_{\rm i}\rightarrow-\infty}\lim\limits_{t_{\rm f}\rightarrow\infty} G_\tau (e,e')= 2\pi\delta(e+e')\frac{ia}{e^2-\omega_{\vec{p}}^2+i\epsilon}=2\pi\delta(e+e')\frac{-ia}{p^2+m^2-i\epsilon}\,,
\end{eqnarray}
where $p^0=e$, $p^2=-e^2+\vec{p}^{,2}$, a redefinition of $\epsilon$ is understood, and the distributional limits
\begin{eqnarray}
\lim\limits_{t_{\rm i}\rightarrow-\infty}\lim\limits_{t_{\rm f}\rightarrow\infty} \frac{{\rm e}^{i(e+e')t_{\rm f}}+{\rm e}^{i(e+e')t_{\rm i}}}{e+e'}\!\!\!&=&\!\!\! \lim\limits_{\tau\rightarrow \infty}2\frac{\cos \big(\frac{e+e'}{2}\tau\big)}{e+e'}=0\,, \\ 
\lim\limits_{t_{\rm i}\rightarrow-\infty}\lim\limits_{t_{\rm f}\rightarrow\infty} \frac{{\rm e}^{i(e+e')t_{\rm f}}-{\rm e}^{i(e+e')t_{\rm i}}}{i(e+e')}\!\!\!&=&\!\!\! \lim\limits_{\tau\rightarrow \infty}2\frac{\sin \big(\frac{e+e'}{2}\tau\big)}{e+e'}= 2\pi \delta(e+e')\,
\end{eqnarray}
have been used. Some remarks are in order.
\begin{itemize}

	\item \textit{Limited energy resolution}: If the boundary conditions, and so the propagator, are defined in a finite interval of time $\tau=t_{\rm f}-t_{\rm i}<\infty$, the invariance under time translation is broken and the Fourier transform of the propagator depends on both Fourier energies $e$ and $e'$. This means that finite-time effects induce an apparent violation of energy conservation controlled by $|e+e'|:$ these are relevant for $|e+e'|\lesssim 1/\tau$ and suppressed for  $|e+e'|\gtrsim 1/\tau$. In reality, this does not mean that we can observe physical processes in which energy is not conserved, but rather that measurements performed in a finite interval of time $\tau$ are characterized by an energy spread inducing a quantum limitation on the energy resolution of~order~$1/\tau$. 
	
	\item \textit{Pole structure}: When $\tau<\infty$, the propagator has no poles as a function of $e$ and $e'$. Indeed, since the free propagator $G(t,t')$ has compact temporal support, its Fourier transform must be an entire function. This property reflects the fact that exact on-shell contributions arise only asymptotically, i.e. in the double limit $t_{\rm f}\rightarrow\infty$ and $t_{\rm i}\rightarrow-\infty$. However, the expressions of the free propagator in~\eqref{space-fourier-propag-position} and~\eqref{time-fourier-propag-position} exhibit singularities at $\omega_{\vec{p}}\tau=n\pi$, $n\in\mathbb{Z}\setminus{0}$, due to the factor $\sin\omega_{\vec{p}}\tau$ in the denominator. These singularities affect the internal sectors of diagrams and enormously complicate the calculation of the spatial part of loop integrals.
	
	\item \textit{Projection on finite-time interval}: The expression~\eqref{time-fourier-propag-position} corresponds to the standard Fourier transform (running from $-\infty$ to $\infty$) of the projected propagator $G_{\tau}(e,e')$. However, as we explain below, the information about the projection can be removed from the internal sector of the diagrams and transferred to the external sector. In this way, all diagrams and loop integrals can be computed using the standard Fourier transform of $G(t,t')$ (without projectors), with an additional external source attached to each vertex.

\end{itemize}

\subsubsection*{Interactions}

To formally work out the structure of the interaction vertices we can write
\begin{eqnarray}
S(\phi,\phi_0)=\int_{t_{\rm i}}^{t_{\rm f}}{\rm d}t\int {\rm d}^3x \left[\mathcal{L}_{\rm int}(\phi_0+\phi)-\mathcal{L}_{\rm int}(\phi_0)\right]=\int{\rm d}^4x \sum\limits_n \sum\limits_\ell \mathcal{J}_{n\ell}(x)V_{n\ell}(\phi(x))\,,
\label{interaction-recast-position}
\end{eqnarray}
where $V_{n\ell}(\phi(x))$ is a monomial of degree $n$ in the field $\phi(x),$ $\mathcal{J}_{n\ell}(x)$ is some function that includes the dependence on $\phi_0(x)$ and contains the projector $\Pi_\tau(t),$ and the index $\ell$ distinguishes different types of $n$-point vertices. If we Fourier transform in both time and space, we can recast~\eqref{interaction-recast-position} as
\begin{eqnarray}
	S(\phi,\phi_0)=\sum\limits_n\sum\limits_\ell \int \left[\prod\limits_{i=1}^n\frac{{\rm d}^4p_i}{(2\pi)^4}\phi(p_i)\right] \mathcal{V}_{n\ell}(p_1,\dots,p_n)\mathcal{J}_{n\ell}(-E,-\vec{P})\,,
	\label{fourier-interaction-recast-position}
\end{eqnarray}
where $\mathcal{V}_{n\ell}(p_1,\cdots,p_n)$ are almost the standard momentum-space vertex functions but with no momentum-conserving delta factor due to the presence of $\mathcal{J}_{n\ell}(-E,-\vec{P})$, which contains the dependence on $\phi_0$ and acts like an additional external source attached to the $(n\ell)$-vertex, where $E=-\sum_{i=1}^n e_i$ and $\vec{P}=-\sum_{i=1}^n\vec{p}_i.$ This means that in a perturbative QFT formulation in a finite interval of time, vertices can be computed just as in the infinite-time formulation, with the only difference being the attachment of an additional external source to each vertex, accounting for the finite-time restrictions. In the infinite-time limit, $\mathcal{J}_{n\ell}$ become momentum-independent except for the factor $\delta^{(4)}(p_1+\cdots+p_n)$, while retaining the dependence on $\phi_0$ that can be grouped together with $\mathcal{V}_{n\ell}$ to recover the standard vertices of the infinite-time formulation.

\subsubsection*{Projected external sources}

If we define the projected external source $J_\tau(x)\equiv \Pi_\tau (t)J(t,\vec{x})$, we can write the source term in the generating functional as 
\begin{eqnarray}
\int_{t_{\rm i}}^{t_{\rm f}}{\rm d}t \int {\rm d}^3x \phi(x)J(x)=\int{\rm d}^4x \phi(x) J_\tau(x)\,.
\label{projected-source-term}
\end{eqnarray}
Consequently, the free functional integral can be recast in the following form:
\begin{eqnarray}
Z_0[J]=Z_0[0]{\rm e}^{\int{\rm d}^4x {\rm d}^4y J_\tau(x) G(x,y) J_\tau(y)}\,.
\label{functional-free-projected-source-term}
\end{eqnarray}
This implies that, by taking functional variations with respect to the projected sources, one can develop the perturbation theory and construct Feynman diagrams in terms of the unprojected propagator $G(t,t')$, which differs from $G_\tau(t,t')$. Together with the discussion of the interaction terms above, this shows that, when the generating functional is defined in terms of projected sources, the standard methods of the infinite-time formulation can still be applied, albeit with two important modifications: the standard free propagator must be replaced by the modified expressions~\eqref{space-fourier-propag-position} and~\eqref{time-fourier-propag-position}, and additional external sources must be attached to the vertices.

It is worth mentioning that finite-time corrections to the unprojected free propagator depend on the choice of field variables in the functional-integral approach, or equivalently on the basis of states in the operator formalism. For instance, in the coherent-state basis the unprojected free propagator $G(t,t')$ retains its infinite-time form due to an order reduction of the differential equations~\cite{Anselmi:2023phm}. In this alternative basis, finite-time effects are encoded entirely in additional external sources attached to each vertex. However, this comes at the cost of introducing spurious spatial nonlocalities and their associated complications.

\subsubsection*{Dressed propagator}

Working in terms of the projected external sources $J_\tau(t,\vec{x})$, the spatial Fourier transform of the dressed propagator before full projection can be written as 
\begin{eqnarray}
\!\!\!\!\!\!\!\!\!\!\!\!\!\!\!\!\hat{G}_\tau(t,t')\!\!\!&=&\!\!\! G(t,t')+ \int {\rm d}t_1 \int {\rm d}t_2\, G(t,t_1)i\Sigma_\tau(t_1,t_2)G(t_2,t')\nonumber\\
\!\!\!\!\!\!\!\!\!\!\!\!\!\!\!&&\!\!\!+ \int {\rm d}t_1 \int {\rm d}t_2 \int {\rm d}t_3 \int {\rm d}t_4\,  G(t,t_1)i\Sigma_\tau(t_1,t_2)G(t_2,t_3)i\Sigma_\tau(t_3,t_4)G(t_4,t')+\cdots\,,
\label{dressed-propag-position-unproj}
\end{eqnarray}
where $\hat{G}_{\rm \tau}(t,t')$ is a partially unprojected dressed propagator and $\Sigma_\tau (t,t')$ is the spatial Fourier transform of the one-particle, irreducible self-energy computed in a finite interval of time, whose dependence on the spatial momentum $\vec{p}$ is implicitly understood. We will limit our discussion to the one-loop order, at which the self-energy before projection is given by the square of the unprojected free propagator at non-coinciding points, i.e. $\Sigma(t,t')=[G(t,t')]^2$. 

Using $\Sigma_\tau(t,t')=\Pi_\tau(t)\Sigma(t,t')\Pi_\tau(t')$, we can write the fully projected dressed propagator $\bar{G}_\tau(t,t')\equiv \Pi_\tau (t)\hat{G}_\tau(t,t')\Pi_\tau(t') $ as
\begin{eqnarray}
\!\!\!\!\!\!\!\!\!\!\!\!\!\!\!\bar{G}_\tau(t,t')\!\!\!&=&\!\!\! G_\tau(t,t')+ \int {\rm d}t_1 \int {\rm d}t_2\, G_\tau(t,t_1)i\Sigma(t_1,t_2)G_\tau(t_2,t')\nonumber\\
\!\!\!\!\!\!\!\!\!\!\!\!\!\!\!&&\!\!\!+ \int {\rm d}t_1 \int {\rm d}t_2 \int {\rm d}t_3 \int {\rm d}t_4\,  G_\tau(t,t_1)i\Sigma(t_1,t_2)G_\tau(t_2,t_3)i\Sigma(t_3,t_4)G_\tau(t_4,t')+\cdots\,.
\label{dressed-propag-position-proj}
\end{eqnarray}
Fourier transforming with respect to $t$ and $t'$, we obtain 
\begin{eqnarray}
&&\!\!\!\!\!\!\!\!\!\!\!\!\bar{G}_\tau(e,e')= G_\tau(e,e')+ \int \frac{{\rm d}e_1}{2\pi} \int \frac{{\rm d}e_2}{2\pi} G_\tau(e,e_1)i\Sigma(-e_1,-e_2)G_\tau(e_2,e')\nonumber\\
&&\!\!\!\!\!+ \int \frac{{\rm d}e_1}{2\pi} \int \frac{{\rm d}e_2}{2\pi} \int \frac{{\rm d}e_3}{2\pi} \int \frac{{\rm d}e_4}{2\pi} G_\tau(e,e_1)i\Sigma(-e_1,-e_2)G_\tau(e_2,e_3)i\Sigma(-e_3,-e_4)G_\tau(e_4,e')+\cdots\,.\nonumber\\
\label{fourier-dressed-propag-proj}
\end{eqnarray}
Let us emphasize that $\Sigma(t,t')$ and its Fourier transform $\Sigma(e,e')$ are not the one-loop self-energy expressions derived in the infinite-time formulation because the unprojected free propagators $G(t,t')$ attached to the internal lines of the bubble diagram receive finite-time corrections.

\subsection{Suitable form of the free propagator}\label{sec:suitable-free}

The complicated form of the free~\eqref{time-fourier-propag-position} and dressed~\eqref{dressed-propag-position-proj} propagators in a finite time interval obscure their physical content. We now show that more suitable expressions can be obtained in a large-$\tau$ approximation. Let~us begin with the~free~propagator.

We are interested in the minimal finite-time modifications, which can be isolated by taking $\tau$ finite but sufficiently large: $\tau \gg 1/\omega_{\vec{p}}$ with $t$ and $t'$ kept fixed. This condition allows us to neglect the homogeneous solution~\eqref{homog-sol} due to the exponential suppression coming from $1/(\sin \omega_{\vec{p}}\tau)$ with $\omega_{\vec{p}}\to \omega_{\vec{p}}-i\epsilon$, so that the projection of~\eqref{space-fourier-propag-position} onto the finite-time interval can be~approximated~as\footnote{The approximation is not spoiled by the presence of poles in $1/\sin(\omega_{\vec{p}}\tau)$, since we can always assume $1 \ll \omega_{\vec{p}}\tau \neq n$, with $n\in\mathbb{Z}$. For instance, in the rest frame ($\vec{p}=0$), this amounts to requiring $\tau$ to be sufficiently large and such that $1 \ll m\tau \neq n$, with $n\in\mathbb{Z}$. It is also worth noting that the expression in~\eqref{space-fourier-propag-position-approx} is exact in the coherent-state basis~\cite{Anselmi:2023phm}. However, since the large-$\tau$ approximation is required in the subsequent steps anyway, no change of field basis is needed for our purposes.}
\begin{eqnarray}
	G_\tau(t,t')\!\!\!&=&\!\!\! \Pi_\tau(t) \left[G^+(t,t')+G^-(t,t')\right] \Pi_\tau (t')=G_\tau^+(t,t')+G_\tau^-(t,t')\,,\nonumber\\[1.5mm]
	G^+_\tau(t,t')\!\!\!&=&\!\!\!G_\tau^-(t',t)\simeq  \frac{a}{2\omega_{\vec{p}}}\Pi_\tau(t) \theta(t-t') \, {\rm e}^{-i(t-t')\omega_{\vec{p}}}\Pi_\tau(t')\,,
	\label{space-fourier-propag-position-approx}
\end{eqnarray}
where the finite-time restriction is now encoded only in the projectors, while additional finite-time corrections coming from the homogeneous solution~\eqref{homog-sol} are subdominant.

The next step is to take the temporal Fourier transform of~\eqref{space-fourier-propag-position-approx}. To further simplify the expression of the propagator, without loss of generality, we consider a symmetric time interval: $t_f=\tau/2$ and $t_i=-\tau/2$. Thus, we get
\begin{eqnarray}
	G^+_{\tau}(e,e')\!\!\!\!&=&\!\!\!\! G^-_{\tau}(e',e)= \int_{-\infty}^{\infty}{\rm d}t \int_{-\infty}^{\infty}{\rm d}t' {\rm e}^{iet} {\rm e}^{ie't'} G^+_\tau(t,t') = \int_{t_{\rm i}}^{t_{\rm f}}{\rm d}t \int_{t_{\rm i}}^{t_{\rm f}}{\rm d}t' {\rm e}^{iet} {\rm e}^{ie't'} G^+(t,t')\nonumber\\[1mm]
	\!\!\!\!&\simeq &\!\!\!\! \frac{a}{2\omega_{\vec{p}}}\frac{{\rm e}^{i( e-e'-2\omega_{\vec{p}})\tau/2}(e+e')-{\rm e}^{i(e+e')\tau/2}(e-\omega)-{\rm e}^{-i(e+e')\tau/2}(e'+\omega_{\vec{p}})}{(e-\omega_{\vec{p}})(e'+\omega_{\vec{p}})}\,.
	\label{time-fourier-propag-position-plus-approx}
\end{eqnarray}
Since $\tau$ is finite, the Dirac delta $2\pi \delta(e+e')$ does not factorize, so the propagator generally depends on two independent energy variables, making its analysis more involved. However, for sufficiently large $\tau$, one can factor out the finite-time analogue of $2\pi \delta(e+e')$, namely
\begin{equation}
2\frac{\sin \big(\frac{e+e'}{2}\tau\big)}{e+e'}\,,
\label{finite-time-delta-analogue}
\end{equation}
and reduce the expression~\eqref{time-fourier-propag-position-plus-approx} to~\cite{Anselmi:2023wjx}
\begin{equation}
G^\pm_\tau(e,e') \simeq 2\frac{\sin \big(\frac{e+e'}{2}\tau\big)}{e+e'} G_\tau^\pm(e)\,,
\label{two-factors}
\end{equation}
where
\begin{equation}
G_\tau^\pm(e)\equiv\lim\limits_{e'\rightarrow -e} \frac{e+e'}{2\sin \big(\frac{e+e'}{2}\tau\big)}G_\tau^\pm (e,e')= \frac{a\tau }{2\omega_{\vec{p}}}\frac{{\rm e}^{i\kappa_\pm}-i\kappa_\pm-1}{(i\kappa_\pm)^2}\,,\qquad \kappa_\pm\equiv (\pm e-\omega_{\vec{p}})\tau\,.
\label{g2-limit-definition}
\end{equation}
The result of the last limit is justified only for sufficiently large $\tau$ and yields a suitable form of the free propagator as a function of a single Fourier energy variable, up to the factor~\eqref{finite-time-delta-analogue}. Therefore, the finite-time restriction of the propagator is encoded in the $(e,e')$-dependent function~\eqref{finite-time-delta-analogue} and in the $\tau$ dependence of $G_\tau^\pm(e)$. All other finite-time corrections are subdominant within the working approximation under consideration.

Summing the positive and negative frequency terms, we obtain the following large-$\tau$ expression for the Fourier transform of the free propagator in a finite interval of time:
\begin{eqnarray}
G_\tau(e,e')\simeq 2\frac{\sin \big(\frac{e+e'}{2}\tau\big)}{e+e'} G_\tau (e)\,,\qquad G_\tau (e)\equiv G_\tau^+(e)+G_\tau^-(e)\,.
\label{free-propag-suitable}
\end{eqnarray}
As a consistency check, we can verify that in the limit $\tau\rightarrow \infty,$ with $\omega_{\vec{p}}\rightarrow \omega_{\vec{p}}-i\epsilon,$ the expression of the free propagator in an infinite interval of time is recovered. Indeed, using
\begin{eqnarray}
\lim\limits_{\tau\rightarrow \infty} 2\frac{\sin \big(\frac{e+e'}{2}\tau\big)}{e+e'} = 2\pi \delta(e+e')\,,\qquad \lim\limits_{\tau\rightarrow \infty} G_\tau^\pm(e)=\frac{1}{2\omega_{\vec{p}}}\frac{ia}{\pm e-\omega_{\vec{p}}+i\epsilon}\,,
\label{limits-tau-inf}
\end{eqnarray}
we obtain
\begin{equation}
	\lim\limits_{\tau\rightarrow \infty} G_\tau(e,e')=2\pi \delta(e+e')\frac{1}{2\omega_{\vec{p}}}\left[\frac{ai}{e-\omega_{\vec{p}}-i\epsilon}+\frac{ai}{-e-\omega_{\vec{p}}-i\epsilon}\right]=2\pi \delta(e+e') \frac{-ia}{p^2+m^2-i\epsilon}\,,
	\label{recovery-feynm-propag}
\end{equation}
where we recall that $p^2+m^2=-e^2+\omega_{\vec{p}}^2$, and a redefinition of $\epsilon$ in the last step is understood. In particular, the term in the square brackets multiplied by $1/(2\omega_{\vec{p}})$ corresponds to the infinite-$\tau$ limit of $G_\tau(e)$. This means that $G_\tau(e)$ defined in~\eqref{free-propag-suitable} is the appropriate object to compare with the Feynman propagator in order to capture the physical implications due to finite-time effects.

The free propagator $G_\tau(e)$ is an entire function; namely, it has neither real nor complex poles as a function of $e\in\mathbb{C}$ for $\omega_{\vec{p}}\in\mathbb{R}$ and $\tau<\infty$. Two real poles at $e=\pm \omega_{\vec{p}}$ emerge only in the asymptotic limit $\tau \to \infty$. The real and imaginary parts respectively read
\begin{eqnarray}
	{\rm Re}\left[G_\tau(e)\right]\!\!\!&=&\!\!\! \frac{a\tau}{2\omega_{\vec{p}}}\left[\frac{1-\cos \kappa_+}{\kappa_+^2}+\frac{1-\cos \kappa_-}{\kappa_-^2}\right]\,,	\label{real-part-free-propag-finite-t}\\[1mm]
	{\rm Im}\left[G_\tau(e)\right]\!\!\!&=&\!\!\! \frac{a\tau}{2\omega_{\vec{p}}}\left[\frac{\kappa_+ -\sin \kappa_+}{\kappa_+^2}+\frac{\kappa_- -\sin \kappa_-}{\kappa_-^2}\right]\,.
	\label{imag-part-free-propag-finite-t}
\end{eqnarray}
The real part is positive for $a=1$ and negative for $a=-1$, in agreement with the fact that unitarity  requires indefinite-norm states for ghost fields. Using the inequalities
\begin{eqnarray}
	0\leq \frac{1-\cos \kappa_\pm}{\kappa_\pm^2}\leq \frac{1}{2}\,,\qquad \left|\frac{\kappa_\pm-\sin \kappa_\pm}{\kappa_\pm^2}\right|\leq \frac{1}{\pi}\,,
\end{eqnarray}
we can notice that the real and imaginary parts of the free propagator are bounded for $\tau<\infty$:
\begin{eqnarray}
	\left|{\rm Re}\left[G_\tau(e)\right]\right|\leq \frac{\tau}{2\omega_{\vec{p}}}\,,\qquad \left|{\rm Im}\left[G_\tau(e)\right]\right|\leq \frac{\tau}{\pi \omega_{\vec{p}}}\,.
	\label{real-imag-bounds}
\end{eqnarray}
%

%%%%%%%%%%%%%%%%%%%%%%%%%%%%%%%%%%%%%%%%

\begin{figure}[t!]
	\centering
	\subfloat[Subfigure 1 list of figures text][]{
		\includegraphics[scale=0.386]{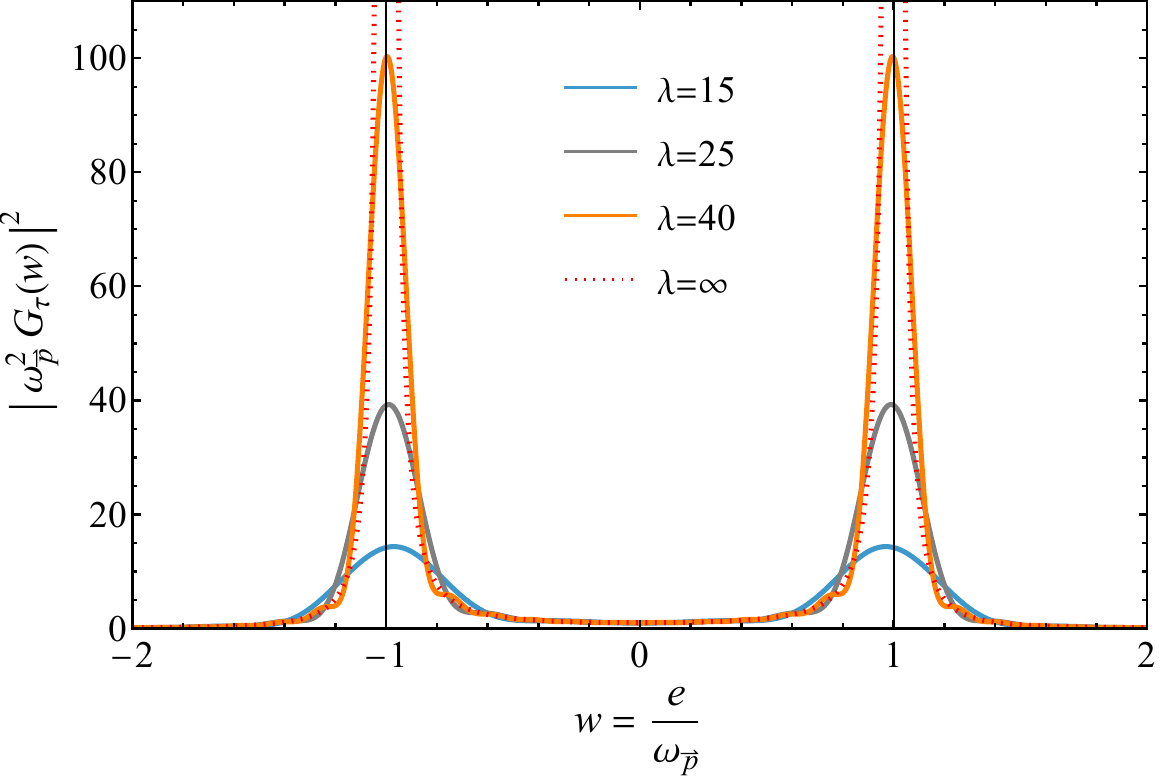}\label{figA1.1}}\qquad\,
	\subfloat[Subfigure 2 list of figures text][]{
		\includegraphics[scale=0.36]{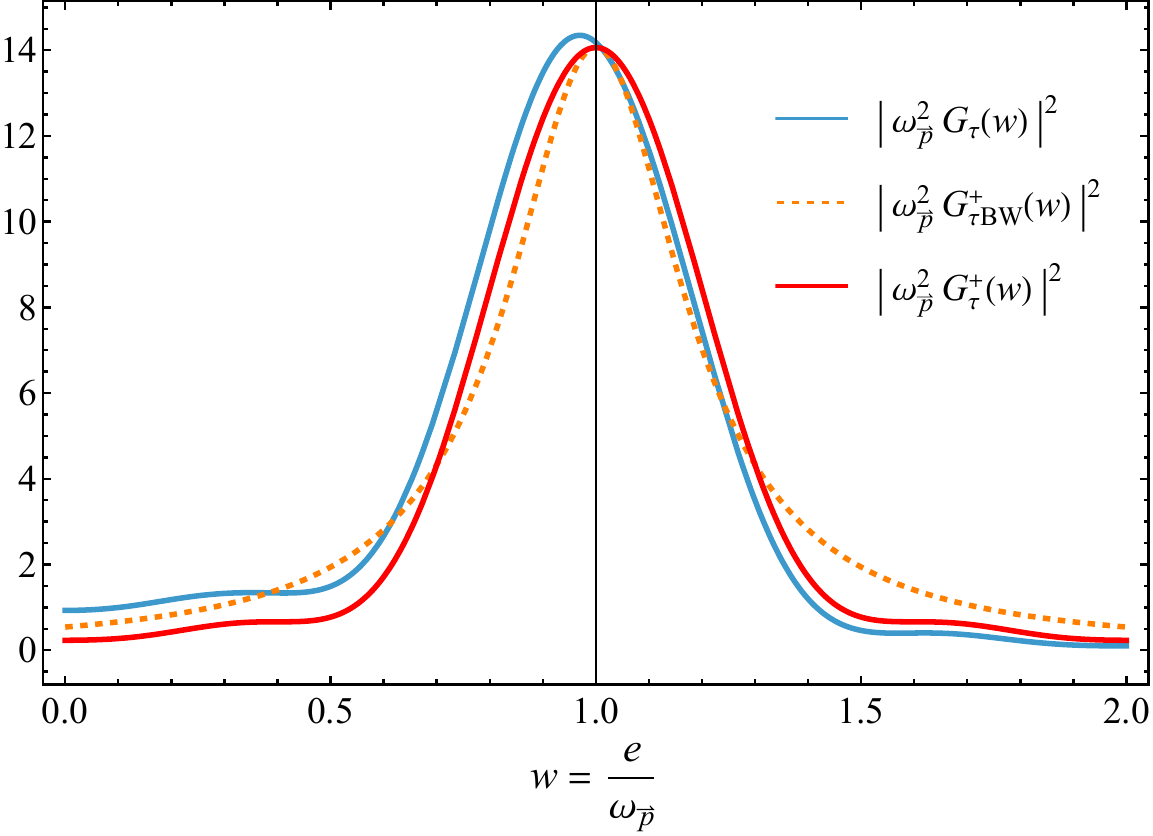}\label{figA1.2}}
	\protect\caption{(a) The behavior of $|\omega_{\vec{p}}^2G_\tau(w)|^2$ as a function of the dimensionless energy variable $w= e/\omega_{\vec{p}}$ is plotted for different values of $\lambda= \omega_{\vec{p}}\tau:$ $\lambda=15$~(blue solid line), $\lambda=25$~(gray solid line), $\lambda=40$~(orange solid line), and $\lambda=\infty$~(red dotted line). (b) The behaviors of $|\omega_{\vec{p}}^2 G_\tau(w)|^2$ (blue solid line), $|\omega_{\vec{p}}^2G^+_{\tau{\rm BW}}(w)|^2$ (orange dashed line), and $|\omega_{\vec{p}}^2\, G_\tau^+(w)|^2$ (red solid line)  as functions of $w= e/\omega_{\vec{p}}$ are compared around the positive-energy peak $w\simeq 1 \Leftrightarrow e\simeq \omega_{\vec{p}}$ for $\lambda=15$.}
	\label{figA1}
\end{figure}

%%%%%%%%%%%%%%%%%%%%%%%%%%%%%%%%%%%%%%%%

In Fig.~\ref{figA1.1} we show the behavior of the dimensionless quantity $|\omega_{\vec{p}}^2 G_\tau(w)|^2$ as a function of the dimensionless energy variable $w \equiv e/\omega_{\vec{p}}$ for different values of $\lambda \equiv \omega_{\vec{p}}\tau$, including $\lambda=\infty$ (i.e. $\tau=\infty$); our working approximation requires $\lambda$ to be sufficiently larger than one.  

We learn that the width of the two peaks increases (decreases) by increasing (decreasing) $1/\tau.$ An estimation of this \textit{effective width} can be obtained by expanding the free propagator around the two energy peaks $e\simeq \pm \omega_{\vec{p}}$. Doing so, we obtain
\begin{eqnarray}
G_\tau (e\simeq \pm\omega_{\vec{p}})\simeq  \frac{3}{4\omega_{\vec{p}}}\frac{\pm i a}{e\mp \omega_{\vec{p}}\pm i\Gamma_{\rm eff}/2}\equiv G^\pm_{\tau{\rm BW}}(e)\,,\qquad \Gamma_{\rm eff}\equiv \frac{6}{\tau}\,,
\label{BW-form-finite-free}
\end{eqnarray}
where $6/\tau$ provides a good estimation of the effective width of the free propagator~\eqref{free-propag-suitable}, i.e.~the width of $|G_\tau(e)|^2$ at its half maximum value. We have defined the Breit-Wigner-like functions $G^\pm_{\tau{\rm BW}}(e)$, whose expressions are valid only near their respective peaks. Note that the sign in front of $\Gamma_{\rm eff}$ is independent of $a$, i.e. it is the same for both ordinary and ghost propagators. See Footnote~\ref{foot-limit} for a discussion on the origin of the additional overall factor of $3/2$ in~\eqref{BW-form-finite-free}.

In Fig.~\ref{figA1.2}, we compare the moduli squared of $G_\tau(w)$, $G_{\tau{\rm BW}}(w)$, and $G_\tau^+(w)$ as functions of $w=e/\omega_{\vec{p}}$, focusing on the region around the positive-energy peak. We find that the Breit-Wigner-like expression~\eqref{BW-form-finite-free} provides a good approximation of $G^\pm_\tau(e)$ in the peak region. Moreover, the positive (negative) energy peak of the free propagator $G_\tau(e)$ is slightly shifted to the left (right) and enhanced due to interference effects between the positive- and negative-energy peaks.

\subsection{Suitable form of the dressed propagator}\label{sec:suitable-dressed}

The large-$\tau$ approximation used to derive a suitable form of the free propagator in a finite interval of time is also useful for obtaining a suitable expression for the dressed propagator~\eqref{fourier-dressed-propag-proj}. In this case as well, we assume that the time interval is sufficiently large, $\tau \gg 1/\omega_{\vec{p}}$, so as to isolate the relevant correction due to the finite value of $\tau$. Under this assumption, the dressed propagator can be recast as the product of the finite-time analogue of $2\pi \delta(e+e')$, i.e. the factor in~\eqref{finite-time-delta-analogue}, and a function depending on a single Fourier energy $e$. In particular, the free propagators appearing in~\eqref{fourier-dressed-propag-proj} are approximated as in~\eqref{free-propag-suitable}, while the one-loop self-energy as
\begin{eqnarray}
	\Sigma_\tau(e,e')\simeq 2\frac{\sin \big(\frac{e+e'}{2}\tau\big)}{e+e'} \Sigma(e^2)\,,
	\label{self-energy-factoriz}
\end{eqnarray}
where $\Sigma(e^2)$ is the standard infinite-$\tau$ expression, with its argument written in terms of $e^2$ instead of $-p^2=e^2-\vec{p}^2$. The internal factors $\frac{2\sin (\frac{e+e'}{2}\tau)}{e+e'}$ arising from the free propagators and the self-energies, and entering the vertices in~\eqref{fourier-dressed-propag-proj}, are approximated by their infinite-$\tau$ limit, i.e. $2\pi \delta(e+e')$, except for one of them, which is kept and factorizes in the final expression.

Hence, resumming the series~\eqref{fourier-dressed-propag-proj} within the large-$\tau$ approximation under consideration yields
\begin{eqnarray}
\bar{G}_\tau(e,e')\simeq 2\frac{\sin \big(\frac{e+e'}{2}\tau\big)}{e+e'} \bar{G}_\tau (e)\,,
\end{eqnarray}
where
\begin{eqnarray}
\bar{G}_\tau (e)\equiv G_\tau (e) \sum\limits_{n=0}^\infty \left[i\Sigma(e^2)G_\tau(e)\right]^n=\frac{G_\tau(e)}{1-iG_\tau(e)\Sigma(e^2)}= \frac{1}{G^{-1}_\tau(e)-i\Sigma(e^2)}\,,
\label{dressed-propag-suitable}
\end{eqnarray}
with $G_\tau(e)$ being the free propagator in~\eqref{free-propag-suitable}. 

In Sec.~\ref{sec:finite-time} of the main text, we discuss the key properties and physical content of~\eqref{dressed-propag-suitable} and present a detailed comparison between the ordinary and ghost cases.

%%%%%%%%%%%%%%%%%%%%%%%%%%%%%%%%%%%%%%%%%%%%%%%%%%%%%%%%%%%%%%%%%%

%----------------------------------------------------------------------------------------------------
\bibliographystyle{utphys}
\bibliography{References}

@article{Anselmi:2018kgz,
      author         = "Anselmi, Damiano",
      title          = "{Fakeons And Lee-Wick Models}",
      journal        = "JHEP",
      volume         = "02",
      year           = "2018",
      pages          = "141",
      doi            = "10.1007/JHEP02(2018)141",
      eprint         = "1801.00915",
      archivePrefix  = "arXiv",
      primaryClass   = "hep-th",
      SLACcitation   = "%%CITATION = ARXIV:1801.00915;%%"
}

@article{Anselmi:2018tmf,
      author         = "Anselmi, Damiano and Piva, Marco",
      title          = "{Quantum Gravity, Fakeons And Microcausality}",
      journal        = "JHEP",
      volume         = "11",
      year           = "2018",
      pages          = "021",
      doi            = "10.1007/JHEP11(2018)021",
      eprint         = "1806.03605",
      archivePrefix  = "arXiv",
      primaryClass   = "hep-th",
      SLACcitation   = "%%CITATION = ARXIV:1806.03605;%%"
}

@article{Anselmi:2018bra,
      author         = "Anselmi, Damiano",
      title          = "{Fakeons, Microcausality And The Classical Limit Of
                        Quantum Gravity}",
      journal        = "Class. Quant. Grav.",
      volume         = "36",
      year           = "2019",
      pages          = "065010",
      doi            = "10.1088/1361-6382/ab04c8",
      eprint         = "1809.05037",
      archivePrefix  = "arXiv",
      primaryClass   = "hep-th",
      SLACcitation   = "%%CITATION = ARXIV:1809.05037;%%"
}

@article{Pais:1950za,
      author         = "Pais, A. and Uhlenbeck, G. E.",
      title          = "{On Field theories with nonlocalized action}",
      journal        = "Phys. Rev.",
      volume         = "79",
      year           = "1950",
      pages          = "145-165",
      doi            = "10.1103/PhysRev.79.145",
      SLACcitation   = "%%CITATION = PHRVA,79,145;%%"
}

@article{Stelle:1976gc,
    author = "Stelle, K. S.",
    title = "{Renormalization of Higher Derivative Quantum Gravity}",
    reportNumber = "PRINT-76-1059 (BRANDEIS)",
    doi = "10.1103/PhysRevD.16.953",
    journal = "Phys. Rev. D",
    volume = "16",
    pages = "953--969",
    year = "1977"
}

@article{Lee:1969fy,
    author = "Lee, T. D. and Wick, G. C.",
    editor = "Feinberg, G.",
    title = "{Negative Metric and the Unitarity of the S Matrix}",
    doi = "10.1016/0550-3213(69)90098-4",
    journal = "Nucl. Phys. B",
    volume = "9",
    pages = "209--243",
    year = "1969"
}

@article{Cutkosky:1969fq,
    author = "Cutkosky, R. E. and Landshoff, P. V. and Olive, David I. and Polkinghorne, J. C.",
    title = "{A non-analytic S matrix}",
    doi = "10.1016/0550-3213(69)90169-2",
    journal = "Nucl. Phys. B",
    volume = "12",
    pages = "281--300",
    year = "1969"
}

@article{Bender:2007wu,
    author = "Bender, Carl M. and Mannheim, Philip D.",
    title = "{No-ghost theorem for the fourth-order derivative Pais-Uhlenbeck oscillator model}",
    eprint = "0706.0207",
    archivePrefix = "arXiv",
    primaryClass = "hep-th",
    reportNumber = "PREPRINT-LA-UR-07-3525",
    doi = "10.1103/PhysRevLett.100.110402",
    journal = "Phys. Rev. Lett.",
    volume = "100",
    pages = "110402",
    year = "2008"
}

@article{Grinstein:2008bg,
    author = "Grinstein, Benjamin and O'Connell, Donal and Wise, Mark B.",
    title = "{Causality as an emergent macroscopic phenomenon: The Lee-Wick O(N) model}",
    eprint = "0805.2156",
    archivePrefix = "arXiv",
    primaryClass = "hep-th",
    reportNumber = "CALT-68-2684, UCSD-PTH-08-03",
    doi = "10.1103/PhysRevD.79.105019",
    journal = "Phys. Rev. D",
    volume = "79",
    pages = "105019",
    year = "2009"
}

@article{tHooft:1973wag,
    author = "'t Hooft, Gerard and Veltman, M. J. G.",
    title = "{DIAGRAMMAR}",
    reportNumber = "CERN-73-09",
    doi = "10.1007/978-1-4684-2826-1_5",
    journal = "NATO Sci. Ser. B",
    volume = "4",
    pages = "177--322",
    year = "1974"
}

@article{Donoghue:2019ecz,
    author = "Donoghue, John F. and Menezes, Gabriel",
    title = "{Arrow of Causality and Quantum Gravity}",
    eprint = "1908.04170",
    archivePrefix = "arXiv",
    primaryClass = "hep-th",
    reportNumber = "ACFI-T19-09",
    doi = "10.1103/PhysRevLett.123.171601",
    journal = "Phys. Rev. Lett.",
    volume = "123",
    number = "17",
    pages = "171601",
    year = "2019"
}

@article{Salvio:2018crh,
    author = "Salvio, Alberto",
    title = "{Quadratic Gravity}",
    eprint = "1804.09944",
    archivePrefix = "arXiv",
    primaryClass = "hep-th",
    reportNumber = "CERN-TH-2018-099",
    doi = "10.3389/fphy.2018.00077",
    journal = "Front. in Phys.",
    volume = "6",
    pages = "77",
    year = "2018"
}

@article{Holdom:2021hlo,
    author = "Holdom, Bob",
    title = "{Ultra-Planckian scattering from a QFT for gravity}",
    eprint = "2107.01727",
    archivePrefix = "arXiv",
    primaryClass = "hep-th",
    doi = "10.1103/PhysRevD.105.046008",
    journal = "Phys. Rev. D",
    volume = "105",
    number = "4",
    pages = "046008",
    year = "2022"
}

@article{Avramidi:1985ki,
    author = "Avramidi, I. G. and Barvinsky, A. O.",
    title = "{ASYMPTOTIC FREEDOM IN HIGHER DERIVATIVE QUANTUM GRAVITY}",
    doi = "10.1016/0370-2693(85)90248-5",
    journal = "Phys. Lett. B",
    volume = "159",
    pages = "269--274",
    year = "1985"
}

@article{Holdom:2023usn,
    author = "Holdom, Bob",
    title = "{Running couplings and unitarity in a 4-derivative scalar field theory}",
    eprint = "2303.06723",
    archivePrefix = "arXiv",
    primaryClass = "hep-th",
    doi = "10.1016/j.physletb.2023.138023",
    journal = "Phys. Lett. B",
    volume = "843",
    pages = "138023",
    year = "2023"
}

@article{Lee:1970iw,
    author = "Lee, T. D. and Wick, G. C.",
    editor = "Feinberg, G.",
    title = "{Finite Theory of Quantum Electrodynamics}",
    doi = "10.1103/PhysRevD.2.1033",
    journal = "Phys. Rev. D",
    volume = "2",
    pages = "1033--1048",
    year = "1970"
}

@inproceedings{Lee:chicago,
    author = "Lee, T. D.",
    title = "{Relativistic complex pole model with indefinite metric}",
    booktitle = "Quanta: Essays in Theoretical Physics Dedicated to G. Wentzel (Chicago Univ. Press)",
    pages = "260-308",
    year = "1970"
}

@article{Tomboulis:1980bs,
    author = "Tomboulis, E.",
    title = "{Renormalizability and Asymptotic Freedom in Quantum Gravity}",
    reportNumber = "Print-80-0537 (PRINCETON)",
    doi = "10.1016/0370-2693(80)90550-X",
    journal = "Phys. Lett. B",
    volume = "97",
    pages = "77--80",
    year = "1980"
}

@article{Kubo:2024ysu,
    author = "Kubo, Jisuke and Kugo, Taichiro",
    title = "{Anti-Instability of Complex Ghost}",
    eprint = "2402.15956",
    archivePrefix = "arXiv",
    primaryClass = "hep-th",
    reportNumber = "YITP-24-20",
    doi = "10.1093/ptep/ptae053",
    journal = "PTEP",
    volume = "2024",
    number = "5",
    pages = "053B01",
    year = "2024"
}

@article{Holdom:2024onr,
    author = "Holdom, Bob",
    title = "{Making sense of ghosts}",
    eprint = "2408.04089",
    archivePrefix = "arXiv",
    primaryClass = "hep-th",
    doi = "10.1016/j.nuclphysb.2024.116696",
    journal = "Nucl. Phys. B",
    volume = "1008",
    pages = "116696",
    year = "2024"
}

@article{Buoninfante:2023ryt,
    author = "Buoninfante, Luca",
    title = "{Massless and partially massless limits in Quadratic Gravity}",
    eprint = "2308.11324",
    archivePrefix = "arXiv",
    primaryClass = "hep-th",
    reportNumber = "NORDITA 2023-038",
    doi = "10.1007/JHEP12(2023)111",
    journal = "JHEP",
    volume = "12",
    pages = "111",
    year = "2023"
}

@article{Kubo:2023lpz,
    author = "Kubo, Jisuke and Kugo, Taichiro",
    title = "{Unitarity violation in field theories of Lee\textendash{}Wick\textquoteright{}s complex ghost}",
    eprint = "2308.09006",
    archivePrefix = "arXiv",
    primaryClass = "hep-th",
    reportNumber = "YITP-23-104",
    doi = "10.1093/ptep/ptad143",
    journal = "PTEP",
    volume = "2023",
    number = "12",
    pages = "123B02",
    year = "2023"
}

@article{Donoghue:2021cza,
    author = "Donoghue, John F. and Menezes, Gabriel",
    title = "{On quadratic gravity}",
    eprint = "2112.01974",
    archivePrefix = "arXiv",
    primaryClass = "hep-th",
    reportNumber = "ACFI-T21-14",
    doi = "10.1393/ncc/i2022-22026-7",
    journal = "Nuovo Cim. C",
    volume = "45",
    number = "2",
    pages = "26",
    year = "2022"
}

@article{Veltman:1963th,
    author = "Veltman, M. J. G.",
    title = "{Unitarity and causality in a renormalizable field theory with unstable particles}",
    doi = "10.1016/S0031-8914(63)80277-3",
    journal = "Physica",
    volume = "29",
    pages = "186--207",
    year = "1963"
}

@inproceedings{Coleman:1969xz,
    author = "Coleman, S.",
    title = "{Acausality}",
    booktitle = "{7th International School of Subnuclear Physics (Ettore Majorana): Subnuclear Phenomena}",
    year = "1969"
}

@book{Brown:1992db,
    author = "Brown, L. S.",
    title = "{Quantum field theory}",
    isbn = "978-0-521-46946-3",
    publisher = "Cambridge University Press",
    month = "7",
    year = "1994"
}

@article{Anselmi:2020lfx,
    author = "Anselmi, Damiano",
    title = "{Dressed propagators, fakeon self-energy and peak uncertainty}",
    eprint = "2201.00832",
    archivePrefix = "arXiv",
    primaryClass = "hep-ph",
    doi = "10.1007/JHEP06(2022)058",
    journal = "JHEP",
    volume = "22",
    pages = "058",
    year = "2020"
}

@article{Holdom:2024cfq,
    author = "Holdom, Bob",
    title = "{UV-complete 4-derivative scalar field theory}",
    eprint = "2402.09223",
    archivePrefix = "arXiv",
    primaryClass = "hep-th",
    doi = "10.1016/j.nuclphysb.2024.116472",
    journal = "Nucl. Phys. B",
    volume = "1000",
    pages = "116472",
    year = "2024"
}

@article{Anselmi:2023wjx,
    author = "Anselmi, Damiano",
    title = "{Propagators and widths of physical and purely virtual particles in a finite interval of time}",
    eprint = "2304.07643",
    archivePrefix = "arXiv",
    primaryClass = "hep-ph",
    doi = "10.1007/JHEP07(2023)099",
    journal = "JHEP",
    volume = "07",
    pages = "099",
    year = "2023"
}

@article{Kuntz:2024rzu,
    author = "Kuntz, Jeffrey",
    title = "{Unitarity through PT symmetry in quantum quadratic gravity}",
    eprint = "2410.08278",
    archivePrefix = "arXiv",
    primaryClass = "hep-th",
    doi = "10.1088/1361-6382/adf606",
    journal = "Class. Quant. Grav.",
    volume = "42",
    number = "17",
    pages = "175003",
    year = "2025"
}

@article{Strumia:2017dvt,
    author = "Strumia, Alessandro",
    title = "{Interpretation of quantum mechanics with indefinite norm}",
    eprint = "1709.04925",
    archivePrefix = "arXiv",
    primaryClass = "quant-ph",
    reportNumber = "IFUP-TH-2016, CERN-TH-2017-191",
    doi = "10.3390/physics1010003",
    journal = "MDPI Physics",
    volume = "1",
    number = "1",
    pages = "17--32",
    year = "2019"
}

@article{Salvio:2015gsi,
    author = "Salvio, Alberto and Strumia, Alessandro",
    title = "{Quantum mechanics of 4-derivative theories}",
    eprint = "1512.01237",
    archivePrefix = "arXiv",
    primaryClass = "hep-th",
    reportNumber = "CERN-PH-TH-2015-290, IFT-UAM-CSIC-15-128",
    doi = "10.1140/epjc/s10052-016-4079-8",
    journal = "Eur. Phys. J. C",
    volume = "76",
    number = "4",
    pages = "227",
    year = "2016"
}

@article{Anselmi:2021hab,
    author = "Anselmi, Damiano",
    title = "{Diagrammar of physical and fake particles and spectral optical theorem}",
    eprint = "2109.06889",
    archivePrefix = "arXiv",
    primaryClass = "hep-th",
    doi = "10.1007/JHEP11(2021)030",
    journal = "JHEP",
    volume = "11",
    pages = "030",
    year = "2021"
}

@article{Buoninfante:2024yth,
    author = "Buoninfante, Luca and others",
    title = "{Visions in quantum gravity}",
    eprint = "2412.08696",
    archivePrefix = "arXiv",
    primaryClass = "hep-th",
journal={SciPost Phys. Comm. Rep.},
	pages={11},
	year={2025},
	publisher={SciPost},
    doi = "10.21468/SciPostPhysCommRep.11",
    year = "2024"
}

@article{Basile:2024oms,
    author = "Basile, Ivano and Buoninfante, Luca and Di Filippo, Francesco and Knorr, Benjamin and Platania, Alessia and Tokareva, Anna",
    title = "{Lectures in quantum gravity}",
    eprint = "2412.08690",
    archivePrefix = "arXiv",
    primaryClass = "hep-th",
    doi = "10.21468/SciPostPhysLectNotes.98",
    journal = "SciPost Phys. Lect. Notes",
    volume = "98",
    pages = "1",
    year = "2025"
}

@article{Buoninfante:2025klm,
    author = "Buoninfante, Luca",
    title = "{Remarks on ghost resonances}",
    eprint = "2501.04097",
    archivePrefix = "arXiv",
    primaryClass = "hep-th",
    doi = "10.1007/JHEP02(2025)186",
    journal = "JHEP",
    volume = "02",
    pages = "186",
    year = "2025"
}

@article{Salvio:2018kwh,
    author = {Salvio, Alberto and Strumia, Alessandro and Veerm\"ae, Hardi},
    title = "{New infra-red enhancements in 4-derivative gravity}",
    eprint = "1808.07883",
    archivePrefix = "arXiv",
    primaryClass = "hep-th",
    reportNumber = "CERN-TH-2018-195",
    doi = "10.1140/epjc/s10052-018-6311-1",
    journal = "Eur. Phys. J. C",
    volume = "78",
    number = "10",
    pages = "842",
    year = "2018"
}

@article{Buoninfante:2025dgy,
    author = "Buoninfante, Luca",
    title = "{Strict renormalizability as a paradigm for fundamental physics}",
    eprint = "2504.05900",
    archivePrefix = "arXiv",
    primaryClass = "hep-th",
    doi = "10.1007/JHEP07(2025)175",
    journal = "JHEP",
    volume = "07",
    pages = "175",
    year = "2025"
}

@article{Collins:2019ozc,
    author = "Collins, John",
    title = "{A new approach to the LSZ reduction formula}",
    eprint = "1904.10923",
    archivePrefix = "arXiv",
    primaryClass = "hep-ph",
    month = "4",
    year = "2019"
}

@article{Anselmi:2023phm,
    author = "Anselmi, Damiano",
    title = "{Quantum field theory of physical and purely virtual particles in a finite interval of time on a compact space manifold: diagrams, amplitudes and unitarity}",
    eprint = "2304.07642",
    archivePrefix = "arXiv",
    primaryClass = "hep-th",
    doi = "10.1007/JHEP07(2023)209",
    journal = "JHEP",
    volume = "07",
    pages = "209",
    year = "2023"
}

@article{Aoki:2025uff,
    author = "Aoki, Shuntaro and Strumia, Alessandro",
    title = "{Testing the arrow of time at the cosmo collider}",
    eprint = "2510.05204",
    archivePrefix = "arXiv",
    primaryClass = "hep-ph",
    reportNumber = "RIKEN-iTHEMS-Report-25",
    doi = "10.1088/1475-7516/2026/03/067",
    journal = "JCAP",
    volume = "03",
    pages = "067",
    year = "2026"
}

@article{Lehmann:1954rq,
    author = "Lehmann, H. and Symanzik, K. and Zimmermann, W.",
    title = "{On the formulation of quantized field theories}",
    doi = "10.1007/BF02731765",
    journal = "Nuovo Cim.",
    volume = "1",
    pages = "205--225",
    year = "1955"
}

@article{Oda:2025buc,
    author = "Oda, Ichiro",
    title = "{Manifestly covariant canonical formalism of quadratic gravity}",
    eprint = "2505.09149",
    archivePrefix = "arXiv",
    primaryClass = "hep-th",
    doi = "10.1088/1475-7516/2025/08/079",
    journal = "JCAP",
    volume = "08",
    pages = "079",
    year = "2025"
}

@article{Kumar:2026qnw,
    author = "Kumar, K. Sravan and Marto, Jo{\~a}o",
    title = "{Unitarity Quadratic Quantum Gravity in 4D}",
    eprint = "2604.19707",
    archivePrefix = "arXiv",
    primaryClass = "hep-th",
    month = "4",
    year = "2026"
}

@article{Buoninfante:2026mve,
    author = "Buoninfante, Luca",
    title = "{Asymptotic Quantum Dynamics of Ghost Fields}",
    eprint = "2605.29047",
    archivePrefix = "arXiv",
    primaryClass = "hep-th",
    month = "5",
    year = "2026"
}
%----------------------------------------------------------------------------------------------------

%%%%%%%%%%%%%%%%%%%%%%%%%%%

\end{document}